\documentclass[aps,prx,twocolumn,superscriptaddress,longbibliography]{revtex4-2}

\usepackage{amsmath}
\usepackage{amssymb}
\usepackage{amsfonts}
\usepackage{bm}
\usepackage{graphicx}
\usepackage{braket}
\usepackage{enumitem}
\usepackage{mathtools}

\usepackage[unicode=true,
 bookmarks=false,
 breaklinks=false,pdfborder={0 0 1},backref=false,colorlinks=true]
 {hyperref}
\hypersetup{
 linkcolor=blue,citecolor=blue,urlcolor=blue,linkcolor=blue,citecolor=blue,urlcolor=blue}
\usepackage{microtype}

\newcommand{\itp}{\affiliation{Institute for Theoretical Physics, University of Innsbruck, Innsbruck, 6020,  Austria}}
\newcommand{\iqoqi}{\affiliation{Institute for Quantum Optics and Quantum Information of the Austrian Academy of Sciences, Innsbruck, 6020, Austria}}
\newcommand{\iisc}{\affiliation{Department of Instrumentation and Applied Physics, Indian Institute of Science,  Bengaluru, 560012, India}}

\DeclareMathOperator{\tr}{\mathrm{Tr}}

\DeclareMathOperator{\Uen} {\mathcal{U}_{\rm en}}
\DeclareMathOperator{\Ude} {\mathcal{U}_{\rm de}}
\DeclareMathOperator{\Uphi} {\mathcal{U}(\bm{\phi})}
\DeclareMathOperator{\prior} {\mathcal{P}(\bm{\phi})}

\DeclareMathOperator{\nen} {n_{\rm en}}
\DeclareMathOperator{\nde} {n_{\rm de}}
\DeclareMathOperator{\psiin} {\psi_{\rm in}}
\DeclareMathOperator{\psiphi} {\psi_{\bm{\phi}}}

\usepackage{dcolumn}
\usepackage{enumitem}
\usepackage{mathtools}
\usepackage{color}
\usepackage{xcolor}
\usepackage{verbatim}
\usepackage{float}
\usepackage{bbold}

\newcommand{\abs}[1]{\lvert #1 \rvert}

\date{\today}

\begin{abstract}
We study multi-parameter sensing of 2D and 3D vector fields within the Bayesian framework for $SU(2)$ quantum interferometry. We establish a method to determine the optimal quantum sensor, which establishes the fundamental limit on the precision of simultaneously estimating multiple parameters with an $N$-atom sensor. Keeping current experimental platforms in mind, we present sensors that have limited entanglement capabilities, and yet, significantly outperform sensors that operate without entanglement and approach the optimal quantum sensor in terms of performance. Furthermore, we show how these sensors can be implemented on current programmable quantum sensors with variational quantum circuits by minimizing a metrological cost function. The resulting circuits prepare tailored entangled states and perform measurements in an appropriate entangled basis to realize the best possible quantum sensor given the native entangling resources available on a given sensor platform. Notable examples include a 2D and 3D quantum ``compass'' and a 2D sensor that provides a scalable improvement over unentangled sensors. Our results on optimal and variational multi-parameter quantum metrology are useful for advancing precision measurements in fundamental science and ensuring the stability of quantum computers, which can be achieved through the incorporation of optimal quantum sensors in a quantum feedback loop.
\end{abstract}

\begin{document}

\title{Optimal and Variational Multi-Parameter Quantum Metrology \\ and Vector Field Sensing}

\author{Raphael Kaubruegger}\itp \iqoqi
\author{Athreya Shankar} \itp \iqoqi \iisc
\author{Denis V. Vasilyev}\itp \iqoqi
\author{Peter Zoller} \itp \iqoqi

\maketitle
\section{Introduction}

The goal of quantum metrology~\cite{Holevo1982, Demkowicz2015, Degen2017, Pezze2018} is to enable and enhance measurements of parameters of physical systems using quantum systems that serve as probes. Quantum physics limits the performance of sensors employing uncorrelated (classical) input states, establishing the so-called standard quantum limit (SQL). At the same time, quantum physics provides us with entanglement as a resource to overcome the SQL and approach the ultimate limits for measurement precision in quantum sensing, which define the Optimal Quantum Sensor (OQS). Identifying and eventually building such OQS is one of the outstanding challenges in quantum metrology. This involves first of all the identification of optimal entangled input states and measurements for a given sensing task. Subsequently, the question arises how quantum sensors operating close to the limits defined by the OQS can be implemented on specific quantum sensor platforms given the experimental (entangling) resources. 

For {\em single-parameter} quantum metrology, 
there is a clear theoretical understanding of the OQS~\cite{Macieszczak2014,Demkowicz2015, Kaubruegger2021}, and recent experiments with programmable quantum sensors have demonstrated close to optimal $N$-atom Ramsey interferometry with low-depth entangling quantum circuits~\cite{Marciniak2021}. In contrast, we will be interested below in {\em multi-parameter} quantum metrology~\cite{Gessner2018, Demkowicz2020, Liu2020, Sidhu2020, Sidhu2021}, which deals with the precise estimation of several parameters \emph{simultaneously}, and is hence relevant for numerous practical applications~\cite{Humphreys2013,Gessner2020, Huang2021, Belliardo2022, Baamara2022, Conlon2023, Eriksson2023}, e.g.~vector field sensing~\cite{Vaneph2013, Baumgratz2016, Gorecki2022}, which is related to magnetometry~\cite{Budker2007, Budker2013, LeSage2013, Behbood2013, Lee2015, Thiele2018, Zheng2020,Wang2021} or electrometry~\cite{Fan2015, Brownutt2015, Sedlacek2013}.

\begin{figure*}
\includegraphics[width=\linewidth]{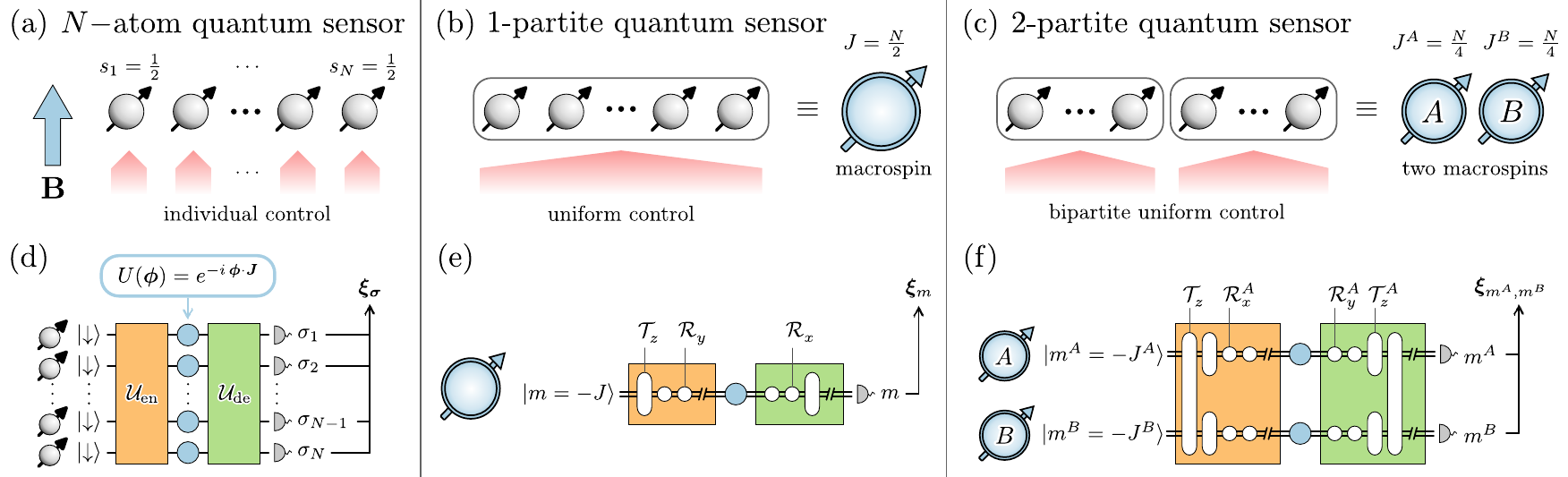}
\caption{\textbf{Quantum sensing of a vector field ${\bf B}$ using $\mathbf{SU(2)}$ interferometry.} We consider Ramsey interferometry with entangled atoms, where the vector of phases $\bm{\phi}$ are encoded in the quantum system via the $SU(2)$ unitary $U(\bm \phi)$ that carries information about the vector field $\textbf{B}$
(see Sec.~\ref{sec:SU2}). Panels (a,b,c) define three interferometers, discussed in more detail in Secs.~\ref{sec:CircuitModel} and \ref{sec:LimitedControl}, which differ in the level of control available for manipulating atomic entanglement in an experimental setup.
Panels (d,e,f) provide operational definitions of the interferometers in terms of quantum circuits and projective measurements. The entangler $\Uen$ and decoder $\Ude$ transform the quantum state of the system before and after the phase encoding. The projective measurement outcomes $\bm{\sigma}$, $m$, $m^{A(B)}$ are converted into an estimate $\bm{\xi}$ of the true phase vector $\bm{\phi}$. (a, d): The most general $N$-atom sensor assumes universal control of atoms, which allows to realize the optimal quantum sensor (see section ~\ref{subsec:OQS}). 
(b, e): The atoms in the 1-partite quantum sensor (see Sec.~\ref{subsec:1p-QS}) are controlled uniformly. The sensor dynamics are thus constrained to the permutation symmetric subspace of $N$ atoms, i.e., the atomic ensemble can be represented by a single macrospin with angular momentum $J=N/2$. 1-partite control is realized by circuits composed of collective rotation $\mathcal{R}_{x(y)}$ and the one-axis twisting (OAT) gate $\mathcal{T}$.  (c, d): The atoms in the 2-partite quantum sensor (see Sec.~\ref{subsec:2p-QS}) are grouped into two partitions $A$ and $B$ and controlled uniformly within each partition, which includes uniform control of the entire atomic ensemble. The system is thus equivalent to two entangled macrospins with angular momenta $J^A=J^B=N/4$. 2-partite control is realized by circuits composed of collective  rotations $\mathcal{R}^{A(B)}_{x(y)}$ and  OAT $\mathcal{T}^{A(B)}_z$ acting on each partition, and  OAT $\mathcal{T}_z$ on the full ensemble.}
\label{fig:Sensors} 
\end{figure*}

A fundamental difference between multi-parameter and single-parameter metrology arises from the potential incompatibility of optimal measurements for different parameters encoded by non-commuting Hamiltonians. This distinctive feature of multi-parameter quantum metrology has been studied primarily within the framework of quantum Fisher information, where the sensor precision is tightly lower bounded by the Holevo Cram\`er-Rao bound~\cite{Holevo1982, Albarelli2019, Demkowicz2020} as the number of measurement repetitions of the same signal tends to infinity. Instead, we are interested here in measurement scenarios that are limited to one or a few measurement repetitions, as is relevant, e.g., for sensing time-varying signals~\cite{Tsang2011}. 
The problem of finding a sensor that maximizes the precision of a single measurement in a given range of parameter values can be formulated in the framework of Bayesian quantum metrology~\cite{Holevo1982, Demkowicz2020}. However, solving Bayesian multi-parameter quantum metrology for non-commuting Hamiltonians remains a longstanding challenge. So far, only a handful of special problems with an underlying group symmetry have been addressed in this framework~\cite{Bagan2001, Chiribella2004a, Bagan2004, Kahn2007}.

In the present work, we study the paradigmatic multi-parameter quantum metrology problem of vector field sensing~\cite{Vaneph2013,Baumgratz2016,Gorecki2022} within the Bayesian framework.
While entangled field sensors for 1D sensing have already been experimentally realized~\cite{Wasilewski2010, Sewell2012, Ockeloen2013, Arias2019}, we consider the open problem of sensing 2D and 3D fields in the context of entangled Ramsey interferometry with $N$ atoms. We establish a method to determine the OQS, that sets the ultimate achievable precision limit for simultaneous estimation of the components of the vector field from a single measurement with a given number of atoms. The latter is of particular importance when the number of atoms cannot be arbitrarily increased, e.g., to maintain a high spatial resolution. Keeping current experimental capabilities in mind, we consider sensors with limited experimental quantum control that nevertheless approach the OQS in terms of performance and significantly outperform the SQL, defined by sensors operating without entanglement. Programmable sensors can be realized on a variety of sensing platforms, including trapped ions~\cite{Marciniak2021, Conlon2023}, laser-excited Rydberg atoms~\cite{Facon2016, Omran2019}, atoms in optical cavities~\cite{Robinson2022}, polar molecules~\cite{Carr2009, Loic2019}, color centers~\cite{Bermudez2011, Yao2012} and photons with orbital angular momentum~\cite{Bouchard2017,Brandt2020,Eriksson2023}. Considering the example of one-axis twisting (OAT) interactions~\cite{Kitagawa1993}, which is available e.g.~for trapped ions, we optimize variational quantum circuits~\cite{Cerezo2021} that point to a practical route to implement entangled quantum sensors~\cite{Marciniak2021}. 

This work is organized as follows. In Sec.~\ref{sec:BayesianMDFS} we introduce $SU(2)$-interferometry to measure the components of a vector field, and we set the stage for OQS's in the framework of Bayesian multi-parameter estimation. Section~\ref{sec:OQS} discusses theoretical aspects and provides (numerical) algorithms for finding the OQS, and defines quantum circuit models. In Sec.~\ref{sec:LimitedControl} we introduce one- and two-partite quantum sensors as quantum sensors with limited  entangling capabilities. This leads to the discussion of variational low-depth quantum circuits approximating the optimal quantum sensor in Sec.~\ref{sec:VariationalCircuits}. While most of the results presented in this work are for 2D field sensing, Sec.~\ref{sec:3D} discusses aspects of 3D field sensing, followed by conclusions. 

\section{Bayesian multi-dimensional field sensing}
\label{sec:BayesianMDFS}

In this section, we review the theory of multi-parameter estimation using multi-dimensional field sensing as an example. We focus on the Bayesian framework of phase estimation, which accounts for prior uncertainty about the parameters being measured. 
We define the OQS as the sensor that minimizes a Bayesian cost function. 

\subsection{Vector field sensing with $SU(2)$-interferometers}
\label{sec:SU2}

We study the problem of estimating all components of a vector field $\mathbf{B}=(\text{B}_x, \text{B}_y, \text{B}_z)^T$ in $SU(2)$ interferometry \footnote{We refer to an interferometer with a unitary phase encoding that belongs to the group of SU(2) as an SU(2) interferometer.}, where we assume that the field is uniformly coupled to an ensemble of $N$ identical two-level atoms. 
The physical sensing scenarios we have in mind can be, e.g., DC magnetic fields or AC magnetic and electric fields. In Appendix~\ref{app:vfs_implement}, we elaborate on exemplary experimental settings that realize the model system consider below.

In our model, the atoms couple to the vector field
according to the Hamiltonian 
\begin{align}
H=\mathbf{B}\cdot\bm{J}\equiv\sum_{\nu}\text{B}_{\nu}J_{\nu}. 
    \label{eq:H}
\end{align}
Here, $J_{\nu}=\sum_{k=1}^{N}\sigma_{k}^{\nu}/2$ are the collective spin operators and $\sigma_{k}^{\nu}$ are the Pauli operators of the $k$-th atoms with $\nu=x,y,z$. 
After an evolution time $T$, the phases $\bm{\phi}=\mathbf{B}\,T$ are imprinted on the input state $\ket{\psi_{{\rm in}}}$ by the unitary $SU(2)$ operator 
\begin{align}
    U(\bm{\phi})=\exp\left[-i\,\bm{\phi}\cdot\bm{J}\right]
    \label{eq:PhaseEncoding}
\end{align}
such that $\ket{\psi_{\bm{\phi}}}=U(\bm{\phi})\ket{\psi_{{\rm in}}}$. Here we assume that the coherence time of the field $\tau \gg T$ is long compared to $T$, i.e.~the field remains effectively constant during the evolution. 
Finally, a measurement of the two-level atoms in the state~$\ket{\psi_{\bm{\phi}}}$ is performed to estimate the phases $\bm{\phi}$ and thus infer the field $\textbf{B}$. While we focus here on measurements at discrete times, an alternative approach based on continuous measurements is discussed in~\cite{Yang2022}.

The most general measurement is described by a positive operator-valued measure (POVM), i.e., a set $\{M_{\mu}\}$ of positive Hermitian operators, $M_{\mu}\succeq0$, such that $\sum_{\mu}\,M_{\mu}=\openone$. In the following, we contrast POVMs with the less general but experimentally more accessible projective measurements, denoted by $\Pi_\mu$. For a given vector of phases $\bm{\phi}$, the measurement outcomes, labeled $\mu$, are realized with probabilities 
\begin{align}
	\label{eq:conditional_probability}
    p(\mu|\bm{\phi})=\mathrm{Tr}\{M_{\mu}\ket{\psi_{\bm{\phi}}}\bra{\psi_{\bm{\phi}}}\}.
\end{align}
Based on the result~$\mu$, one estimates the phases using an estimator $\bm{\xi}_{\mu}=(\xi_{\mu}^{x},\xi_{\mu}^{y},\xi_{\mu}^{z})^T$. The deviation of the estimation from the actual value $\bm\phi$ is characterized by an error~(or loss) function $\epsilon(\bm{\phi}, \bm{\xi}_{\mu})$, commonly chosen to be the squared error 
\begin{equation}
\label{eq:error_function}
\epsilon(\bm{\phi}, \bm{\xi}_{\mu}) = \left(\bm{\phi}-\bm{\xi}_{\mu}\right)\cdot \left(\bm{\phi}-\bm{\xi}_{\mu}\right).
\end{equation}
Finally, the figure of merit, which accounts for measurement statistics, is given by the average estimation error~(or risk) for a given $\bm{\phi}$. In our case, it takes the form of the mean squared error (MSE)
\begin{equation}
    \text{MSE}(\bm{\phi})=\sum_{\mu}\epsilon(\bm{\phi}, \bm{\xi}_{\mu})\, p(\mu|\bm{\phi}).
\label{eq:MSE_POVM}
\end{equation}

Note that the $SU(2)$ phase encoding $U(\bm{\phi})$ is periodic which means that $\abs{\bm{\phi}}$ can only be estimated modulo $2\pi$. Since we want to infer the vector field from the estimated phases according to ${\textbf{B}^{\rm est} \sim \bm{\xi}/ T}$, it is crucial to take this ambiguity into account. Therefore, we consider phases in the interval ${-\infty < \bm{\abs{\phi}} < \infty}$, and we explicitly choose the squared error~\eqref{eq:error_function}, which is a non-periodic error function to strongly penalize phases that slip outside the $d$-dimensional sphere $|\bm{\phi}|\leq \pi$ of unambiguous phase estimation.

\subsection{Bayesian multi-parameter estimation}
\label{sec:BMPE}

In the Bayesian approach, the parameter vector $\bm{\phi}$ to be estimated is assumed to be a random variable. Knowledge about the phases to be estimated is represented by the prior probability density ${\cal P}(\bm{\phi})$. By estimating the phases, we infer information, and the prior density is updated to the posterior probability density. 

The quantity central to our discussion is the Bayes Mean Squared Error (BMSE), defined as the MSE~\eqref{eq:MSE_POVM} averaged over the prior probability density,
 \begin{align}
    \text{BMSE} & \equiv {\cal C} = \int d\bm{\phi}\, \text{MSE}(\bm{\phi})\, \mathcal{P}(\bm{\phi}).
\label{eq:BMSE_POVM}
\end{align}
Below, the BMSE will play the role of a cost function ${\cal C}$, which depends on the input state $\ket{\psiin}$, POVMs $\{M_{\mu}\}$, and estimators $\{\bm{\xi}_{\mu}\}$. 

We use Bayes theorem ${p(\bm{\phi}|\mu) = p(\mu|\bm{\phi}) \mathcal{P}(\bm{\phi}) / p(\mu)}$ to interpret $\mathcal{C}$ as the posterior expected error of the phase estimation. In particular, $\mathcal{C}$ can be rewritten as 
\begin{align} 
    \mathcal{C} = \sum_{\mu} p(\mu)\, \int d\bm{\phi} \, \epsilon(\bm{\phi}, \bm{\xi}_{\mu})\, p(\bm{\phi}|\mu), 
\end{align}
where $p(\mu) = \int d\bm{\phi}\, p(\mu| \bm{\phi}) \mathcal{P}(\bm{\phi})$ is the probability of measuring $\mu$. The integral over the phase vector $\bm{\phi}$ represents the expected squared error, i.e., the squared error averaged with respect to the posterior probability density $p(\bm{\phi}|\mu)$, where the latter represents our knowledge of $\bm{\phi}$ given that the outcome of the measurement is $\mu$. 

The minimum MSE estimator, defined by minimizing the expected squared error and thus the BMSE, is given by 
\begin{align}
\label{eq:MMSEe}
	\bm{\xi}_{\mu}^{*} = \int d\bm{\phi}\,\bm{\phi}\, p(\bm{\phi}|\mu) = \langle \bm{\phi} \rangle _{p(\bm{\phi}|\mu)},
\end{align}
which corresponds to the mean value of the posterior probability density $p(\bm{\phi}|\mu)$.

If the estimators $\{\bm{\xi}_\mu\}$ are chosen to be the minimum MSE estimators $\{\bm{\xi}_\mu^*\}$, the BMSE is the posterior variance 
\begin{align}
    \Delta^{2}&\equiv\left.{\cal C}\right|_{\bm{\xi}=\bm{\xi}^{*}} = \sum_{\mu }p(\mu)\,\Big\langle \epsilon\left(\bm{\phi}, \langle\bm{\phi}\rangle_{p(\bm{\phi}|\mu)}\right) \Big\rangle_{p(\bm{\phi}|\mu)},
\end{align}
i.e.~the variances of the posterior densities averaged according to the probability $p(\mu)$ to measure $\mu$. 

\begin{figure}
\includegraphics[]{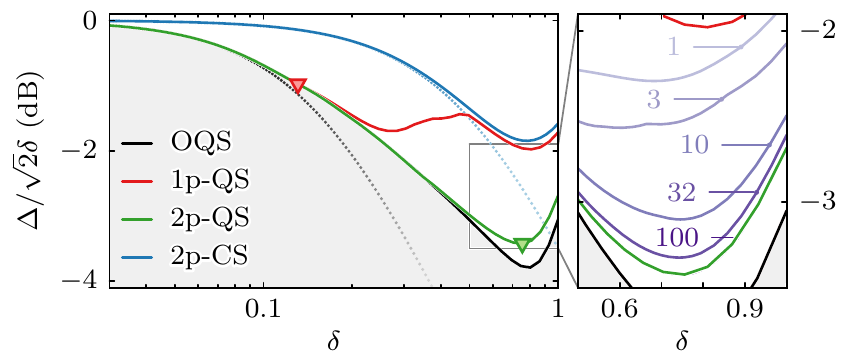}
\caption{
\textbf{Bayesian quantum sensing of a 2D field.} (a) Ratio of the posterior width $\Delta$ to the prior width $\sqrt{2}\delta$ vs. $\delta$ for $N = 8 $ atoms. The OQS  sets the ultimate performance limit; the gray-shaded region is unattainable for any quantum sensor with $N$ atoms. The optimal 1-partite quantum sensor, the 2-partite quantum sensor, and the 2-partite classical sensor are shown for comparison. The asymptotic bounds for the classical (blue dotted line) and quantum (black dotted line) sensors are predicted by the Bayesian Cram\'er-Rao bounds ~\eqref{eq:BayesianQL} and \eqref{eq:BayesianSQL}. In Figs.~\ref{fig:1p-QS} and \ref{fig:2p-QS} we illustrate the working of the optimal sensors indicated here by the colored triangles. Magnified panel: Performance of variational circuits (purple lines) defined in Sec.~\ref{sec:VariationalCircuits} for an increasing number of decoding layers $\nde$, and a fixed entangler depth of $\nen = 2$.}
\label{fig:2DCost}
\end{figure}

In the following, the posterior uncertainty $\Delta^2$ will be the quantity of interest that evaluates performance and allows comparison of various quantum sensors for a given prior density $\mathcal{P}(\bm{\phi})$. By minimizing the metrological cost function~\eqref{eq:BMSE_POVM}, we implicitly minimize the posterior variance of the sensor. Therefore, our results will be presented as plots, where we compare the ratio between posterior uncertainty $\Delta$ and prior uncertainty for the different sensor models (see Figs.~\ref{fig:2DCost}, \ref{fig:2DScaling},
\ref{fig:3DCost}). The more information we can gain about the phases $\bm{\phi}$ in a single measurement, the smaller the value of this ratio.

Before proceeding, we comment on our choice of the prior probability density $\mathcal{P}(\bm{\phi})$. In Bayesian quantum metrology, the $\textbf{B}$ field to be sensed has a prior density which translates into a density for the phases $\mathbf{\phi}$. To give a specific example, we assume below a prior for the field in the form of a Gaussian density centered around the origin with prior width $({\delta {\abs{\mathbf{B}}}})$. At the same time, there is no knowledge of the orientation of the field. The corresponding prior probability density for the phases is thus
\begin{eqnarray}
\mathcal{P}_{\delta}(\bm{\phi}) = \frac{1}{\left(\sqrt{2 \pi}\, \delta\right) ^d } \exp \left[-\frac{\bm{\phi}\cdot \bm{\phi}}{2\delta^2}\right], 
\label{eq:GaussianPrior}
\end{eqnarray}
where $d$ is the dimension of the vector $\bm{\phi}$. For 2D field sensing we take $d=2$ and assume that $\text{B}_z = {\phi}_z = 0$. In the present example, the prior is thus parameterized by a single parameter $\delta\sim (\delta {\abs{\mathbf{B}}})\,T$.  However, the following discussion can be adapted (see Appendix~\ref{app:PriorDensities}) to more general priors motivated by particular quantum sensor settings and tasks. As noted before, in mapping from fields to phases via $\bm{\phi}=\mathbf{B}\,T$, we must ensure with a proper choice of $T$, that there is a one-to-one map of estimated phases to the vector field, i.e.~there are no phase slips. Implicit in our discussion is the assumption that the field $\mathbf{B}$ to be measured is essentially static during time $T$.

\subsection{Optimal quantum sensors}
\label{subsec:OQS}
We define the OQS as the minimum of the metrological cost function BMSE~\eqref{eq:BMSE_POVM} over all input states $\ket{\psi_{{\rm in}}}$, measurements $\{M_{\mu}\}$, and estimators $\{\bm{\xi}_{\mu}\}$  for a given prior density $\mathcal{P}_\delta(\bm{\phi})$. Thus, the posterior variance of the OQS,
\begin{equation}
\Delta^2_{\rm OQS}=\min_{\ket{\psi_{{\rm in}}}, M_{\mu}, \bm{\xi}_{\mu}} \mathcal{C}, 
\label{eq:Delta2_OQS}
\end{equation}
represents the fundamental limit on the posterior variance achievable by any $N$-atom sensor for a given prior $\mathcal{P}_\delta(\bm{\phi})$.

As a first result, in Sec.~\ref{sec:OQS} we will develop a method that allows us to efficiently find the OQS for vector field sensing, i.e. to perform a numerical optimization that scales only polynomially with the number of particles $N$. 

While the OQS is defined via unconstrained optimization over (entangled) input states and measurements, the practical realization of the `best possible' quantum sensors will be constrained by limited experimental quantum resources and imperfections. This will lead us to discuss optimal quantum sensors with limited (experimental) entangling capabilities, i.e. identifying the best possible quantum sensors given experimental constraints. Similar to the OQS, the performance of these sensors is characterized by the minimum posterior variance $\Delta^2$ defined as in Eq.~\eqref{eq:Delta2_OQS}, but evaluated by imposing relevant constraints on the input states and measurements.

Figure~\ref{fig:Sensors} gives an overview of quantum sensors with different levels of control and entanglement capabilities, whose performance we compare in Fig.~\ref{fig:2DCost}. An operational definition of these quantum sensors is given in terms of a quantum circuit model, see Sec.~\ref{sec:CircuitModel}. This includes in Fig.~\ref{fig:Sensors}(a) a quantum sensor with universal ($N$-partite) control of the $N$-particle quantum system, which subsumes the OQS. We contrast this in Fig.~\ref{fig:Sensors}(b) to a 1-partite quantum sensor (1p-QS) defined by uniform control of the $N$ atoms and in Fig.~\ref{fig:Sensors}(c) to a 2-partite quantum sensor (2p-QS) with uniform control on the level of two partitions of the atomic ensemble (for details see Secs.~\ref{sec:1pQS} and~\ref{sec:2pQS}, respectively). We will see below that the optimal 2p-QS operates close to the OQS for 2D field sensing independent of the width of the prior density, and this will motivate our search for variational approximations to the relevant quantum circuits.

Finally, we comment on the relationship between the Bayesian and Fisher information approaches to multi-parameter quantum metrology~\cite{Demkowicz2020} in relation to the present work, see Appendix~\ref{app:Fisher} for a more detailed discussion. The metrological cost function in the latter approach is the MSE of a locally unbiased estimator, which is lower bounded by the Cram\'er-Rao bound. In other words, the sensitivity is considered {\em locally} around a given value $\bm{\phi}_0$ of the phase vector, which is in contrast to the BMSE, which averages the MSE over a {\em range} of phase values. However, in the limiting case of small prior widths $\sqrt{d}\delta$, when the Bayesian approach becomes {\it local}, the two approaches can be connected via the van Trees inequality \eqref{eq:VanTrees}. In the Fisher information framework, the sensitivity of any $N$-atom multi-dimensional field sensor is lower bounded by a Heisenberg-like limit, which in turn limits the BMSE through the van Trees inequality; see Appendix~\ref{app:Fisher}. For the examples considered in the following, the OQS saturates these bounds~\cite{Kolenderski2008, Vaneph2013, Gorecki2022} in the small~$\delta$ limit, which highlights that the OQS is also optimal from the Fisher information point of view.

\section{Theory of the optimal quantum sensor}
\label{sec:OQS}

In this section, we solve the mathematical problem of minimizing the BMSE~\eqref{eq:BMSE_POVM} as the metrological cost function, which defines the OQS and underlies the numerical results for the OQS presented in Figs.~\ref{fig:2DCost} and~\ref{fig:2DScaling}. We conclude the section by introducing a quantum circuit model that describes practical realizations of (optimal) quantum sensors.

\subsection{Identifying the optimal quantum sensor}\label{subsec:identify_OQS}

To find the OQS, one minimizes the metrological cost function~\eqref{eq:BMSE_POVM} over an unconstrained set of input states, measurements, and estimators, as in Eq.~\eqref{eq:Delta2_OQS}. In general, solving such a mathematical problem is difficult, given the exponentially large number of degrees of freedom associated with a quantum system of $N$ particles, the corresponding measurements, and estimators. Here, we show that optimal Bayesian solutions can be found using computational resources that scale only polynomially in the size of the system $N$. Our approach is based on three key observations presented in the following and further detailed in the Appendices~\ref{subsec:Irreducible_Hilbert_space}--\ref{subsec:covariant_POVM}.

First, we note that the dynamics of a sensor comprising $N$ atoms, which is uniformly coupled to the field $\mathbf{B}$ according to Eq.~\eqref{eq:H}, is restricted to an effective Hilbert space with dimension scaling as~$\sim N^3$. This is in contrast to the full $2^N$-dimensional Hilbert space describing $N$ spin-1/2 atoms. The simplification originates in the $SU(2)$ and particle permutation symmetry of the unitaries $U(\bm\phi)$ that encode the phases. The idea is to consider the $N$-qubit space using irreducible representations of the $SU(2)$ group. For example, by adding angular momenta one can represent two spin-1/2's as a direct sum of two orthogonal subspaces corresponding to a spin-1 (triplet) and a spin-0 (scalar), $\bm{\frac12}\otimes\bm{\frac12}=\bm1\oplus\bm0$. The addition of more than 2 angular momenta leads to multiple orthogonal subspaces, or equivalent representations, corresponding to the same spin-$j$, e.g., for three qubits we have 2 equivalent spin-1/2 representations, $\bm{\frac12}^{\otimes 3}=\bm{\frac32}\oplus\bm{\frac12}\oplus\bm{\frac12}$. The number of equivalent representations $\ell_j$ of spin-$j$ grows exponentially with the number of added qubits $N$ and thus exceeds the number of degrees of freedom of a single spin-$j$ subspace, $d_j=2j+1$. Since the $SU(2)$ unitaries do not couple different angular momenta $j$ and their equivalent representations, we can utilize not more than $d_j\leq\ell_j$ equivalent representations by entangling them. The other $\ell_j-d_j$ equivalent representations remain unpopulated and cannot be used for sensing, thus reducing the effective Hilbert space dimension of a $SU(2)$ sensor to $\sim N^3$~\cite{Bagan2004, Chiribella2004a, Chiribella2004b}. We present details in Appendix~\ref{subsec:Irreducible_Hilbert_space}.

The second observation is the multi-convex nature of the sensor optimization problem. As we show in Appendix~\ref{subsec:Minimization-of-the-Cost}, this means that minimization of the metrological cost Eq.~\eqref{eq:BMSE_POVM} is a convex optimization problem~\footnote{Convexity of optimization problem guarantees that every local minimum is a global minimum.} with respect to each of the three sets ($\ket{\psi_{{\rm in}}}$,
$\{M_{\mu}\}$, and $\{\bm{\xi}_{\mu}\}$) of variables, if the other two are fixed. The hardest of the corresponding subproblems is the optimization over POVMs for fixed input state and estimators. It can be recast as a semidefinite program, which is solvable in polynomial time for our problem of polynomial size in $N$~\footnote{The upper bound on the POVM complexity for $d_N\sim N^3$ dimensional Hilbert space is given by a POVM comprising $d_N^2\sim N^6$ measurement operators of full rank. We find the OQS solution to be significantly simpler than the upper bound.}, as discussed above.

The optimization algorithm is summarized as follows: Starting from a random initial state, POVM, and estimators, one iteratively solves the convex subproblems until the solution converges to the optimal sensor (with $\ket{\psi_{\rm in}^{\star}}$, $\{M^{\star}_\mu\}$, and $\{\bm{\xi}_{\mu}^{\star}\}$) that minimizes the metrological cost~\eqref{eq:BMSE_POVM}. We present formal definitions of the corresponding convex subproblems in Appendix~\ref{subsec:Minimization-of-the-Cost}.

Finally, on the more technical side, we observe that a reliable convergence to the optimum solution can be achieved when symmetric degrees of freedom are eliminated from the search space. To this end, we have identified a covariant ansatz for input states, generalized measurements, and estimators, which describes the optimal sensor in the case of {\em a priori} unknown field direction as reflected in the isotropic prior density, Eq.~\eqref{eq:GaussianPrior}, that we consider. The covariant ansatz is detailed in Appendix~\ref{subsec:covariant_POVM} for the optimal 2-partite 2D field sensor. The general case of the OQS for 2D and 3D field sensing will be considered in~\cite{Vasilyev23}.

\subsection{Quantum circuit model of a quantum sensor}
\label{sec:CircuitModel}

Practical realizations of a quantum sensor will rely on unitary quantum circuits that manipulate the quantum state of the sensor, and projective measurements performed on the atoms, as illustrated in Fig.~\ref{fig:Sensors}. In our quantum circuit model of a sensor, the initial state 
\begin{align}
    \ket{\psiin} = \Uen \, \ket{\psi_{0}}
\end{align}
is prepared by an entangling unitary $\Uen$ from a product state, e.g., $\ket{\psi_0} = \ket{\downarrow}^{\otimes N}$. On the other hand, the decoding unitary $\Ude$ can be viewed as transforming the projective measurement basis given by $\ket{\bm{\sigma}}=\ket{\sigma_1, \cdots, \sigma_N}$, where each atom $j$ is projected into $\sigma_j=\ket{\uparrow}$ or $\sigma_j=\ket{\downarrow}$. Therefore, the conditional probability $p(\bm{\sigma}|\bm{\phi}) = \tr\{\Pi_{\bm{\sigma}} \ket{\psiphi}\bra{\psiphi}\}$ to obtain the measurement outcome $\bm{\sigma}$ can be expressed in terms of the effective measurement projectors
\begin{align}
    \Pi_{\bm{\sigma}} = \Ude^{\dagger} \ket{\bm{\sigma}}\bra{\bm{\sigma}} \Ude.
\end{align}

A general POVM underlying the discussion of OQS can be implemented in the circuit model with additional ancillary atoms that are not involved in the actual sensing process. Interestingly, we observe that, for sufficiently large $N$, the optimal POVM acting on the OQS subspace of polynomial dimension, $\sim N^3$, can be realized as projective measurements on the full $N$-atom space. The $N$-atom Hilbert space has an exponential number of degrees of freedom, $2^N$, which can be used instead of the ancillary atoms. This suggests that the OQS can be faithfully represented as a quantum circuit if we assume universal quantum control at the level of individual atoms, as shown in Fig.~\ref{fig:Sensors}(a).
Our numerical results suggest that the OQS for 2D fields can be realized without using ancillary atoms for sensors comprising $N\gtrsim8$ atoms.

\section{Optimal few-partite quantum sensors}
\label{sec:LimitedControl}

In the previous section, we discussed general aspects of the OQS for multi-dimensional field sensing in $SU(2)$-interferometry with $N$ atoms. 
In the present section, we will introduce and discuss few-partite quantum sensors, and compare their performance to the OQS. These sensors are defined by partitioning $N$ atoms representing the sensor into subensembles. In particular, we consider the 1p-QS and the 2p-QS indicated in Figs.~\ref{fig:Sensors}(b,e) and (c, f), respectively. 
For the 1p-QS, we assume that the atomic ensemble is uniformly controlled, while the 2p-QS assumes uniform control at the level of partitions $A$ and $B$.

The motivation for introducing few-partite quantum sensors is two-fold: First, we observe that the wave function of the OQS does not always occupy the full Hilbert space of an $N$-atom $SU(2)$ interferometer, see Appendix~\ref{appendix:1p_and_2p_subspaces}. The specific structure of the occupied subspace suggests that the OQS for 2D field sensing can be approximated by few-partite quantum sensors.
Second, few-partite quantum sensors require limited quantum control to implement entangling and decoding unitaries in our quantum circuit model [see Fig.~\ref{fig:Sensors}(e,f)]. We will discuss such implementations in Sec.~\ref{sec:VariationalCircuits}, where we study variational approximations to the corresponding quantum circuits which can be realized with current experimental techniques.

\subsection{Few-partite quantum sensors}

\subsubsection{One-partite quantum sensor (1p-QS)}
\label{subsec:1p-QS}

The 1p-QS and the corresponding quantum circuit model are sketched in Figs.~\ref{fig:Sensors}(b, e), respectively. We assume global control so that the dynamics of the 1p-QS, i.e. the action of the 1p-QS unitaries $\mathcal{U}_{\rm en(de)}^{\rm 1p}$ and the phase encoding $U(\bm{\phi})$ on $\ket{\psi_0}$, are confined to the permutation invariant subspace $\mathcal{H}_{\text{1p}}= \mathbb{C}^{N + 1}$ of the $N$-atom Hilbert space $(\mathbb{C}^2)^{\otimes N}$. Therefore, the system is equivalent to a single `macrospin' with angular momentum $J = N /2$~\cite{Chalopin2018}. The eigenstates $\ket{m}$ of $J_z$, which satisfy $J_z \ket{m} = m \ket{m}$ with $m = -J,-J+1,\dots,J$, form a basis of the 1p-QS Hilbert space. 

In this basis, the initial state and the measurement projectors in Fig.~\ref{fig:Sensors}(e) are defined as 
\begin{align}
    \ket{\psiin} &= \mathcal{U}_{\rm en}^{\text{1p}} \ket{m = - J}, 
    \label{eq:1pQS_psi}\\
    \Pi_m &= \left. \mathcal{U}_{\rm de}^{\text{1p}}\right.^{\dagger}\ket{m }\bra{m} \mathcal{U}_{\rm de}^{\text{1p}}.
    \label{eq:1pQS_pi}
\end{align}
The optimal 1p-QS is characterized by the posterior variance
\begin{equation}
\Delta^2_{\text{1p-QS}}=\min_{\mathcal{U}_{\rm en}^{\text{1p}},\, \mathcal{U}_{\rm de}^{\text{1p}},\, \bm{\xi}_{m}} \mathcal{C}.
\label{eq:PosteriorVariance1pQS}
\end{equation}
Details of the optimization are presented in Appendix~\ref{app:projective_optimization}.

\subsubsection{Two-partite quantum sensor (2p-QS)}
\label{subsec:2p-QS}

The 2p-QS and the corresponding quantum circuit are displayed in Figs.~\ref{fig:Sensors}(c, f), respectively. Here the sensor dynamics are confined to a subspace ${\mathcal{H}_{\text{2p}} = \mathbb{C}^{N/2 + 1}\otimes \mathbb{C}^{N/2 + 1}}$. This setup is equivalent to a pair of macrospins $J^{A} = N/4$ and $J^{B} = N/4$, which can be entangled through bi-partite uniform control. 

Here, the basis states $\ket{m^A, m^B}$,  are defined as the simultaneous eigenstates of $J_z^A,J_z^B$ satisfying $ J_z^{A(B)}\ket{m^A,m^B} = m^{A(B)}\ket{m^A, m^B}$ with $m^{A(B)}=-J^{A(B)},\dots,J^{A(B)}$. The initial state and the measurement projectors are now 
\begin{align}
    \ket{\psiin} &= \mathcal{U}_{\rm en}^{\text{2p}} \ket{m^A \!=\! -\! J^A,m^B\! =\! -\! J^B}, 
        \label{eq:2pQS_psi}\\
    \Pi_{m^A, m^B} &= \left.\mathcal{U}_{\rm de}^{\text{2p}}\right.^{\dagger} \ket{ m^A, m^B}\bra{m^A, m^B} \mathcal{U}_{\rm de}^{\text{2p}}.
        \label{eq:2pQS_pi}
\end{align}

The optimal 2p-QS is characterized by the posterior variance (see Appendix~\ref{app:projective_optimization})
\begin{equation}
\Delta^2_{\text{2p-QS}}=\min_{\mathcal{U}_{\rm en}^{\text{2p}},\, \mathcal{U}_{\rm de}^{\text{2p}},\, \bm{\xi}_{m^A, m^B}} \mathcal{C}.
\label{eq:PosteriorVariance2pQS}
\end{equation}

\subsubsection{Two-partite classical sensor (2p-CS)}

We contrast the 2p-QS to a 2-partite classical sensor (2p-CS), defined by further restricting $\mathcal{U}_{\rm en (de)}^{\text{2p}}$ to allow only collective rotations of all atoms within each ensemble. This implies that there is no entanglement between the atoms. We use the corresponding $\Delta^2_{\text{2p-CS}}$ to define the SQL for two-partite sensors. 

\subsection{Results: 2D field sensing with few-partite quantum sensors} 

Our main results on 2D sensing, where $\mathbf B=(\text{B}_x,\text{B}_y, 0)^T$, are summarized in Fig.~\ref{fig:2DCost}. We plot the ratio of posterior and prior widths, $\Delta/(\sqrt{2}\,\delta)$, as a function of the prior width $\sqrt{2}\delta$ for the OQS representing the ultimate quantum limit, the optimal 2p-CS defining the SQL, and the optimal 1p-QS and 2p-QS. The results are presented for $N=8$ atoms, which is large enough to demonstrate a reasonable sensitivity gain \footnote{Results for small atom numbers are relevant for field sensing in scenarios where only few atoms can be used for the sensing task, e.g., in order to maintain a good spatial resolution.}. 

In this figure, there are two limits where the Bayesian update following the measurement does not provide information about the signal, i.e., the ratio $\Delta/(\sqrt{2}\,\delta)$ goes to~$1$. The first is the limit $\delta\to 0$, where quantum measurement fluctuations overwhelm the signal, and the second is the regime $\delta\gg1$, where the signal is dwarfed by a large estimation error due to phases that slip outside the unambiguous estimation region $|\bm{\phi}|\lesssim\pi$. As a result, each sensor has an optimal operating point $\delta^*$ at which the information gain per measurement is maximized.

The results in Fig.~\ref{fig:2DCost}(a) show that the  OQS reaches its optimal operating point in the large prior width regime $\delta \sim1$, corresponding to an interrogation time $T\sim(\delta{\abs{\mathbf{B}}})^{-1}$. Another relevant regime is that of small prior widths, $\smash{\delta\lesssim 1/N}$, where the OQS performance is limited by the Bayesian Cram\`er-Rao bound (black dotted line), which follows from the van Trees inequality (see Appendix~\ref{app:Fisher}). The corresponding small evolution times $T\sim(\delta{\abs{\mathbf{B}}}N)^{-1}$ allow us to monitor more rapidly time-varying signals since the requirement $\tau \gg T$ on the coherence time $\tau$ of the fields is consequently relaxed.

Although the 1p-QS can realize the OQS for small prior widths, it fails to improve over the classical sensor (blue curve) in the regime of large prior widths. This is in contrast to the 2p-QS, which can perform close to the OQS for the full range of prior widths. The near-optimal 2D field-sensing performance of the 2p-QS and the limited experimental control it requires motivate a practical implementation on a programmable quantum sensor, which we discuss in Sec.~\ref{sec:VariationalCircuits} in terms of variational quantum circuits. 

In the following, we will take a closer look at the interplay between the optimal input states, measurement projectors, and estimators to understand how these sensors achieve (near-)optimal sensitivity at various prior widths. In particular, we will rely on the Wigner quasi-probability distributions~\cite{Dowling1994}
to represent states $\ket{\psiin}\bra{\psiin}$ and measurement projectors $\Pi_m$ on a generalized Bloch sphere corresponding to the macrospin representation of the sensor. Wigner distributions rotate as rigid objects under the action of $\Uphi$, and the overlap between the Wigner distribution of the rotated state $\ket{\psiphi}\bra{\psiphi}$ and the projective measurement $\Pi_m$ integrated over the Bloch sphere corresponds to the probability $p(m|\bm{\phi})$ of observing the corresponding measurement outcome $m$.

\subsubsection{Optimal 1p-QS}
\label{sec:1pQS}

\begin{figure}[t]
    \centering
    \includegraphics[]{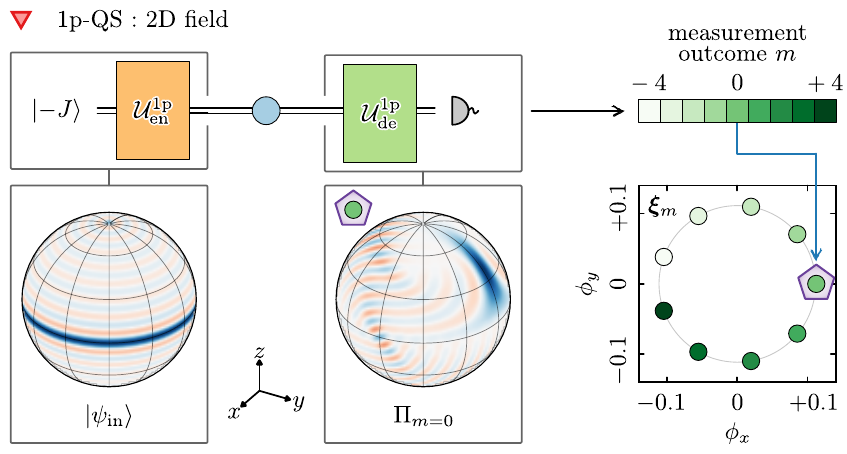}
    \caption{
    \textbf{Illustration of the optimal 1p-QS for 2D fields.} The sensor is optimal for the prior width $\delta$ indicated by the red triangle in Fig.~\ref{fig:2DCost}.  We show the Wigner distributions of the input state $\ket{\psiin}$ [prepared by the unitary $\mathcal{U}_{\rm en}^{\text{1p}}$~\eqref{eq:1pQS_psi}] and a representative projective measurement operator $\Pi_m$ [realized by the unitary $\mathcal{U}_{\rm de}^{\text{1p}}$~\eqref{eq:1pQS_pi}]. For visual clarity, here we show the Wigner distributions for $N = 32$, which are qualitatively the same for any particle number in the regime $\delta \lesssim 1 /N$. The measurement outcomes $m$ are mapped to estimators $\bm{\xi}_{m}$ in the $\phi_x\phi_y$-plane, and the one corresponding to the displayed $\Pi_m$ is highlighted by a purple frame.}
    \label{fig:1p-QS}
\end{figure}

The 1p-QS is optimal for sensing in the regime of small prior widths $\delta\lesssim 1/N$. In this regime, the optimal entangled input state is $\ket{m=0}$, i.e.~a permutation symmetric state with an equal number of atoms in $\ket{\uparrow}$ and $\ket{\downarrow}$ (we consider even $N$).
The Wigner distribution of this state is shown in Fig.~\ref{fig:1p-QS} and represents a ring around the equator of the generalized Bloch sphere, which implies that the state is invariant under $z$-rotations and is equally sensitive to rotations around any axis in the $xy$-plane. Remarkably, the optimal measurement in this regime of 2D field sensing is projective. Figure~\ref{fig:1p-QS} shows a representative projector $\Pi_m$, which has the greatest overlap with the input state rotated about the $x$-axis. The corresponding measurement outcome $m$ is consequently mapped to a phase estimate $\bm\xi_m$ that indicates a field oriented along the $x$ axis. 

The $N+1$ possible measurement outcomes are mapped to phase estimates that are uniformly spaced in a circle. This means that, in the small prior width limit, the 1p-QS operates like a ``compass'', that is, a \emph{single} measurement provides information only on the direction but not the magnitude of $\bm{\phi}$, and hence also of $\mathbf{B}$. The high sensitivity of the 2D ``compass'' to small rotations is particularly suitable for monitoring rapidly changing signals. This is analogous to the GHZ state interferometer~\cite{Bollinger1996} in 1D field sensing, which in a \emph{single} measurement can estimate only the sign but not the strength of a 1D field. Note that the field magnitude is encoded in the probabilities of the different measurement outcomes. Thus, repeated measurements with the ``compass'' will enable us to access full information about the vector field.

\subsubsection{Optimal 2p-QS}
\label{sec:2pQS}

\begin{figure}[t]
    \centering
    \includegraphics[]{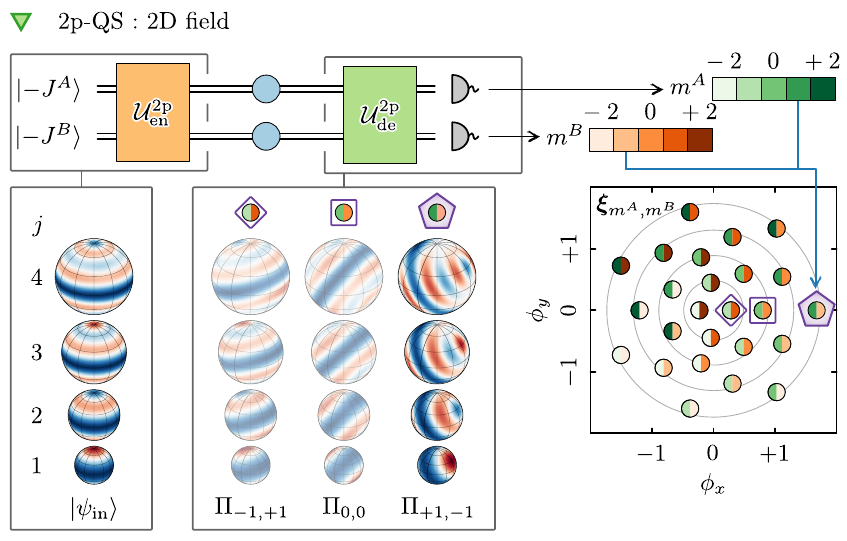}
    \caption{
    \textbf{Illustration of the optimal 2p-QS for 2D fields.} The sensor is optimal for the prior width $\delta$ indicated by the green triangle in Fig.~\ref{fig:2DCost}.  We show
    the Wigner distributions of the input state $\ket{\psiin}$ [prepared by the unitary $\mathcal{U}_{\rm en}^{\text{2p}}$~\eqref{eq:2pQS_psi}] and representative projective measurement operators $\Pi_{m^A, m^B}$ [realized by the unitary $\mathcal{U}_{\rm de}^{\text{2p}}$~\eqref{eq:2pQS_pi}]. We visualize projections on to  the spin-$j$ representations of the 2p-QS space, with  $j=0, 1, \dots, N/2$ (for further details see Sec.~\ref{sec:2pQS}). The measurement outcomes $m^A$ and $m^B$ are mapped to estimators $\bm{\xi}_{m^A, m^B}$ in the $\phi_x\phi_y$-plane, and the ones corresponding to the displayed $\Pi_{m^A, m^B}$'s are highlighted by purple frames.}
    \label{fig:2p-QS}
\end{figure}

In Fig.~\ref{fig:2p-QS}, we visualize the state, projectors and estimators at the optimal operating point of the 2p-QS, i.e., at the minimum of the green curve in Fig.~\ref{fig:2DCost}, around $\delta\approx 0.75$. To show the Wigner distributions, we use the addition of angular momentum to represent the Hilbert space of the two equal-sized macrospins as the direct sum of spin-$j$ representations, i.e. $\mathcal{H}_{\rm 2p} = {\mathbb{C}^{N/2 + 1} \otimes \mathbb{C}^{N/2 + 1}=\bigoplus_{j = 0}^{N /2} \mathbb{C}^{2j +1}}$. We project the operators onto the different angular momentum manifolds and show Wigner distributions corresponding to each of the $j$-manifolds. This means that coherences between different $j$-manifolds are not represented in the resulting Wigner distributions. Nevertheless, the distributions give an intuitive picture of how the sensor works. The optimal input state of the 2p-QS consists of equatorial rings in all spin representations, which is an approximation of the optimal covariant 2p-QS (see Appendix~\ref{subsec:covariant_POVM}) that is a superposition of $\ket{j, m = 0}$ states. The use of different spin manifolds enables the the 2p-QS to extend the range of detectable field strengths. 

The projective measurement outcomes $m^A$ and $m^B$ corresponding to the entangled projectors $\Pi_{m^A, m^B}$ are assigned to one of the evenly distributed estimators, which approximate $N/2 + 1$ concentric rings. The rings shown in the estimator plot in Fig.~\ref{fig:2p-QS} are predicted from the optimal covariant 2p-QS. In contrast to the 2D ``compass'', a single measurement by the optimal 2p-QS provides information on both field strength and direction. The three representative projectors visualized in Fig.~\ref{fig:2p-QS} detect rotations around the $x$-axis with increasing strength, i.e., they have a large overlap with $\ket{\psiin}\bra{\psiin}$ after it is increasingly rotated around the $x$-axis. 

\section{Variational quantum circuits}
\label{sec:VariationalCircuits}

Today's programmable quantum sensors are NISQ devices implemented on various platforms that
support a reduced instruction set of native gates, which can nevertheless be executed with high fidelity and are typically scalable to tens, and potentially hundreds of particles. A practical implementation of a quantum sensor, and in particular a quantum circuit model, must therefore rely on a decomposition of the entangling and decoding unitaries into this native set of gates. This can be achieved in an approximate manner via variational quantum circuits. Here, an ansatz is made for ${\cal U}_{\rm en}$ and ${\cal U}_{\rm de}$ in terms of low-depth quantum circuits, which are parametrized by a set of variational parameters $\bm{\theta}$ ($\bm{\vartheta}$) for the entangler (decoder). Given such a variational ansatz, the optimal entangler and decoder within a class of variational quantum circuits correspond to the minimum of the metrological cost function
\begin{align}\label{eq:varCost}
    \Delta^2_{\rm var } = \min_{\bm{\theta}, \bm{\vartheta}, \bm{\xi}} \mathcal{C}\quad (\ge \Delta ^2_{\rm OQS}),
\end{align}
thus defining the best possible quantum sensor for given (entangling) resources, i.e. resource gate set, circuit design, and circuit depth. The practical aspects of the variational minimization are discussed in Appendix~\ref{app:VQ_optimization}.
This optimization can be performed within the theoretical model of the sensor or by optimizing the variational parameters in a quantum feedback loop on the actual physical device~\cite{Marciniak2021}. The latter case provides us with the opportunity to optimize the quantum sensor even when the corresponding quantum circuits are challenging to simulate classically, and in the presence of control errors and noise.

Below we present a study of variational quantum circuits of increasing depth, which allow us to approach the performance of the OQS in multi-parameter quantum metrology. As a native gate set, we assume one-axis twisting (OAT) operations as entangling gates and uniform spin rotations as single qubit gates. This is motivated by the feasibility of realizing programmable quantum sensors using strings of trapped ions, where OAT is natively implemented as a M\o{}lmer-S\o{}rensen gate 
\cite{Leibfried2005, Monz2011}, and spin rotations are performed via laser or microwave driving. The 1p-QS and 2p-QS can be implemented directly as circuits of the natively available gates in such a setup. We note that a similar set of quantum resources is also available in Cavity QED setups~\cite{Leroux2010} and spin mixtures of Bose-Einstein condensates~\cite{Riedel2010, Gross2010}.

To be specific, our discussion will focus on the 2p-QS illustrated in Fig.~\ref{fig:Sensors}c. In the previous section, we have shown that the 2p-QS has close to optimal performance for 2D field sensing. Here, we show that low-depth variational circuits built from OAT and spin rotations provide good approximations to the optimal 2p-QS, and enable a significant enhancement over unentangled sensors (see Fig.~\ref{fig:2DCost}).

\subsection{Resource gate set for 2p-QS}

The 2p-QS assumes that we have a native set of gate operations available that act independently on partitions $A$ and $B$ [see Fig.~\ref{fig:Sensors}(c)]. This includes, first of all, uniform rotation of spins in the partitions, 
\begin{align}
    \mathcal{R}^{A(B)}_{\nu}(\theta) = e^{-i \theta J^{A(B)}_{\nu}}\qquad \text{with}\ \nu = x,y,z.
\end{align}
Furthermore, we assume entanglement operations in the form of OAT. These include
\begin{align}
    \mathcal{T}^{A(B)}_z(\theta) = e^{-i \theta \left(J_{z}^{A(B)}\right)^2 } 
    \label{eq:TwistingPartition}
\end{align}
which respectively entangle the atoms within the partitions $A$ and $B$, while the entanglement between $A$ and $B$ is generated with
\begin{align}
    \mathcal{T}_z(\theta) = e^{-i \theta \left(J_{z}^A + J_{z}^{B}\right)^2 }, 
    \label{eq:TwistingGate}
\end{align}
which completes the gate set~\footnote{Note that extensions of this gate set, i.e. allowing for twisting around arbitrary axes~\cite{Schulte2020, Thurtell2022} can lead to more efficient parameterization in terms of the number of gates required to achieve a certain performance.}. Such gates can be realized on trapped ion platforms, such as the one used to demonstrate optimal 1D field sensing~\cite{Marciniak2021}. 

\subsection{Circuit design for 2p-QS}

\begin{figure}[t]
    \centering
    \includegraphics[]{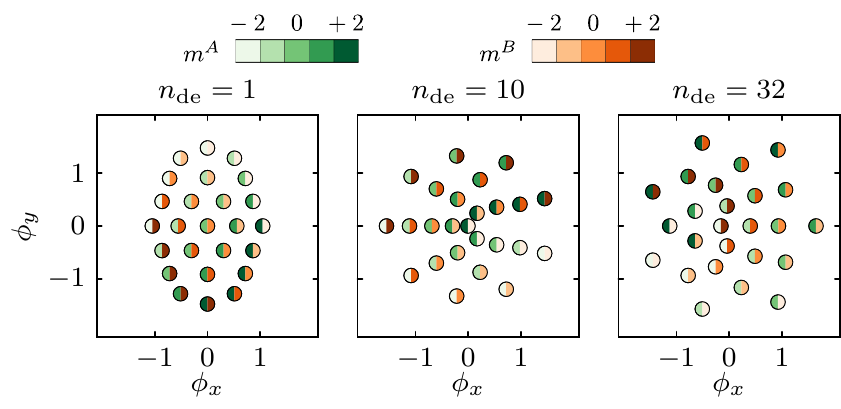}
    \caption{\textbf{Convergence of estimators of optimized variational circuits.} Estimators $\bm{\xi}_{m^A, m^B}$ of the optimized $N = 8$ atoms variational circuits for $\delta \approx 0.75$, with $\nen = 2$ and $\nde = 1, 10, 32$. The estimators are colored according to the corresponding measurement outcomes $m^A$ and $m^B$. The corresponding sensitivity is shown in Fig.~\ref{fig:2DCost}. } 
    \label{fig:2DLargePriorEstimators}
\end{figure}

Variational circuits approximating the optimal entangler and decoder of the 2p-QS (see Fig.~\ref{fig:Sensors}c)
 are compiled from the above elementary gates. In particular, we consider circuits that are constructed by stacking multiple layers. Each layer is described by a unitary 
\begin{align}
    \mathcal{L}(\bm{\theta_k}) = &\,
    \mathcal{R}^A_y\left(\theta_{k}^{(4)}\right)
    \mathcal{R}^B_y\left(-\theta_{k}^{(4)}\right)
    \mathcal{R}^A_x\left(\theta_{k}^{(3)}\right)
    \mathcal{R}^B_x\left(\theta_{k}^{(3)}\right)
    \notag \\ &\, \times
    \mathcal{T}^A_z\left(\theta_{k}^{(2)}\right)
    \mathcal{T}^B_z\left(\theta_{k}^{(2)}\right)
    \mathcal{T}_z\left(\theta_{k}^{(1)}\right),
\end{align}
that depends on the four parameters contained in the vector $\bm{\theta}_{k}= \left(\theta_{k}^{(1)}, \cdots, \theta_{k}^{(4)}\right)^T$. The parameters within each layer are correlated (i.e. $\theta_{k}^{(4)}, \theta_{k}^{(3)}, \text {and } \theta_{k}^{(2)}$ respectively appear in two consecutive gates), so that the resulting circuits preserve a symmetry that we identified in the optimal 2p-QS solution. Therefore, the unitaries will not explore the entire Hilbert space $\mathcal{H}_{\text{2p}}$ of the 2p-QS, which reduces the number of variational parameters required for a given circuit depth of the 2p-QS. 

We parameterize entangling unitaries consisting of $\nen$ layers through a $4\nen$ dimensional parameter vector $\bm{\theta}= \left(\bm{\theta}_{1}, \cdots, \bm{\theta}_{\nen}\right)^T$ as
\begin{align}
	\mathcal{U}_{\rm en}^{\text{2p}}(\bm{\theta}) = \mathcal{L}\left(\bm{\theta}_{\nen}\right) \cdots \mathcal{L}\left(\bm{\theta}_{n_{\text{1}}}\right)
 \mathcal{R}^A_y\left(\tfrac{\pi}{2}\right)
 \mathcal{R}^B_y\left(\tfrac{\pi}{2}\right),
\label{eq:2p_Uen}
\end{align}
and decoding unitaries consisting of $\nde$ layers through a $4\nde+2$ dimensional parameter vector $\bm{\vartheta}= \left(\vartheta_0^{(1)}, \vartheta_0^{(2)}, \bm{\vartheta}_{1} ,\cdots, \bm{\vartheta}_{\nen}\right)^T$ as
\begin{align}
    \mathcal{U}_{\rm de}^{\rm 2p}(\bm{\vartheta}) = &\,
    \mathcal{R}^A_x\big(\vartheta_0^{(1)}\big)\,
    \mathcal{R}^B_x\big(\vartheta_0^{(1)}\big)\,
    \mathcal{R}^A_y\big(\vartheta_0^{(2)}\big)\,
    \mathcal{R}^B_y\big(- \vartheta_0^{(2)}\big)\notag \\\
    &\, \times\mathcal{L^{\dagger}}(\bm{\vartheta}_{1}) \cdots \mathcal{L}^{\dagger}(\bm{\vartheta}_{n_{\text{de}}}).
	\label{eq:2p_Ude}
\end{align}
Therefore, a sensor with an entangler and decoder depth of $(\nen, \nde)$ depends on a total number of $4(\nen + \nde) + 2$ variational parameters. The parameters of the circuit are variationally optimized using the cost function Eq.~\eqref{eq:varCost} to find the best-performing sensor for a given circuit with depth $(\nen,\nde)$.

\subsection{Results: variational optimization of 2p-QS}

In the magnified panel of Fig.~\ref{fig:2DCost}, we examine how closely the optimized variational circuits approximate the performance of the optimal 2p-QS comprising $N=8$ atoms. With increasing decoder depth, the variational circuits consistently approach the optimal 2p-QS and eventually converge to it for large $\nde$. Interestingly, optimal sensitivity is achieved for the shallow depth of the entangler $\nen = 2$, indicating that it is significantly more challenging to implement optimal decoding unitaries than to implement optimal entangling unitaries. This is due to the fact that the entangler unitary only needs to transform $\ket{\psi_0}$ to $\ket{\psiin}$, while the decoder needs to transform all measurement bases to the desired bases simultaneously, imposing significantly more constraints on the optimal unitary. 

To understand how the variational circuits approach the 2p-QS, it is insightful to study the optimal estimators and see how they change as the depth of the decoder increases. In Fig.~\ref{fig:2DLargePriorEstimators}, we show the estimators of the optimized variational circuits. The minimal depth circuit $\nde = 1$ results in an asymmetric distribution of estimators in the $\phi_x\phi_y$-plane, indicating that the sensor is not equally sensitive to $\phi_x$ and $\phi_y$. Increasing the depth to $\nde = 10$, the estimators become more evenly distributed, whereas the estimator pattern for $\nde = 32$ is very close to the optimal 2p-QS shown in Fig.~\ref{fig:2p-QS}.

\begin{figure}[t]
    \centering
    \includegraphics[]{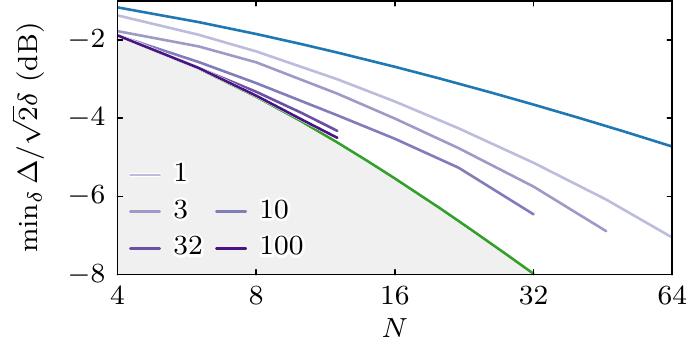}
    \caption{\textbf{Performance scaling of variational 2D field sensors.}
        Scaling of the minimum value of the posterior to prior uncertainty ratio as a function of system size $N$. The purple lines correspond to variational quantum sensors with different numbers of decoding layers $\nde$ at a fixed entangler depth of $\nen = 2$. For comparison, we also show the performance of the 2p-CS (blue line) and the optimal 2p-QS (green line).}
    \label{fig:2DScaling}
\end{figure}

An important question for currently available quantum sensors is how much sensitivity gain we can expect for a fixed circuit depth. We address this by determining the minimum value of the ratio $\Delta/\sqrt{2} \delta$ at the optimal operating value of $\delta$ for a given circuit depth. In Fig.~\ref{fig:2DScaling}, this value is shown as a function of the system size $N$. Interestingly, we find that even for $\nde=1$, there is an increase in performance that appears to scale with $N$. As the depth of the circuit increases, the sensitivity and scaling improve steadily.

\section{3D field sensing}
\label{sec:3D}

While so far our focus has been on 2D field sensing, the techniques developed in the previous sections are readily extended to 3D. Here, we present first results and observations for optimal and variational 3D field quantum sensors.

We observe that the OQS solution in 3D makes use of the full effective Hilbert space of an $SU(2)$ sensor~($\text{dimension}\sim N^3$), see Appendix~\ref{appendix:1p_and_2p_subspaces}. This suggests that sensors made of a few partitions do not closely approximate the OQS for 3D sensing of fields with arbitrary prior widths. Nevertheless, few-partite sensors can provide a significant improvement over classical sensors and even perform optimally at small prior widths, as we demonstrate below.

\begin{figure}
    \centering
    \includegraphics[]{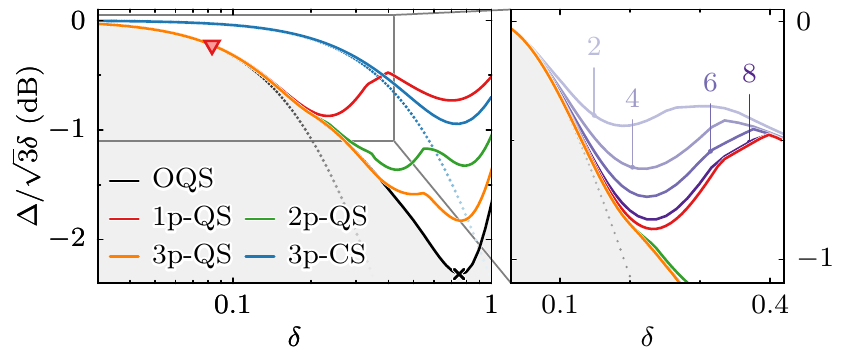}
    \caption{\textbf{Bayesian quantum sensing of a 3D field.}
    Left panel: Ratio of the posterior width $\Delta$ to the prior width $\sqrt{3}\delta$ vs. $\delta$ for $N=6$ atoms. The black marker shows one preliminary data point obtained for the OQS at $\delta=0.75$. This point matches well with the sensitivity of a $N$p-QS (dark gray line). The asymptotic bound on quantum (black dotted line) and classical (blue dotted line) sensors are predicted by the Bayesian Cram\'er-Rao bounds~\eqref{eq:BayesianQL} and \eqref{eq:BayesianSQL}. In Fig.~\ref{fig:3D_1p-QS}, we illustrate the working of the optimal sensor indicated here by the red triangle.
    Magnified panel (right): Performance of variational circuits (purple lines) with an increasing total number of layers $n_{\rm l} = n_{\rm en} + n_{\rm de}$. }
    \label{fig:3DCost}
\end{figure}

Figure~\ref{fig:3DCost} shows the performance of various sensors with $N=6$ atoms, quantified by the ratio $\Delta/\sqrt{3}\delta$, as  $\delta$ is scanned.  The 1p-QS and 2p-QS sensors, are both significantly less sensitive than the OQS in the large prior regime and only the full $N$-partite quantum sensor ($N$p-QS), relying on universal control, approaches the OQS. Here, we extend the discussion to a 3-partition quantum sensor (3p-QS), with uniform control at the level of three partitions. However, the 3p-QS performance is still appreciably different from that of the OQS, highlighting the added complexity of the 3D OQS near its optimal operating point. To compare the few-partite sensors with a classical reference, we study a 3-partite classical sensor, where the atoms are unentangled. Here, we find that each partition is primarily sensitive to one component. In comparison to the 3p-CS, the 2p-QS and 3p-QS show a considerable improvement by using entanglement between the sensor atoms.

\subsubsection{Optimal 1p-QS}
\begin{figure}
    \centering
    \includegraphics[]{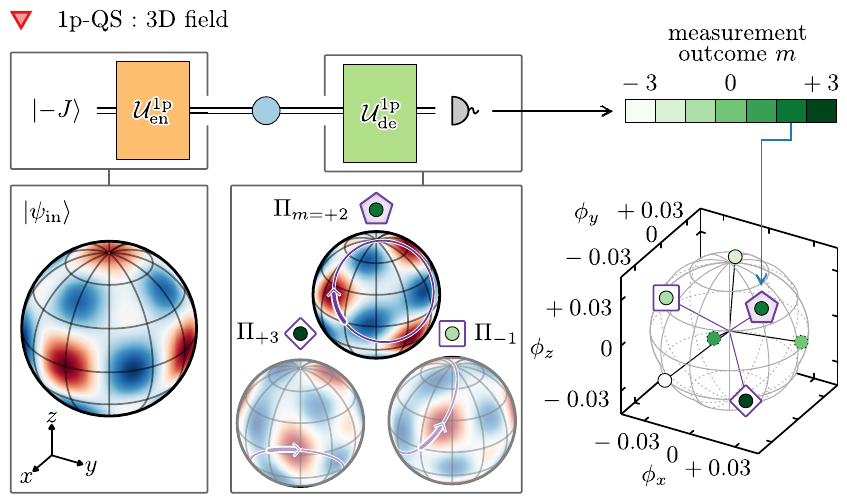}
    \caption{\textbf{Illustration of the optimal 1p-QS for 3D fields.} The sensor is optimal for the prior width $\delta$ indicated by the red triangle in Fig.~\ref{fig:3DCost}.  We show the Wigner distributions of the input state $\ket{\psiin}$ [prepared by the unitary $\mathcal{U}_{\rm en}^{\text{1p}}$~\eqref{eq:1pQS_psi}] and representative projective measurement operators $\Pi_m$ [realized by the unitary $\mathcal{U}_{\rm de}^{\text{1p}}$~\eqref{eq:1pQS_pi}]. The measurement outcomes $m$ are mapped to estimators $\bm{\xi}_{m}$ in the $\phi_x\phi_y$-plane, and the ones corresponding to the displayed $\Pi_m$ are highlighted by colored frames. The purple arrows illustrate that the Wigner distribution of the projectors resemble approximate rotated versions of the Wigner distribution of the input state, when it is rotated about the axis along the vector specified by the corresponding estimator $\bm{\xi}_{m}$.}
    \label{fig:3D_1p-QS}
\end{figure}

In the regime of small prior widths $\delta\lesssim 1/N$, the performance of the 1p-QS is indistinguishable from the optimal performance. To gain insight into the 1p-QS solution, we visualize the optimal state, measurement projectors, and estimators in Fig.~\ref{fig:3D_1p-QS}. The optimal state shows a regular distribution of maxima and minima on the generalized Bloch sphere, highlighting the high degree of symmetry in this state. Notably, the estimators are found to lie on a sphere. As a result, they provide only directional information and therefore the sensor acts as a 3D ``compass''. The visualized measurement projectors resemble the input state rotated around the axis specified by the respective estimator. In other words, each projector is sensitive to rotations around a different axis. 

We conclude that the optimal sensor at small prior widths displays a universal compass-like behavior for 1D, 2D, and 3D fields. However, a key difference is that while GHZ ($\ket{m=0}$) states are optimal independent of $N$ in 1D (2D), the optimal state for 3D field sensing changes with $N$ similar to the states discussed in~\cite{Zimba2006, Kolenderski2008, Giraud2010, Bjork2015,Bouchard2017}. 

\subsubsection{Variational optimization of 1p-QS}
The performance of the 1p-QS motivates us to study variational circuits to approximate the optimal 1p-QS. As sketched in Fig.~\ref{fig:Sensors}(e), we consider circuits that parameterize $\mathcal{U}_{\rm en(de)}^{\rm 1p}$ according to 
\begin{align}
    \mathcal{U}_{\rm en}^{\text{1p}}(\bm{\theta}) &= \mathcal{L}\left(\bm{\theta}_{\nen}\right) \cdots \mathcal{L}\left(\bm{\theta}_{n_{\text{1}}}\right)\\
    \mathcal{U}_{\rm de}^{\rm 1p}(\bm{\vartheta}) &= \mathcal{L^{\dagger}}(\bm{\vartheta}_{1}) \cdots \mathcal{L}^{\dagger}(\bm{\vartheta}_{n_{\text{de}}}). 
\label{eq:1p_U}
\end{align}
Here, the unitaries are constructed by repeated application of layers containing three elementary gates
\begin{align}
\mathcal{L}(\bm{\theta}) = \mathcal{T}_z\left(\bm{\theta}^{(3)}\right)\,
    \mathcal{R}_y\left(\bm{\theta}^{(2)}\right)\,
    \mathcal{R}_{x}\left(\theta^{(1)}\right).
\end{align}
The number of variational parameters at a given circuit depth is therefore $3(\nen + \nde)$. In the magnified panel of Fig.~\ref{fig:3DCost}, we plot the performance of variational circuits with varying depths and indeed find that circuits with modest depths can approach the optimal 1p-QS performance. These results highlight the potential for implementing a 3D quantum ``compass'' with currently available resources. 

\section{Conclusion and Outlook} 

In this work, we have identified fundamental bounds and studied the performance of practically feasible sensors for the sensing of 2D and 3D vector fields within the Bayesian framework for $SU(2)$-quantum interferometry. We established a procedure to determine the OQS that sets the fundamental bound on the performance of an $N$-atom sensor. Furthermore, we introduced sensors with limited experimental control, namely the 1p-QS and 2p-QS, and studied their ability to approach the OQS. As a further step towards practical realization, we also simulated the variational optimization of sensors with limited entangling capabilities, considering the resources available on state-of-the-art trapped ion systems.

Our study of 2D field sensing reveals important similarities and differences between single- and multi-parameter metrology. In the case of 1D sensing, the OQS operates in the particle permutation symmetric subspace of the $N$ atoms, and hence the 1p-QS is optimal at all prior widths. Here, we find that the 1p-QS is optimal for 2D sensing only in the regime of small prior widths. In this regime, it operates like a 2D ``compass'', providing information only on the field direction. This is similar to the GHZ-state interferometer, optimal for 1D sensing with small prior widths, which only provides information on the sign of the 1D field. In contrast to the 1D case, we find that the 2p-QS approaches the OQS performance for 2D field sensing at large prior widths. 

Although such a sensor requires more experimental control than a 1p-QS, it can nevertheless be realized on state-of-the-art sensing platforms, as we demonstrate by simulating a variational circuit optimization of the 2p-QS. Our simulation reveals that sensors made of low-depth circuits provide appreciable improvements over classical (unentangled) sensors, while deeper circuits may be required to achieve optimal performance. This is in contrast to 1D sensing, where low-depth variational circuits have previously been shown theoretically and experimentally to achieve optimal performance. Nevertheless, these results have promising implications for small-scale sensors as well as for those that involve a macroscopic number of atoms; current technology provides excellent coherent control of a small number of atoms, enabling the realization of deep circuits  that can reach the 2p-QS limit for small quantum sensors. Such sensors may be relevant for practical applications where spatial resolution is important. On the other hand, for sensors operating with a large number of atoms, our results demonstrate that shallow-depth circuits can significantly improve performance compared to unentangled sensors. 

Although our study has focused on using infinite-range OAT as an entangling resource, many promising quantum sensing platforms host finite-range interactions, e.g. solid-state systems 
 or neutral atoms in tweezer arrays
\cite{Young2020, Graham2019, Madjarov2020, Schine2022}. Variational optimization of the metrological cost function can be directly generalized to finite-range interactions, as done in~\cite{Kaubruegger2019, Koczor2020, Kaubruegger2021, Zheng2022} for other metrological cost functions. However, in that case, the classification into few-partite quantum sensors does not apply, and simulating the full many-body problem is challenging for classical algorithms, which makes feedback loop optimization on the physical quantum device even more desirable. Furthermore, the $N$-atom 1p-QS and 2p-QS can also be realized, respectively, as a single large macrospin \cite{Chalopin2018} of length $J=N/2$ or two entangled macrospins \cite{Kruckenhauser2023} of length $J=N/4$ each, allowing for a route to optimal sensing with qudit-based quantum sensors. 

We also briefly extended our study to the problem of 3D field sensing. Here, we found an appreciable difference in the performance of sensors with a few partitions and the optimal sensor. Therefore, it is an open challenge to find circuits that efficiently approximate the OQS for 3D field sensing. However, in the regime of small prior widths, we once again find that the 1p-QS is essentially optimal. Similar to the 1D and 2D cases, it operates as a 3D ``compass'' in this regime, with the estimators arranged on the surface of a sphere. Furthermore, we showed that such a 3D ``compass'' can be well approximated with low-depth variational circuits, paving the way for their experimental realization.

In conclusion, we have highlighted the importance of finding Optimal Quantum Sensors, which operate at the ultimate precision allowed by quantum physics, and implementing them as Variational Quantum Sensors. This is crucial not just for uncovering new physics through precision measurements \cite{Gilmore2021} but also for practical applications such as quantum computing. Programmable atomic sensors, discussed in this article, can be integrated with atomic quantum computing systems by utilizing common quantum resources, like entangling quantum gates. In this way, an optimal quantum sensor for electromagnetic fields embedded in the system can precisely detect and monitor noise fields, contributing to the stability of quantum devices through an optimal feedback mechanism.

\begin{acknowledgments} We thank C. Marciniak, and Q. Liu for comments on the manuscript. The computational results were obtained using the University of Innsbruck's LEO HPC infrastructure.  Research  in  Innsbruck  is supported by the US Air Force Office of Scientific Research (AFOSR) via IOE Grant No.~FA9550-19-1-7044 LASCEM, 
the European Union’s Horizon 2020 research and innovation program under PASQuanS 2, and by the Simons Collaboration on Ultra-Quantum Matter, which is a grant from the Simons Foundation (651440, P.Z.), 
and by the Institut f\"ur Quanteninformation. Innsbruck theory is a member of the NSF Quantum Leap Challenge. AS acknowledges the support of a CV Raman Post-Doctoral Fellowship, IISc.
\end{acknowledgments}

\appendix

\section{Physical implementation of vector field sensors}
\label{app:vfs_implement} Sec.~\ref{sec:SU2} provides a formal description of an $SU(2)$-interferometer to sense vector fields. There, the discussion is phrased as an ensemble of $N$ two-level atoms, described by collective angular momentum operators $\bm{\mathcal{J}}$,  coupled uniformly to a static (DC) vector field $\bm{\mathcal{B}}$ given by the Hamiltonian \eqref{eq:H}. In Appendix~\ref{app:MFS_atoms}, we summarize various physical scenarios where such a Hamiltonian arises in the context of magnetometry. Furthermore, in Sec.~\ref{sec:LimitedControl}, we have defined limited control sensors where the $N$ two-level atoms are equivalent to one or two `macrospins' with large angular momenta (see Fig.~\ref{fig:Sensors}). Accordingly, in Appendix~\ref{app:macro-spins}, we provide examples of vector field sensors implemented directly using macrospins.

\subsection{Vector field sensing with ensembles of $N$ two-level atoms}
\label{app:MFS_atoms}

\emph{DC magnetometry} -- For a single atom, the magnetic field couples to Zeeman levels provided by a fine (or hyperfine) structure manifold $J$ (or $F$) according to $H_{\rm M} = -\bm{\mu}_{\mathcal{B}} \cdot \bm{\mathcal{B}}$. Here, the magnetic dipole moment $\bm{\mu}_{\mathcal{B}} = \mu_{\rm B} g \bm{\mathcal{J}}/{\hbar}$ is given by the Bohr magneton $\mu_{\rm B}$, the Land\'e g-factor $g$ and the total  angular momentum operator $\bm{\mathcal{J}}$ (or $\bm{\mathcal{F}}$). This Zeeman Hamiltonian is valid within degenerate perturbation theory in the given angular momentum manifold, resulting in a Zeeman splitting linear in the magnetic field.

An example of a Zeeman manifold representing a two-level atom is the ${}^2 S_{1/2}$ ground state of ${}^{40}$Ca ions, as used in trapped ion programmable quantum sensor experiments \cite{Marciniak2021}. For $N$ atoms, summing over these single-atom contributions leads to the Hamiltonian \eqref{eq:H}, and thus sensing the three components of a (static) magnetic field.

\emph{AC magnetometry} -- Experiments commonly apply a bias field $\mathcal{B}_0$ that defines a quantization axis and lifts the degeneracy of Zeeman states, as in NMR. Continuing with the example of the ${}^2 S_{1/2}$ manifold, a static bias field will introduce a Zeeman splitting $\omega$ defining a $z$-axis. This Zeeman transition can be driven by a transverse AC magnetic field. Within the validity of the rotating wave approximation, and going to a rotating frame leads to a static Hamiltonian, which for $N$ atoms is of the form of Eq.~\eqref{eq:H}. In this case the three components of the magnetic field to be sensed are the static magnetic field deviation from the bias field, and the in-phase and out-phase components of the driving magnetic field (Rabi frequencies).

In the AC field measurement example, the origins of the effective longitudinal and transverse components (with respect to the bias field) are different, and thus our prior knowledge of these terms will generally not be the same. Therefore, the assumption of an isotropic prior density \eqref{eq:GaussianPrior} used in the main text may not be accurate. In appendix~\ref{app:PriorDensities}, we discuss how the results described in the main text can be generalized to anisotropic prior probability densities.

The above discussion can be generalized to pairs of atomic levels belonging to different Zeeman manifolds with different Land\'e g-factors, which however leads to anisotropic coupling Hamiltonians.

\subsection{Vector field sensing with macroscopic spins}
\label{app:macro-spins}

In Sec.~\ref{sec:LimitedControl}, we have defined limited control sensors, i.e. the 1p-QS and 2p-QS (see Fig.~\ref{fig:Sensors}). In the 1p-QS, the $N$ atoms are uniformly controlled and hence it is equivalent to a sensor consisting of a single macrospin with angular momentum $J=N/2$. In the case of the 2p-QS, the $N$ atoms are partitioned into two equal-sized partitions $A$ and $B$ with uniform control at the level of each partition and over the entire ensemble. Thus, the 2p-QS is equivalent to a sensor consisting of two physical, entangled `macro-spins' each with angular momenta $J^A=J^B=N/4$. Here, we briefly elaborate on the direct physical implementation of macrospins in large angular momentum manifolds. These are available as large angular momentum Zeeman manifolds, $SO(4)$-symmetric manifolds in Rydberg atoms for electric field sensing, and in photon interferometry with large orbital angular momentum states.

\emph{Large Zeeman manifolds for magnetometry} -- As already noted in Appendix~\ref{app:MFS_atoms}, large Zeeman fine and hyperfine manifolds available in trapped ion and atomic tweezer experiments lead to a Hamiltonian of the form  \eqref{eq:H}. An example is provided by the $3^2\text{D}_{5/2}$ manifold of $\,^{40}\text{Ca}^{+}$ \cite{Ringbauer2022}, which can be manipulated and entangled in qudit operations. Other platforms that qualify for proof-of-principle experiments for AC magnetometry with qudits are 
Dysprosium atoms~\cite{Chalopin2018} with an electronic spin of $J=8$, or Holmium \cite{Saffman2008, Robicheaux2018} and Erbium\cite{Patscheider2020} with large hyperfine splitting of the ground state.

\emph{Electrometry with Rydberg atoms--}  Highly excited, large orbital angular momentum Rydberg states display a linear Stark effect. This is related to the SO(4) symmetry of the hydrogen atom and the conservation of the Runge-Lenz vector. The states belonging to a given $n$-manifold, where $n$ is the principal atom number, are degenerate and highly sensitive to electric fields~\cite{Facon2016}. The resulting Hamiltonian for coupling of the electric field to Rydberg states can again be written in the form~\eqref{eq:H}, where the effective angular momentum of the manifold scales as $J\sim n$ with the principal quantum number $n$. Ref.~\cite{Kruckenhauser2023} outlines a method to entangle the Runge-Lentz vectors of multiple atoms. 

\emph{Photons with orbital angular momentum--} Another platform that can be used as a test bed for proof-of-principle multidimensional SU(2) sensors are photons, where a qudit of tunable dimension can be encoded into the orbital angular momentum degree of freedom of each photon~\cite{Eriksson2023}. State preparation, unitary phase encoding and measurement can be realized~\cite{Brandt2020}. However, it is a challenge to create entanglement between different photons.

\section{Generalization to other prior probability densities}
\label{app:PriorDensities}
In the main text, we consider a prior probability density~\eqref{eq:GaussianPrior} of the phase vector, that is isotropic in $\bm{\phi}$. In many practical applications, knowledge of the field components $\text{B}_{\nu}$, and hence the phases, might not be isotropic. Note that 1D and 2D sensing are anisotropic cases of a 3D field, where we have complete knowledge of two and one components of the field, respectively. Therefore, depending on the degree of anisotropy, the 1D or 2D solution can be used as a starting point to find optimal solutions for a given anisotropic prior density. Variational circuits for 1D sensing are discussed in Ref.~\cite{Kaubruegger2021}. 

Another assumption we make is that the prior density is centered around zero. In the presence of a known offset field, this assumption is still valid if we can apply an additional external field that exactly compensates for the offset \emph{during} the interrogation time.  However, when this is not possible, the more general case with a prior density centered around a phase vector $\bm{\phi}_0$ must be treated separately since $\bm{\phi}_0 \cdot \bm{J}$ and $\bm{\phi} \cdot \bm{J}$ do not commute, which distinguishes 2D and 3D from 1D field sensing. In the 1D case, we can apply a unitary \emph{after} the interrogation time to compensate for $\bm{\phi}_0$, which is not possible for 2D and 3D fields because of the non-commutativity. Instead, one can try to identify alternative entangler and decoder circuits that take the effect of $\bm{\phi}_0$ into account.

Finally, for probability densities that are isotropic and have a single maximum at the origin, we expect the Gaussian prior solution to be a good approximation. 

\section{Bayesian Cram\'er-Rao like bound for multi-dimensional field sensing}
\label{app:Fisher}
In this appendix, we elaborate on a connection (see Sec.~\ref{subsec:OQS}) between the Fisher information (FI) framework and the Bayesian framework adopted in this work. We first recall the Cram\'er-Rao (CR) inequality, which lower bounds the performance of multi-parameter estimation in the FI approach~\cite{Demkowicz2020}. Then we elaborate on the assumptions that underlie the FI approach and introduce the van Trees inequality, which connects the FI matrix with the BMSE. Finally, we provide the corresponding bounds on the BMSE for multi-dimensional field sensing.

Parameter estimation in the FI approach considers estimators ${\bm{\xi}}_{\mu}$ that are locally unbiased around some phase value ${\bm\phi}_0$~\cite{Demkowicz2020}. The corresponding MSE~\eqref{eq:MSE_POVM}, is then lower-bounded by the CR inequality (bound)
\begin{align}
	\label{eq:CRB}
	\mathrm{MSE}(\bm{\phi}_0) &\geq \tr[F(\bm{\phi}_0)^{-1}],
	\\
    F(\bm{\phi}) &= \sum_{\mu} \frac{\bm{\nabla} p(\mu | \bm{\phi})\bm{\nabla}^T p(\mu | \bm{\phi})}{p(\mu | \bm{\phi})},
    \label{eq:FisherMatrix}
\end{align}
where $F(\bm{\phi})$ is the (classical) Fisher information matrix of the conditional probability $p(\mu | \bm{\phi})$~\eqref{eq:conditional_probability}, at $\bm\phi$. Here, $\bm{\nabla}  = (\partial / \partial_{\phi_x}, \partial /\partial_{\phi_y}, \partial /\partial_{\phi_z})^T$ is the gradient operator that contains partial derivatives with respect to the components of $\bm{\phi}$.

The FI matrix depends on the input state and the measurement and, therefore, is upper bounded by the quantum FI (QFI) matrix, $F_Q(\bm{\phi})$, which is obtained by optimizing the MSE for all measurements $\{M_{\mu}\}$ allowed by quantum physics~\cite{Demkowicz2020}. Note that in the multi-parameter case, the corresponding quantum CR bound, $\mathrm{MSE}(\bm{\phi}_0) \geq \tr[F_Q(\bm{\phi}_0)^{-1}]$, cannot be saturated in general due to the potential incompatibility of optimal measurements for different parameters encoded by non-commuting Hamiltonians. The issue of measurement incompatibility is addressed by the Holevo CR bound~\cite{Holevo1982,Demkowicz2020}. However, in the case of multi-dimensional field sensing, the quantum CR and the Holevo CR bounds coincide for optimal sensors comprising an even number of atoms~$N$~\cite{Vaneph2013, Baumgratz2016}. This is the case we consider below, and thus the quantum CR inequality provides a tight bound.

In general, not only the unbiased estimators but also the optimal measurement projectors depend explicitly on $\bm{\phi}_0$ so that an exact value of $\bm{\phi}_0$ must be known in advance to optimally measure around this value. This drawback can be overcome if the measurement can be repeated many times for the same unknown parameter vector or if many ensembles are available to measure the same parameter vector. Under these conditions, an adaptive protocol can be performed in which a preliminary estimate is first obtained using which the optimal measurement strategy can be executed~\cite{Barndorff2000}. 

In a Bayesian sense, the preliminary estimate can be understood as prior knowledge about the phases being measured, which allows for a connection between the two frameworks in the limiting case of narrow prior density. This connection is given by the van Trees inequality~\cite{Trees1968}, which defines CR-like bounds for the Bayesian cost. The multi-parameter generalization~\cite{Demkowicz2020} lower bounding the BMSE~\eqref{eq:BMSE_POVM}, is given by 
\begin{align}
    \mathcal{C} \geq \tr \left[ \left(\overline F + \mathcal I\right)^{-1}\right], 
    \label{eq:VanTrees}
\end{align}
where $\overline F = \int d\bm{\phi}\, F(\bm{\phi})\, \mathcal{P}(\bm{\phi})$ is the FI matrix averaged over the prior density and $\mathcal I$ is the FI matrix of the prior density 
\begin{align}
    \mathcal I = \int d\bm{\phi}\,\frac{ \bm{\nabla} \mathcal{P}( \bm{\phi})\bm{\nabla}^T \mathcal{P}(\bm{\phi})}{\mathcal{P}(\bm{\phi})}.
\end{align}
The inequality~\eqref{eq:VanTrees} provides us with a Heisenberg-limit-like bound on the sensitivity of a multi-dimensional field sensor in the Bayesian setting.

In particular, for an $N$-atom quantum sensor with Gaussian prior~\eqref{eq:GaussianPrior}, the inequality~\eqref{eq:VanTrees} is defined using the following parameters. The FI matrix of the prior $\mathcal{P}_{\delta}(\bm{\phi})$ reads
\begin{equation}
\label{eq:prior_FIM}
\mathcal I = \frac{1}{\delta^2}\mathbb{1}_d,
\end{equation}
where $\mathbb{1}_d$ is the $d$-dimensional identity matrix with $d$ the $\mathbf{B}$ field dimension. The average FI matrix is upper bounded by the QFI matrix (for even $N$)~\cite{Kolenderski2008, Vaneph2013, Gorecki2022}
\begin{equation}
\label{eq:QFIM}
\overline F \leq F_Q = \frac{N(N+2)}{d}\mathbb{1}_d.
\end{equation}
Substituting~\eqref{eq:prior_FIM} and \eqref{eq:QFIM} into the inequality~\eqref{eq:VanTrees}, we obtain the Heisenberg-limit-like bound
\begin{equation}
\Delta^2_{\rm OQS}\geq\frac{d}{N(N+2)/d + 1/\delta^{2}},
\label{eq:BayesianQL}
\end{equation}
which defines the black dotted line in Figs.~\ref{fig:2DCost} and~\ref{fig:3DCost} of the main text.

Finally, the standard quantum limit for the multi-dimensional field sensing is set by the FI matrix corresponding to the best classical input state
\begin{equation}
\label{eq:QFIM_Classical}
\overline F \leq F_{\rm SQL} = \frac{N}{d}\mathbb{1}_d.
\end{equation}
The corresponding Bayesian SQL bound reads
\begin{equation}
\Delta^2_{\rm SQL}\geq\frac{d}{N/d + 1/\delta^{2}}.
\label{eq:BayesianSQL}
\end{equation}
The SQL bound is shown with the blue dotted line in Figs.~\ref{fig:2DCost} and~\ref{fig:3DCost} of the main text.

\section{Irreducible Hilbert space of the $SU(2)$ sensor}
\label{subsec:Irreducible_Hilbert_space}
In this appendix, we provide details on the Hilbert space describing an $SU(2)$ sensor~(see Sec.~\ref{subsec:identify_OQS}).
A general quantum system of $N$
qubits is described by a quantum state living in an exponentially
large Hilbert space $\left(\mathbb{C}^{2}\right)^{\otimes N}$. Despite this, the symmetry and invariance under qubit permutations
of the generators $J_{x,y,z}$ reduce the dimensionality of the
effective Hilbert space explored by the sensor to polynomial in $N$.

More precisely, according to Weyl’s Theorem~\citep{Fulton2004}, the
full $N$-qubits space can be decomposed into orthogonal subspaces:
\begin{equation}
\label{eq:Hilber_space_decomposition}
\left(\mathbb{C}^{2}\right)^{\otimes N}=\bigoplus_{j=0(\frac{1}{2})}^{N/2}\mathbb{C}^{2j+1}\otimes\mathbb{C}^{\ell_{j}},
\end{equation}
where $j$ labels the irreducible representations of $SU(2)$ with
the lower limit in the direct sum given by $0$ or $1/2$ for even
or odd $N$, respectively. Each spin-$j$ representation occurs
with multiplicity $\ell_{j}=\binom{N}{N/2-j}\frac{2j+1}{N/2+j+1}$~\citep{Cirac1999}.
Importantly, the number $\ell_{j}$ of equivalent spin-$j$ representations
exceeds the number of spin degrees of freedom, $\ell_{j}\ge(2j+1)$,
for all but the largest spin representation $j=J\equiv N/2$. Thus, for any pure quantum state, one can always find a decomposition in terms of equivalent spin-$j$ representations in which the state populates only $2j + 1$ instead of all the $\ell_j$ representations associated with each~$j$~\cite{Bagan2004, Chiribella2004a, Chiribella2004b}. Since phases $\bm{\phi}$ are imprinted through unitaries $U(\bm{\phi})\in SU(2)$ that do not couple different angular momenta $j$ and their equivalent representations, we can, without loss of generality, describe each $j$-subspace as a pair of spin-$j$'s living in the space $\mathbb{C}^{2j+1}\otimes\mathbb{C}^{2j+1}$. Here, only the first spin of each pair is affected by the $SU(2)$ unitary $U(\bm{\phi})$. 

Summing up all spin-$j$ pairs and the single symmetric subspace ($j=J$), we obtain the irreducible Hilbert space of an $SU(2)$ sensor.
\begin{equation}
\label{eq:H_OQS}
\mathcal{H}_{\rm OQS} \equiv \mathbb{C}^{2J+1}\oplus\bigoplus_{j=0(\frac{1}{2})}^{J-1}\mathbb{C}^{2j+1}\otimes\mathbb{C}^{2j+1}.
\end{equation}
The dimension of the space scales only as the \emph{cube} of the system size~$N$,i.e., $d_{N}=(N+1)+\sum_{j=0(\frac{1}{2})}^{N/2-1}(2j+1)^{2}=1 + N(N^{2}+5)/6.$
The polynomial growth of the size of the irreducible Hilbert space is crucial for the efficient minimization of the metrological cost function~\eqref{eq:BMSE_POVM} with the numerical approach presented in the main text (see Sec.~\ref{subsec:identify_OQS}) and detailed in Appendix~\ref{subsec:Minimization-of-the-Cost}.

\section{Cost minimization as a multi-convex problem}
\label{subsec:Minimization-of-the-Cost}

In this appendix, we discuss the multi-convexity of the sensor optimization problem and present a corresponding efficient algorithm for solving Bayesian sensing problems~(see Sec.~\ref{subsec:identify_OQS}).
We start by recasting the cost~\eqref{eq:BMSE_POVM} in a form that highlights its multi-convexity and which is suitable for numerical optimization. To this end, it is convenient to define linear maps:
\begin{align}
\label{eq:Lambda_0}
    \Lambda_0\left[\bullet\right] = & \int d\bm{\phi}\, U(\bm\phi) \bullet U(\bm\phi)^{\dagger} \prior, \\
    \label{eq:Lambda_1}
    \bm{\varLambda}_1\left[\bullet\right] = & \int d\bm{\phi}\, \bm{\phi}\, U(\bm\phi) \bullet U(\bm\phi)^{\dagger} \prior,
\end{align}
where the unitary $U(\bm{\phi})=\exp\left[-i\bm{\phi}\cdot\bm{J}\right]$ imprints phases $\bm\phi$. In terms of these maps, Eq.~\eqref{eq:BMSE_POVM} can be expressed as
\begin{align}
    \mathcal{C} = \mathrm{var}(\bm{\phi}) - \sum_{\mu} \Big(& 2\, \bm{\xi}_{\mu}^T\cdot \tr\big\{ M_{\mu} \bm{\varLambda}_1\left[\rho_{\rm in}\right] \big\}\notag \\
    &- \bm{\xi}_{\mu}^T\cdot \bm{\xi}_{\mu} \tr\big\{ M_{\mu} \Lambda_0\left[\rho_{\rm in}\right]\big\}\Big),   \label{eq:cost_multi_convex}
\end{align}
where $\mathrm{var}(\bm{\phi}) = \int d \bm{\phi}\, \bm{\phi}^T \bm{\phi}\, \mathcal{P}(\bm{\phi})$ is the prior phase variance and $\rho_{\rm in } = \ket{\psiin}\bra{\psiin}$.

Equation~\eqref{eq:cost_multi_convex} shows that the minimization of the metrological cost function
is a convex optimization problem with respect to each of the three sets of variables ($\ket{\psi_{{\rm in}}}$,
$\{M_{\mu}\}$, and $\{\bm{\xi}_{\mu}\}$) if the other two are kept fixed. More precisely:

(i) For fixed $\ket{\psi_{{\rm in}}}$ and $\{M_{\mu}\}$ we have
a quadratic minimization problem for estimators $\{\bm{\xi}_{\mu}\}$.
The optimal solution is explicitly expressed as 
the minimum MSE estimator~\eqref{eq:MMSEe}, corresponding to the average value of the phases with respect to the
posterior probability density $p(\bm{\phi}|\mu)$.

(ii) For fixed $\ket{\psi_{{\rm in}}}$ and $\{\bm{\xi}_{\mu}\}$,
the cost~\eqref{eq:cost_multi_convex} is linear with respect to the measurement operators $\{M_{\mu}\}$, which are positive semidefinite ($M_{\mu}\succeq0$) and subjected to the POVM condition $\sum_{\mu}\,M_{\mu}=\openone$. Thus, the cost minimization with fixed state and estimators can be
recast as a semidefinite program (SDP). This
is an optimization problem that can be efficiently implemented with
polynomial-time solvers provided the size of the problem is polynomial, as is the case here.

(iii) For fixed $\{M_{\mu}\}$ and $\{\bm{\xi}_{\mu}\}$,
the optimization problem is quadratic with respect to the quantum
state. This can be seen by writing the cost~\eqref{eq:cost_multi_convex} as
\[
\mathcal{C}=\braket{\psi_{{\rm in}}|\Upsilon|\psi_{{\rm in}}},
\]
where the Hermitian operator $\Upsilon$ reads
\begin{equation*}
    \Upsilon = \mathrm{var}(\bm{\phi})
    -\sum_{\mu} \Big( 2\,\bm{\xi}_{\mu}^{T} \cdot \bm{\varLambda}_1^{\#}\left[ M_{\mu}\right] - \bm{\xi}_{\mu}^T\cdot \bm{\xi}_{\mu}\, \Lambda_0^{\#}\left[M_{\mu}\right]\Big),
    \label{eq:EigenvaluePsiin}
\end{equation*}
where $\Lambda_0^{\#}$ and $\bm{\varLambda}_1^{\#}$ are linear maps adjoint to Eqs.~\eqref{eq:Lambda_0} and \eqref{eq:Lambda_1}, respectively. Thus, the optimum state $\ket{\psi_{{\rm in}}^{*}}$ is given by an eigenstate of $\Upsilon$ corresponding to the lowest eigenvalue. 

The multi-convexity of the cost~\eqref{eq:cost_multi_convex} with respect
to the three sets of parameters, (i)--(iii), makes the problem of
finding the optimal multi-parameter quantum sensor approachable with
numerical algorithms. Even though multi-convex problems
are in general hard to solve globally, local methods based on block coordinate
descent work well and are widely used in practice~\cite{Shen2017}. For example, in
the case of single parameter metrology, optimization of the cost~\eqref{eq:cost_multi_convex}
becomes a biconvex problem (points (i) and (ii) are combined into a single quadratic optimization of projective measurements), leading to a
fast and reliable convergence to a unique solution~\citep{Macieszczak2014}.

In the general case, the optimization procedure can be summarized as follows: the minimization problem is initialized with a random input state, measurement, and estimators. After that, we iteratively solve subproblems (i)-(iii) until reaching convergence to the optimal sensor solution with $\ket{\psi_{\rm in}^{\star}}$, $\{M^{\star}_\mu\}$, and $\{\bm{\xi}_{\mu}^{\star}\}$ minimizing the cost~\eqref{eq:cost_multi_convex}~\footnote{We note that due to the resulting low rank of the optimal measurement operators $\{M^{\star}_q\}$, it is beneficial to use the nonlinear programming with augmented Lagrangian algorithm~\citep{Burer2003} for solving the SDP in subproblem~(ii).}.
This algorithm combined with a covariant ansatz for input states and measurements (see Appendix~\ref{subsec:covariant_POVM}) is used to obtain the OQS results presented in the main text.

\section{Covariant states and POVMs for 2p-QS}
\label{subsec:covariant_POVM}

Here, we present covariant~\citep{Holevo1982,Demkowicz2020} states and measurements for 2D field sensing, $\mathbf B=\{\text{B}_x,\text{B}_y\}$, with the 2p-QS~(see Sec.~\ref{subsec:identify_OQS}). This not only illustrates the central ideas, which can be generalized to the case of the OQS including 3D field sensing, but it also enables us to efficiently determine the 2p-QS performance limit for large system sizes, as shown in Fig.~\ref{fig:2DScaling}. 

Using the method of Appendix~\ref{subsec:Minimization-of-the-Cost} for the case of 2D fields with a priori unknown field direction (i.e. $\mathcal{P}(\bm{\phi})=\mathcal{P}(|\bm{\phi}|)$), we obtain the following optimal input state minimizing the cost~\eqref{eq:cost_multi_convex} for the 2p-QS:
\begin{equation}
\ket{\psi_{\rm in}^{({\text{2p}})}}=\bigoplus_{j=0}^{J}\alpha_{j}\ket{j,0}.
\label{eq:BQS_psi_in}
\end{equation}
The state~\eqref{eq:BQS_psi_in} features a direct sum of the eigenstates of $J_z$ with eigenvalue zero (we consider even $N$) of all $j$-representations, with corresponding amplitudes~$\alpha_j$. Consequently, the state is invariant under rotations around the $z$-axis. The numerical solution for the isotropic prior density suggests that the combination of the input state~\eqref{eq:BQS_psi_in} and the corresponding optimal measurements results in an MSE~\eqref{eq:MSE_POVM} that is also invariant under $z$-rotations. We can impose the rotational symmetry by using measurements parametrized as
\begin{equation}
M_{q\bar{\varphi}}=e^{-i\bar{\varphi}J_z}M_{q}e^{i\bar{\varphi}J_z},
\label{eq:BQS_mesurement}
\end{equation}
such that all POVM operators are determined by seed operators
$M_{q}$. The corresponding rotationally invariant estimators read
$\bm{\xi}_{q\bar{\varphi}}=\{r_{q}\cos\bar{\varphi},r_{q}\sin\bar{\varphi}\}$. Here, $r_q$ and $\bar\varphi$ are polar coordinates of the phase estimates which correspond to the field strength and direction, respectively. The estimators of the optimal 2p-QS are illustrated as concentric rings in Figs.~\ref{fig:1p-QS} and~\ref{fig:2p-QS} of the main text.

Using the ansatz~\eqref{eq:BQS_psi_in} for the input state and for the measurements~\eqref{eq:BQS_mesurement}, and the corresponding estimators and $\varphi$-invariant prior [$\mathcal{P}(r)$ in polar coordinates], the cost~\eqref{eq:cost_multi_convex} can be expressed as
\begin{equation}
\mathcal{C}=\mathrm{var}(r)
-\sum_{q}\left(2r_{q}\mathrm{Tr}\left\{ M_{q}\varrho_{1}\right\} -r_{q}^{2}\mathrm{Tr}\left\{ M_{q}\varrho_{0}\right\} \right),
\label{eq:cost_covariant}
\end{equation}
where $\mathrm{var}(r) = 2\pi\int r^3dr\,\mathcal{P}(r)$ is the prior variance
and 
\begin{align*}
\varrho_{0} & =\int rdr\,\mathcal{P}(r)\frac{d\varphi}{2\pi}e^{-i\varphi J_z}\ket{\psi_{r}}\bra{\psi_{r}}e^{i\varphi J_z},\\
\varrho_{1} & =\int r^{2}dr\,\mathcal{P}(r)\frac{d\varphi}{2\pi}\cos\varphi\,e^{-i\varphi J_z}\ket{\psi_{r}}\bra{\psi_{r}}e^{i\varphi J_z}.
\end{align*}
Here, we define the state $\ket{\psi_{r}}\equiv e^{-irJ_{x}}\ket{\psi_{{\rm in}}}$.

The cost~\eqref{eq:cost_covariant} is linear in $\{M_{q}\}$
which are semidefinite operators, $M_{q}\succeq0$, with a modified
POVM condition $\sum_{q}\int\frac{d\varphi}{2\pi}e^{-i\varphi J_z}M_{q}e^{i\varphi J_z}=\openone$.
Thus, as discussed in Appendix~\ref{subsec:Minimization-of-the-Cost},
minimization of~\eqref{eq:cost_covariant} for a fixed state~$\ket{\psiin}$ and estimators~$\{r_{q}\}$ is an SDP program.

As a result, the problem of estimation of two parameters
$\{\phi_x,\phi_y\}$ with non-commuting generators $J_x$, $J_y$ has been decoupled into two separate problems of estimation of the field strength $r$ with generator $J_x$, and the field direction $\varphi$ with generator $J_z$. The latter is explicitly solved by the corresponding covariant POVM~\eqref{eq:BQS_mesurement} and the former single-parameter problem is amenable to the numerical method of Appendix~\ref{subsec:Minimization-of-the-Cost}. Note that the resulting single-parameter problem of estimating $r$ requires optimization over POVMs. This makes it qualitatively different from single-parameter metrology problems which, in general, are solved with projective measurements~\citep{Demkowicz2020}.

The remarkable simplification of the original two-parameter problem to a single-parameter estimation task allows us to efficiently obtain the optimal 2-partite 2D field sensor. In particular, the optimal solution comprises at most $N/2+1$ measurement seed operators $M_q$ of rank~$1$. 
A generalization of the covariant ansatz approach to obtain the OQS for 2D and 3D field sensing will be presented in~\cite{Vasilyev23}.

\section{Hilbert space of 1p-QS and 2p-QS}\label{appendix:1p_and_2p_subspaces}

Here, we identify two important subspaces occupied by the OQS in different regimes, which correspond to the Hilbert spaces of the 1p-QS and 2p-QS introduced in the main text (see Sec.~\ref{sec:LimitedControl}). First, we find that in the limit of small prior widths, $\delta\lesssim1/N$, the OQS for 2D and 3D fields occupies only the maximum spin representation ($J=N/2$) in $\mathcal{H}_{\rm OQS}$~\eqref{eq:H_OQS}. The permutation symmetric subspace describes the 1p-QS, such that
\begin{equation}
\mathcal{H}_{\text{1p}} \equiv \mathbb{C}^{2J+1}.
\end{equation}

In the case of 2D field sensing at arbitrary prior widths, we observe that the OQS does not utilize the full $\mathcal{H}_{\rm OQS}$,~\eqref{eq:H_OQS}, as the optimal input state consists of weakly entangled pairs of spins in each of the $j$-subspaces. Thus, the effective Hilbert space of the 2D field OQS can be very well approximated by the 2p-QS which contains $j$-representations with multiplicity 1:
\begin{align}
\label{eq:H_1p}
\mathcal{H}_{\text{2p}} \equiv \mathbb{C}^{J+1}\otimes\mathbb{C}^{J+1} = \bigoplus_{j=0(\frac{1}{2})}^{J}\mathbb{C}^{2j+1}.
\end{align}
It follows from the discussion in Ref.~\cite{Gorecki2022} that in the limit of a large number of atoms, $N\to\infty$, the 2p-QS saturates the fundamental bound on 2D field sensing and, thus, converges to the OQS. We will discuss this convergence in more detail in Ref.~\cite{Vasilyev23}.

In contrast, the OQS for 3D fields makes use of the full $\mathcal{H}_{\rm OQS}$~\eqref{eq:H_OQS}, by employing strongly entangled pairs of spins in the $j$-subspaces for the optimal input state. Hence, in general, it cannot be efficiently represented by a sensor comprising a small and predefined number of partitions. The optimal partitioning of the 3D field sensor should depend on the sensor size $N$ and represents an interesting topic for further study.

\section{Optimization of sensors with limited entangling capabilities}
\label{app:projective_optimization}
In Sec.~\ref{sec:LimitedControl}, we discuss the performance of sensors with limited entangling capabilities, which live in $\mathcal{H}_\mathrm{1p}, \mathcal{H}_\mathrm{2p}$ and are restricted to operate with projective measurements. Here, we describe how the optimization of these sensors is carried out. 

To constrain the optimization to projective measurements, we express the measurement projectors $\Pi_m$ and $\Pi_{m^A, m^B}$ according to Eqs.~\eqref{eq:1pQS_pi} and ~\eqref{eq:2pQS_pi} for the 1p-QS and 2p-QS. We further parametrize the decoding unitary as $\mathcal{U}_{\mathrm{de}}^\mathrm{1p} = e^{-i H^\mathrm{1p}_\mathrm{de}}$ ($\mathcal{U}_{\mathrm{de}}^\mathrm{2p} = e^{-i H^\mathrm{2p}_\mathrm{de}}$), where $H^\mathrm{1p}_\mathrm{de}$ ($H^\mathrm{2p}_\mathrm{de}$) is a Hermitian matrix. Finally, the iterative procedure discussed in Appendix~\ref{subsec:Minimization-of-the-Cost} is performed, with the key difference that the optimization over the POVM set $\{M_\mu\}$ is replaced with optimization over the set of Hermitian matrices $\{H^\mathrm{1p}_\mathrm{de}\}$ [$\{H^\mathrm{2p}_\mathrm{de}\}$]. 

Unlike the case of general POVMs, the optimization problem is not convex with respect to the Hermitian matrix that parameterizes the projective measurements. Typically, the optimization is performed starting from several random initial conditions, iteratively converging to a local minimum for each starting point and then selecting the global minimum from the converged solutions. The ``brute-force'' optimization over the elements of the Hermitian matrix restricts the use of this approach to small systems. However, the results from this approach provide a saturable lower limit to the performance of variational quantum circuits, which are also restricted to projective measurements.

\section{Variational optimization}
\label{app:VQ_optimization}
In Secs.~\ref{sec:VariationalCircuits} and \ref{sec:3D}, we present results that are obtained by optimizing variational quantum circuits  where the unitaries $\mathcal{U}_{\rm en(de)}$ are parameterized by the parameter vectors $\bm{\theta}$ and $\bm{\vartheta}$. The gates considered in this work are all periodic; nonetheless, the parameter space grows exponentially in the number of parameters, that is, convergence to the global minimum~\eqref {eq:varCost} of the cost function cannot be guaranteed if the depth of the circuit grows. 

The results in this work were obtained using gradient-descent optimization based on exact gradient calculations. To approach the global minimum, we optimized starting from a number of initial conditions and kept the best result. In a first step, we sampled random initial parameters and later refined the variational solution by choosing parameters that were optimal for other settings of the sensor as starting points. This involved starting with optimal parameters for different $N$ and/or $\delta$, adding layers to shorter circuits, and removing closest-to-identity layers from longer circuits.

\begin{thebibliography}{96}%
\makeatletter
\providecommand \@ifxundefined [1]{%
 \@ifx{#1\undefined}
}%
\providecommand \@ifnum [1]{%
 \ifnum #1\expandafter \@firstoftwo
 \else \expandafter \@secondoftwo
 \fi
}%
\providecommand \@ifx [1]{%
 \ifx #1\expandafter \@firstoftwo
 \else \expandafter \@secondoftwo
 \fi
}%
\providecommand \natexlab [1]{#1}%
\providecommand \enquote  [1]{``#1''}%
\providecommand \bibnamefont  [1]{#1}%
\providecommand \bibfnamefont [1]{#1}%
\providecommand \citenamefont [1]{#1}%
\providecommand \href@noop [0]{\@secondoftwo}%
\providecommand \href [0]{\begingroup \@sanitize@url \@href}%
\providecommand \@href[1]{\@@startlink{#1}\@@href}%
\providecommand \@@href[1]{\endgroup#1\@@endlink}%
\providecommand \@sanitize@url [0]{\catcode `\\12\catcode `\$12\catcode
  `\&12\catcode `\#12\catcode `\^12\catcode `\_12\catcode `\%12\relax}%
\providecommand \@@startlink[1]{}%
\providecommand \@@endlink[0]{}%
\providecommand \url  [0]{\begingroup\@sanitize@url \@url }%
\providecommand \@url [1]{\endgroup\@href {#1}{\urlprefix }}%
\providecommand \urlprefix  [0]{URL }%
\providecommand \Eprint [0]{\href }%
\providecommand \doibase [0]{https://doi.org/}%
\providecommand \selectlanguage [0]{\@gobble}%
\providecommand \bibinfo  [0]{\@secondoftwo}%
\providecommand \bibfield  [0]{\@secondoftwo}%
\providecommand \translation [1]{[#1]}%
\providecommand \BibitemOpen [0]{}%
\providecommand \bibitemStop [0]{}%
\providecommand \bibitemNoStop [0]{.\EOS\space}%
\providecommand \EOS [0]{\spacefactor3000\relax}%
\providecommand \BibitemShut  [1]{\csname bibitem#1\endcsname}%
\let\auto@bib@innerbib\@empty
\bibitem [{\citenamefont {Holevo}(1982)}]{Holevo1982}%
  \BibitemOpen
  \bibfield  {author} {\bibinfo {author} {\bibfnamefont {A.}~\bibnamefont
  {Holevo}},\ }\href@noop {} {\emph {\bibinfo {title} {Probabilistic and
  Statistical Aspects of Quantum Theory}}},\ North-Holland series in statistics
  and probability\ (\bibinfo  {publisher} {North-Holland Publishing Company},\
  \bibinfo {year} {1982})\BibitemShut {NoStop}%
\bibitem [{\citenamefont {Demkowicz-Dobrza{\'n}ski}\ \emph
  {et~al.}(2015)\citenamefont {Demkowicz-Dobrza{\'n}ski}, \citenamefont
  {Jarzyna},\ and\ \citenamefont {Ko{\l}ody{\'n}ski}}]{Demkowicz2015}%
  \BibitemOpen
  \bibfield  {author} {\bibinfo {author} {\bibfnamefont {R.}~\bibnamefont
  {Demkowicz-Dobrza{\'n}ski}}, \bibinfo {author} {\bibfnamefont
  {M.}~\bibnamefont {Jarzyna}},\ and\ \bibinfo {author} {\bibfnamefont
  {J.}~\bibnamefont {Ko{\l}ody{\'n}ski}},\ }\href
  {https://doi.org/https://doi.org/10.1016/bs.po.2015.02.003} {\emph {\bibinfo
  {title} {Quantum Limits in Optical Interferometry}}},\ \bibinfo {series}
  {Progress in Optics}, Vol.~\bibinfo {volume} {60}\ (\bibinfo  {publisher}
  {Elsevier},\ \bibinfo {year} {2015})\ pp.\ \bibinfo {pages} {345 --
  435}\BibitemShut {NoStop}%
\bibitem [{\citenamefont {Degen}\ \emph {et~al.}(2017)\citenamefont {Degen},
  \citenamefont {Reinhard},\ and\ \citenamefont {Cappellaro}}]{Degen2017}%
  \BibitemOpen
  \bibfield  {author} {\bibinfo {author} {\bibfnamefont {C.~L.}\ \bibnamefont
  {Degen}}, \bibinfo {author} {\bibfnamefont {F.}~\bibnamefont {Reinhard}},\
  and\ \bibinfo {author} {\bibfnamefont {P.}~\bibnamefont {Cappellaro}},\
  }\bibfield  {title} {\bibinfo {title} {Quantum sensing},\ }\href
  {https://doi.org/10.1103/RevModPhys.89.035002} {\bibfield  {journal}
  {\bibinfo  {journal} {Rev. Mod. Phys.}\ }\textbf {\bibinfo {volume} {89}},\
  \bibinfo {pages} {035002} (\bibinfo {year} {2017})}\BibitemShut {NoStop}%
\bibitem [{\citenamefont {Pezz\`e}\ \emph {et~al.}(2018)\citenamefont
  {Pezz\`e}, \citenamefont {Smerzi}, \citenamefont {Oberthaler}, \citenamefont
  {Schmied},\ and\ \citenamefont {Treutlein}}]{Pezze2018}%
  \BibitemOpen
  \bibfield  {author} {\bibinfo {author} {\bibfnamefont {L.}~\bibnamefont
  {Pezz\`e}}, \bibinfo {author} {\bibfnamefont {A.}~\bibnamefont {Smerzi}},
  \bibinfo {author} {\bibfnamefont {M.~K.}\ \bibnamefont {Oberthaler}},
  \bibinfo {author} {\bibfnamefont {R.}~\bibnamefont {Schmied}},\ and\ \bibinfo
  {author} {\bibfnamefont {P.}~\bibnamefont {Treutlein}},\ }\bibfield  {title}
  {\bibinfo {title} {Quantum metrology with nonclassical states of atomic
  ensembles},\ }\href {https://doi.org/10.1103/RevModPhys.90.035005} {\bibfield
   {journal} {\bibinfo  {journal} {Rev. Mod. Phys.}\ }\textbf {\bibinfo
  {volume} {90}},\ \bibinfo {pages} {035005} (\bibinfo {year}
  {2018})}\BibitemShut {NoStop}%
\bibitem [{\citenamefont {Macieszczak}\ \emph {et~al.}(2014)\citenamefont
  {Macieszczak}, \citenamefont {Fraas},\ and\ \citenamefont
  {Demkowicz-Dobrza{\'{n}}ski}}]{Macieszczak2014}%
  \BibitemOpen
  \bibfield  {author} {\bibinfo {author} {\bibfnamefont {K.}~\bibnamefont
  {Macieszczak}}, \bibinfo {author} {\bibfnamefont {M.}~\bibnamefont {Fraas}},\
  and\ \bibinfo {author} {\bibfnamefont {R.}~\bibnamefont
  {Demkowicz-Dobrza{\'{n}}ski}},\ }\bibfield  {title} {\bibinfo {title}
  {Bayesian quantum frequency estimation in presence of collective dephasing},\
  }\href {https://doi.org/10.1088/1367-2630/16/11/113002} {\bibfield  {journal}
  {\bibinfo  {journal} {New Journal of Physics}\ }\textbf {\bibinfo {volume}
  {16}},\ \bibinfo {pages} {113002} (\bibinfo {year} {2014})}\BibitemShut
  {NoStop}%
\bibitem [{\citenamefont {Kaubruegger}\ \emph {et~al.}(2021)\citenamefont
  {Kaubruegger}, \citenamefont {Vasilyev}, \citenamefont {Schulte},
  \citenamefont {Hammerer},\ and\ \citenamefont {Zoller}}]{Kaubruegger2021}%
  \BibitemOpen
  \bibfield  {author} {\bibinfo {author} {\bibfnamefont {R.}~\bibnamefont
  {Kaubruegger}}, \bibinfo {author} {\bibfnamefont {D.~V.}\ \bibnamefont
  {Vasilyev}}, \bibinfo {author} {\bibfnamefont {M.}~\bibnamefont {Schulte}},
  \bibinfo {author} {\bibfnamefont {K.}~\bibnamefont {Hammerer}},\ and\
  \bibinfo {author} {\bibfnamefont {P.}~\bibnamefont {Zoller}},\ }\bibfield
  {title} {\bibinfo {title} {Quantum variational optimization of ramsey
  interferometry and atomic clocks},\ }\href
  {https://doi.org/10.1103/PhysRevX.11.041045} {\bibfield  {journal} {\bibinfo
  {journal} {Phys. Rev. X}\ }\textbf {\bibinfo {volume} {11}},\ \bibinfo
  {pages} {041045} (\bibinfo {year} {2021})}\BibitemShut {NoStop}%
\bibitem [{\citenamefont {Marciniak}\ \emph {et~al.}(2022)\citenamefont
  {Marciniak}, \citenamefont {Feldker}, \citenamefont {Pogorelov},
  \citenamefont {Kaubruegger}, \citenamefont {Vasilyev}, \citenamefont {van
  Bijnen}, \citenamefont {Schindler}, \citenamefont {Zoller}, \citenamefont
  {Blatt},\ and\ \citenamefont {Monz}}]{Marciniak2021}%
  \BibitemOpen
  \bibfield  {author} {\bibinfo {author} {\bibfnamefont {C.~D.}\ \bibnamefont
  {Marciniak}}, \bibinfo {author} {\bibfnamefont {T.}~\bibnamefont {Feldker}},
  \bibinfo {author} {\bibfnamefont {I.}~\bibnamefont {Pogorelov}}, \bibinfo
  {author} {\bibfnamefont {R.}~\bibnamefont {Kaubruegger}}, \bibinfo {author}
  {\bibfnamefont {D.~V.}\ \bibnamefont {Vasilyev}}, \bibinfo {author}
  {\bibfnamefont {R.}~\bibnamefont {van Bijnen}}, \bibinfo {author}
  {\bibfnamefont {P.}~\bibnamefont {Schindler}}, \bibinfo {author}
  {\bibfnamefont {P.}~\bibnamefont {Zoller}}, \bibinfo {author} {\bibfnamefont
  {R.}~\bibnamefont {Blatt}},\ and\ \bibinfo {author} {\bibfnamefont
  {T.}~\bibnamefont {Monz}},\ }\bibfield  {title} {\bibinfo {title} {Optimal
  metrology with programmable quantum sensors},\ }\href
  {https://doi.org/10.1038/s41586-022-04435-4} {\bibfield  {journal} {\bibinfo
  {journal} {Nature}\ }\textbf {\bibinfo {volume} {603}},\ \bibinfo {pages}
  {604} (\bibinfo {year} {2022})}\BibitemShut {NoStop}%
\bibitem [{\citenamefont {Gessner}\ \emph {et~al.}(2018)\citenamefont
  {Gessner}, \citenamefont {Pezz\`e},\ and\ \citenamefont
  {Smerzi}}]{Gessner2018}%
  \BibitemOpen
  \bibfield  {author} {\bibinfo {author} {\bibfnamefont {M.}~\bibnamefont
  {Gessner}}, \bibinfo {author} {\bibfnamefont {L.}~\bibnamefont {Pezz\`e}},\
  and\ \bibinfo {author} {\bibfnamefont {A.}~\bibnamefont {Smerzi}},\
  }\bibfield  {title} {\bibinfo {title} {Sensitivity bounds for multiparameter
  quantum metrology},\ }\href {https://doi.org/10.1103/PhysRevLett.121.130503}
  {\bibfield  {journal} {\bibinfo  {journal} {Phys. Rev. Lett.}\ }\textbf
  {\bibinfo {volume} {121}},\ \bibinfo {pages} {130503} (\bibinfo {year}
  {2018})}\BibitemShut {NoStop}%
\bibitem [{\citenamefont {Demkowicz-Dobrza{\'{n}}ski}\ \emph
  {et~al.}(2020)\citenamefont {Demkowicz-Dobrza{\'{n}}ski}, \citenamefont
  {G{\'{o}}recki},\ and\ \citenamefont {Gu{\c{t}}{\u{a}}}}]{Demkowicz2020}%
  \BibitemOpen
  \bibfield  {author} {\bibinfo {author} {\bibfnamefont {R.}~\bibnamefont
  {Demkowicz-Dobrza{\'{n}}ski}}, \bibinfo {author} {\bibfnamefont
  {W.}~\bibnamefont {G{\'{o}}recki}},\ and\ \bibinfo {author} {\bibfnamefont
  {M.}~\bibnamefont {Gu{\c{t}}{\u{a}}}},\ }\bibfield  {title} {\bibinfo {title}
  {Multi-parameter estimation beyond quantum {Fisher} information},\ }\href
  {https://doi.org/10.1088/1751-8121/ab8ef3} {\bibfield  {journal} {\bibinfo
  {journal} {Journal of Physics A: Mathematical and Theoretical}\ }\textbf
  {\bibinfo {volume} {53}},\ \bibinfo {pages} {363001} (\bibinfo {year}
  {2020})}\BibitemShut {NoStop}%
\bibitem [{\citenamefont {Liu}\ \emph {et~al.}(2019)\citenamefont {Liu},
  \citenamefont {Yuan}, \citenamefont {Lu},\ and\ \citenamefont
  {Wang}}]{Liu2020}%
  \BibitemOpen
  \bibfield  {author} {\bibinfo {author} {\bibfnamefont {J.}~\bibnamefont
  {Liu}}, \bibinfo {author} {\bibfnamefont {H.}~\bibnamefont {Yuan}}, \bibinfo
  {author} {\bibfnamefont {X.-M.}\ \bibnamefont {Lu}},\ and\ \bibinfo {author}
  {\bibfnamefont {X.}~\bibnamefont {Wang}},\ }\bibfield  {title} {\bibinfo
  {title} {Quantum fisher information matrix and multiparameter estimation},\
  }\href {https://doi.org/10.1088/1751-8121/ab5d4d} {\bibfield  {journal}
  {\bibinfo  {journal} {Journal of Physics A: Mathematical and Theoretical}\
  }\textbf {\bibinfo {volume} {53}},\ \bibinfo {pages} {023001} (\bibinfo
  {year} {2019})}\BibitemShut {NoStop}%
\bibitem [{\citenamefont {Sidhu}\ and\ \citenamefont {Kok}(2020)}]{Sidhu2020}%
  \BibitemOpen
  \bibfield  {author} {\bibinfo {author} {\bibfnamefont {J.~S.}\ \bibnamefont
  {Sidhu}}\ and\ \bibinfo {author} {\bibfnamefont {P.}~\bibnamefont {Kok}},\
  }\bibfield  {title} {\bibinfo {title} {Geometric perspective on quantum
  parameter estimation},\ }\href {https://doi.org/10.1116/1.5119961} {\bibfield
   {journal} {\bibinfo  {journal} {AVS Quantum Science}\ }\textbf {\bibinfo
  {volume} {2}},\ \bibinfo {pages} {014701} (\bibinfo {year} {2020})},\ \Eprint
  {https://arxiv.org/abs/https://doi.org/10.1116/1.5119961}
  {https://doi.org/10.1116/1.5119961} \BibitemShut {NoStop}%
\bibitem [{\citenamefont {Sidhu}\ \emph {et~al.}(2021)\citenamefont {Sidhu},
  \citenamefont {Ouyang}, \citenamefont {Campbell},\ and\ \citenamefont
  {Kok}}]{Sidhu2021}%
  \BibitemOpen
  \bibfield  {author} {\bibinfo {author} {\bibfnamefont {J.~S.}\ \bibnamefont
  {Sidhu}}, \bibinfo {author} {\bibfnamefont {Y.}~\bibnamefont {Ouyang}},
  \bibinfo {author} {\bibfnamefont {E.~T.}\ \bibnamefont {Campbell}},\ and\
  \bibinfo {author} {\bibfnamefont {P.}~\bibnamefont {Kok}},\ }\bibfield
  {title} {\bibinfo {title} {Tight bounds on the simultaneous estimation of
  incompatible parameters},\ }\href
  {https://doi.org/10.1103/PhysRevX.11.011028} {\bibfield  {journal} {\bibinfo
  {journal} {Phys. Rev. X}\ }\textbf {\bibinfo {volume} {11}},\ \bibinfo
  {pages} {011028} (\bibinfo {year} {2021})}\BibitemShut {NoStop}%
\bibitem [{\citenamefont {Humphreys}\ \emph {et~al.}(2013)\citenamefont
  {Humphreys}, \citenamefont {Barbieri}, \citenamefont {Datta},\ and\
  \citenamefont {Walmsley}}]{Humphreys2013}%
  \BibitemOpen
  \bibfield  {author} {\bibinfo {author} {\bibfnamefont {P.~C.}\ \bibnamefont
  {Humphreys}}, \bibinfo {author} {\bibfnamefont {M.}~\bibnamefont {Barbieri}},
  \bibinfo {author} {\bibfnamefont {A.}~\bibnamefont {Datta}},\ and\ \bibinfo
  {author} {\bibfnamefont {I.~A.}\ \bibnamefont {Walmsley}},\ }\bibfield
  {title} {\bibinfo {title} {Quantum enhanced multiple phase estimation},\
  }\href {https://doi.org/10.1103/PhysRevLett.111.070403} {\bibfield  {journal}
  {\bibinfo  {journal} {Phys. Rev. Lett.}\ }\textbf {\bibinfo {volume} {111}},\
  \bibinfo {pages} {070403} (\bibinfo {year} {2013})}\BibitemShut {NoStop}%
\bibitem [{\citenamefont {Gessner}\ \emph {et~al.}(2020)\citenamefont
  {Gessner}, \citenamefont {Smerzi},\ and\ \citenamefont
  {Pezz{\`e}}}]{Gessner2020}%
  \BibitemOpen
  \bibfield  {author} {\bibinfo {author} {\bibfnamefont {M.}~\bibnamefont
  {Gessner}}, \bibinfo {author} {\bibfnamefont {A.}~\bibnamefont {Smerzi}},\
  and\ \bibinfo {author} {\bibfnamefont {L.}~\bibnamefont {Pezz{\`e}}},\
  }\bibfield  {title} {\bibinfo {title} {Multiparameter squeezing for optimal
  quantum enhancements in sensor networks},\ }\href
  {https://doi.org/10.1038/s41467-020-17471-3} {\bibfield  {journal} {\bibinfo
  {journal} {Nature Communications}\ }\textbf {\bibinfo {volume} {11}},\
  \bibinfo {pages} {3817} (\bibinfo {year} {2020})}\BibitemShut {NoStop}%
\bibitem [{\citenamefont {Huang}\ \emph {et~al.}(2021)\citenamefont {Huang},
  \citenamefont {Lupo},\ and\ \citenamefont {Kok}}]{Huang2021}%
  \BibitemOpen
  \bibfield  {author} {\bibinfo {author} {\bibfnamefont {Z.}~\bibnamefont
  {Huang}}, \bibinfo {author} {\bibfnamefont {C.}~\bibnamefont {Lupo}},\ and\
  \bibinfo {author} {\bibfnamefont {P.}~\bibnamefont {Kok}},\ }\bibfield
  {title} {\bibinfo {title} {Quantum-limited estimation of range and
  velocity},\ }\href {https://doi.org/10.1103/PRXQuantum.2.030303} {\bibfield
  {journal} {\bibinfo  {journal} {PRX Quantum}\ }\textbf {\bibinfo {volume}
  {2}},\ \bibinfo {pages} {030303} (\bibinfo {year} {2021})}\BibitemShut
  {NoStop}%
\bibitem [{\citenamefont {Belliardo}\ \emph {et~al.}(2022)\citenamefont
  {Belliardo}, \citenamefont {Cimini}, \citenamefont {Polino}, \citenamefont
  {Hoch}, \citenamefont {Piccirillo}, \citenamefont {Spagnolo}, \citenamefont
  {Giovannetti},\ and\ \citenamefont {Sciarrino}}]{Belliardo2022}%
  \BibitemOpen
  \bibfield  {author} {\bibinfo {author} {\bibfnamefont {F.}~\bibnamefont
  {Belliardo}}, \bibinfo {author} {\bibfnamefont {V.}~\bibnamefont {Cimini}},
  \bibinfo {author} {\bibfnamefont {E.}~\bibnamefont {Polino}}, \bibinfo
  {author} {\bibfnamefont {F.}~\bibnamefont {Hoch}}, \bibinfo {author}
  {\bibfnamefont {B.}~\bibnamefont {Piccirillo}}, \bibinfo {author}
  {\bibfnamefont {N.}~\bibnamefont {Spagnolo}}, \bibinfo {author}
  {\bibfnamefont {V.}~\bibnamefont {Giovannetti}},\ and\ \bibinfo {author}
  {\bibfnamefont {F.}~\bibnamefont {Sciarrino}},\ }\bibfield  {title} {\bibinfo
  {title} {Optimizing quantum-enhanced {Bayesian} multiparameter estimation in
  noisy apparata},\ }\bibfield  {journal} {\bibinfo  {journal} {arXiv}\ }\href
  {https://doi.org/10.48550/ARXIV.2211.04747} {10.48550/ARXIV.2211.04747}
  (\bibinfo {year} {2022})\BibitemShut {NoStop}%
\bibitem [{\citenamefont {Baamara}\ \emph {et~al.}(2022)\citenamefont
  {Baamara}, \citenamefont {Gessner},\ and\ \citenamefont
  {Sinatra}}]{Baamara2022}%
  \BibitemOpen
  \bibfield  {author} {\bibinfo {author} {\bibfnamefont {Y.}~\bibnamefont
  {Baamara}}, \bibinfo {author} {\bibfnamefont {M.}~\bibnamefont {Gessner}},\
  and\ \bibinfo {author} {\bibfnamefont {A.}~\bibnamefont {Sinatra}},\
  }\bibfield  {title} {\bibinfo {title} {Quantum-enhanced multiparameter
  estimation and compressed sensing of a field},\ }\bibfield  {journal}
  {\bibinfo  {journal} {arXiv}\ }\href
  {https://doi.org/10.48550/ARXIV.2208.00640} {10.48550/ARXIV.2208.00640}
  (\bibinfo {year} {2022})\BibitemShut {NoStop}%
\bibitem [{\citenamefont {Conlon}\ \emph {et~al.}(2023)\citenamefont {Conlon},
  \citenamefont {Vogl}, \citenamefont {Marciniak}, \citenamefont {Pogorelov},
  \citenamefont {Yung}, \citenamefont {Eilenberger}, \citenamefont {Berry},
  \citenamefont {Santana}, \citenamefont {Blatt}, \citenamefont {Monz},
  \citenamefont {Lam},\ and\ \citenamefont {Assad}}]{Conlon2023}%
  \BibitemOpen
  \bibfield  {author} {\bibinfo {author} {\bibfnamefont {L.~O.}\ \bibnamefont
  {Conlon}}, \bibinfo {author} {\bibfnamefont {T.}~\bibnamefont {Vogl}},
  \bibinfo {author} {\bibfnamefont {C.~D.}\ \bibnamefont {Marciniak}}, \bibinfo
  {author} {\bibfnamefont {I.}~\bibnamefont {Pogorelov}}, \bibinfo {author}
  {\bibfnamefont {S.~K.}\ \bibnamefont {Yung}}, \bibinfo {author}
  {\bibfnamefont {F.}~\bibnamefont {Eilenberger}}, \bibinfo {author}
  {\bibfnamefont {D.~W.}\ \bibnamefont {Berry}}, \bibinfo {author}
  {\bibfnamefont {F.~S.}\ \bibnamefont {Santana}}, \bibinfo {author}
  {\bibfnamefont {R.}~\bibnamefont {Blatt}}, \bibinfo {author} {\bibfnamefont
  {T.}~\bibnamefont {Monz}}, \bibinfo {author} {\bibfnamefont {P.~K.}\
  \bibnamefont {Lam}},\ and\ \bibinfo {author} {\bibfnamefont {S.~M.}\
  \bibnamefont {Assad}},\ }\bibfield  {title} {\bibinfo {title} {Approaching
  optimal entangling collective measurements on quantum computing platforms},\
  }\bibfield  {journal} {\bibinfo  {journal} {Nature Physics}\ }\href
  {https://doi.org/10.1038/s41567-022-01875-7} {10.1038/s41567-022-01875-7}
  (\bibinfo {year} {2023})\BibitemShut {NoStop}%
\bibitem [{\citenamefont {Eriksson}\ \emph {et~al.}(2023)\citenamefont
  {Eriksson}, \citenamefont {Goldberg}, \citenamefont {Hiekkam{\"a}ki},
  \citenamefont {Bouchard}, \citenamefont {Leuchs}, \citenamefont {Fickler},\
  and\ \citenamefont {Sanchez-Soto}}]{Eriksson2023}%
  \BibitemOpen
  \bibfield  {author} {\bibinfo {author} {\bibfnamefont {M.}~\bibnamefont
  {Eriksson}}, \bibinfo {author} {\bibfnamefont {A.~Z.}\ \bibnamefont
  {Goldberg}}, \bibinfo {author} {\bibfnamefont {M.}~\bibnamefont
  {Hiekkam{\"a}ki}}, \bibinfo {author} {\bibfnamefont {F.}~\bibnamefont
  {Bouchard}}, \bibinfo {author} {\bibfnamefont {G.}~\bibnamefont {Leuchs}},
  \bibinfo {author} {\bibfnamefont {R.}~\bibnamefont {Fickler}},\ and\ \bibinfo
  {author} {\bibfnamefont {L.~L.}\ \bibnamefont {Sanchez-Soto}},\ }\bibfield
  {title} {\bibinfo {title} {Sensing rotations with multiplane light
  conversion},\ }\bibfield  {journal} {\bibinfo  {journal} {arXiv}\ }\href
  {https://doi.org/10.48550/ARXIV.2301.10265} {10.48550/ARXIV.2301.10265}
  (\bibinfo {year} {2023})\BibitemShut {NoStop}%
\bibitem [{\citenamefont {Vaneph}\ \emph {et~al.}(2013)\citenamefont {Vaneph},
  \citenamefont {Tufarelli},\ and\ \citenamefont {Genoni}}]{Vaneph2013}%
  \BibitemOpen
  \bibfield  {author} {\bibinfo {author} {\bibfnamefont {C.}~\bibnamefont
  {Vaneph}}, \bibinfo {author} {\bibfnamefont {T.}~\bibnamefont {Tufarelli}},\
  and\ \bibinfo {author} {\bibfnamefont {M.~G.}\ \bibnamefont {Genoni}},\
  }\bibfield  {title} {\bibinfo {title} {Quantum estimation of a two-phase spin
  rotation},\ }\href {https://doi.org/doi:10.2478/qmetro-2013-0003} {\bibfield
  {journal} {\bibinfo  {journal} {Quantum Measurements and Quantum Metrology}\
  }\textbf {\bibinfo {volume} {1}},\ \bibinfo {pages} {12} (\bibinfo {year}
  {2013})}\BibitemShut {NoStop}%
\bibitem [{\citenamefont {Baumgratz}\ and\ \citenamefont
  {Datta}(2016)}]{Baumgratz2016}%
  \BibitemOpen
  \bibfield  {author} {\bibinfo {author} {\bibfnamefont {T.}~\bibnamefont
  {Baumgratz}}\ and\ \bibinfo {author} {\bibfnamefont {A.}~\bibnamefont
  {Datta}},\ }\bibfield  {title} {\bibinfo {title} {{Quantum Enhanced
  Estimation of a Multidimensional Field}},\ }\href
  {https://doi.org/10.1103/PhysRevLett.116.030801} {\bibfield  {journal}
  {\bibinfo  {journal} {Physical Review Letters}\ }\textbf {\bibinfo {volume}
  {116}},\ \bibinfo {pages} {030801} (\bibinfo {year} {2016})}\BibitemShut
  {NoStop}%
\bibitem [{\citenamefont {G{\'{o}}recki}\ and\ \citenamefont
  {Demkowicz-Dobrza{\'{n}}ski}(2022)}]{Gorecki2022}%
  \BibitemOpen
  \bibfield  {author} {\bibinfo {author} {\bibfnamefont {W.}~\bibnamefont
  {G{\'{o}}recki}}\ and\ \bibinfo {author} {\bibfnamefont {R.}~\bibnamefont
  {Demkowicz-Dobrza{\'{n}}ski}},\ }\bibfield  {title} {\bibinfo {title}
  {{Multiparameter quantum metrology in the Heisenberg limit regime:
  Many-repetition scenario versus full optimization}},\ }\href
  {https://doi.org/10.1103/PhysRevA.106.012424} {\bibfield  {journal} {\bibinfo
   {journal} {Physical Review A}\ }\textbf {\bibinfo {volume} {106}},\ \bibinfo
  {pages} {012424} (\bibinfo {year} {2022})}\BibitemShut {NoStop}%
\bibitem [{\citenamefont {Budker}\ and\ \citenamefont
  {Romalis}(2007)}]{Budker2007}%
  \BibitemOpen
  \bibfield  {author} {\bibinfo {author} {\bibfnamefont {D.}~\bibnamefont
  {Budker}}\ and\ \bibinfo {author} {\bibfnamefont {M.}~\bibnamefont
  {Romalis}},\ }\bibfield  {title} {\bibinfo {title} {Optical magnetometry},\
  }\href {https://doi.org/10.1038/nphys566} {\bibfield  {journal} {\bibinfo
  {journal} {Nature Physics}\ }\textbf {\bibinfo {volume} {3}},\ \bibinfo
  {pages} {227} (\bibinfo {year} {2007})}\BibitemShut {NoStop}%
\bibitem [{\citenamefont {Budker}\ and\ \citenamefont
  {Kimball}(2013)}]{Budker2013}%
  \BibitemOpen
  \bibinfo {editor} {\bibfnamefont {D.}~\bibnamefont {Budker}}\ and\ \bibinfo
  {editor} {\bibfnamefont {D.~F.~J.}\ \bibnamefont {Kimball}},\ eds.,\ \href
  {https://doi.org/10.1017/CBO9780511846380} {\emph {\bibinfo {title} {Optical
  Magnetometry}}}\ (\bibinfo  {publisher} {Cambridge University Press},\
  \bibinfo {year} {2013})\BibitemShut {NoStop}%
\bibitem [{\citenamefont {Le~Sage}\ \emph {et~al.}(2013)\citenamefont
  {Le~Sage}, \citenamefont {Arai}, \citenamefont {Glenn}, \citenamefont
  {DeVience}, \citenamefont {Pham}, \citenamefont {Rahn-Lee}, \citenamefont
  {Lukin}, \citenamefont {Yacoby}, \citenamefont {Komeili},\ and\ \citenamefont
  {Walsworth}}]{LeSage2013}%
  \BibitemOpen
  \bibfield  {author} {\bibinfo {author} {\bibfnamefont {D.}~\bibnamefont
  {Le~Sage}}, \bibinfo {author} {\bibfnamefont {K.}~\bibnamefont {Arai}},
  \bibinfo {author} {\bibfnamefont {D.~R.}\ \bibnamefont {Glenn}}, \bibinfo
  {author} {\bibfnamefont {S.~J.}\ \bibnamefont {DeVience}}, \bibinfo {author}
  {\bibfnamefont {L.~M.}\ \bibnamefont {Pham}}, \bibinfo {author}
  {\bibfnamefont {L.}~\bibnamefont {Rahn-Lee}}, \bibinfo {author}
  {\bibfnamefont {M.~D.}\ \bibnamefont {Lukin}}, \bibinfo {author}
  {\bibfnamefont {A.}~\bibnamefont {Yacoby}}, \bibinfo {author} {\bibfnamefont
  {A.}~\bibnamefont {Komeili}},\ and\ \bibinfo {author} {\bibfnamefont {R.~L.}\
  \bibnamefont {Walsworth}},\ }\bibfield  {title} {\bibinfo {title} {Optical
  magnetic imaging of living cells},\ }\href
  {https://doi.org/10.1038/nature12072} {\bibfield  {journal} {\bibinfo
  {journal} {Nature}\ }\textbf {\bibinfo {volume} {496}},\ \bibinfo {pages}
  {486} (\bibinfo {year} {2013})}\BibitemShut {NoStop}%
\bibitem [{\citenamefont {Behbood}\ \emph {et~al.}(2013)\citenamefont
  {Behbood}, \citenamefont {Martin~Ciurana}, \citenamefont {Colangelo},
  \citenamefont {Napolitano}, \citenamefont {Mitchell},\ and\ \citenamefont
  {Sewell}}]{Behbood2013}%
  \BibitemOpen
  \bibfield  {author} {\bibinfo {author} {\bibfnamefont {N.}~\bibnamefont
  {Behbood}}, \bibinfo {author} {\bibfnamefont {F.}~\bibnamefont
  {Martin~Ciurana}}, \bibinfo {author} {\bibfnamefont {G.}~\bibnamefont
  {Colangelo}}, \bibinfo {author} {\bibfnamefont {M.}~\bibnamefont
  {Napolitano}}, \bibinfo {author} {\bibfnamefont {M.~W.}\ \bibnamefont
  {Mitchell}},\ and\ \bibinfo {author} {\bibfnamefont {R.~J.}\ \bibnamefont
  {Sewell}},\ }\bibfield  {title} {\bibinfo {title} {Real-time vector field
  tracking with a cold-atom magnetometer},\ }\href
  {https://doi.org/10.1063/1.4803684} {\bibfield  {journal} {\bibinfo
  {journal} {Applied Physics Letters}\ }\textbf {\bibinfo {volume} {102}},\
  \bibinfo {pages} {173504} (\bibinfo {year} {2013})},\ \Eprint
  {https://arxiv.org/abs/https://doi.org/10.1063/1.4803684}
  {https://doi.org/10.1063/1.4803684} \BibitemShut {NoStop}%
\bibitem [{\citenamefont {Lee}\ \emph {et~al.}(2015)\citenamefont {Lee},
  \citenamefont {Niethammer},\ and\ \citenamefont {Wrachtrup}}]{Lee2015}%
  \BibitemOpen
  \bibfield  {author} {\bibinfo {author} {\bibfnamefont {S.-Y.}\ \bibnamefont
  {Lee}}, \bibinfo {author} {\bibfnamefont {M.}~\bibnamefont {Niethammer}},\
  and\ \bibinfo {author} {\bibfnamefont {J.}~\bibnamefont {Wrachtrup}},\
  }\bibfield  {title} {\bibinfo {title} {Vector magnetometry based on
  $s=\frac{3}{2}$ electronic spins},\ }\href
  {https://doi.org/10.1103/PhysRevB.92.115201} {\bibfield  {journal} {\bibinfo
  {journal} {Phys. Rev. B}\ }\textbf {\bibinfo {volume} {92}},\ \bibinfo
  {pages} {115201} (\bibinfo {year} {2015})}\BibitemShut {NoStop}%
\bibitem [{\citenamefont {Thiele}\ \emph {et~al.}(2018)\citenamefont {Thiele},
  \citenamefont {Lin}, \citenamefont {Brown},\ and\ \citenamefont
  {Regal}}]{Thiele2018}%
  \BibitemOpen
  \bibfield  {author} {\bibinfo {author} {\bibfnamefont {T.}~\bibnamefont
  {Thiele}}, \bibinfo {author} {\bibfnamefont {Y.}~\bibnamefont {Lin}},
  \bibinfo {author} {\bibfnamefont {M.~O.}\ \bibnamefont {Brown}},\ and\
  \bibinfo {author} {\bibfnamefont {C.~A.}\ \bibnamefont {Regal}},\ }\bibfield
  {title} {\bibinfo {title} {Self-calibrating vector atomic magnetometry
  through microwave polarization reconstruction},\ }\href
  {https://doi.org/10.1103/PhysRevLett.121.153202} {\bibfield  {journal}
  {\bibinfo  {journal} {Phys. Rev. Lett.}\ }\textbf {\bibinfo {volume} {121}},\
  \bibinfo {pages} {153202} (\bibinfo {year} {2018})}\BibitemShut {NoStop}%
\bibitem [{\citenamefont {Zheng}\ \emph {et~al.}(2020)\citenamefont {Zheng},
  \citenamefont {Sun}, \citenamefont {Chatzidrosos}, \citenamefont {Zhang},
  \citenamefont {Nakamura}, \citenamefont {Sumiya}, \citenamefont {Ohshima},
  \citenamefont {Isoya}, \citenamefont {Wrachtrup}, \citenamefont
  {Wickenbrock},\ and\ \citenamefont {Budker}}]{Zheng2020}%
  \BibitemOpen
  \bibfield  {author} {\bibinfo {author} {\bibfnamefont {H.}~\bibnamefont
  {Zheng}}, \bibinfo {author} {\bibfnamefont {Z.}~\bibnamefont {Sun}}, \bibinfo
  {author} {\bibfnamefont {G.}~\bibnamefont {Chatzidrosos}}, \bibinfo {author}
  {\bibfnamefont {C.}~\bibnamefont {Zhang}}, \bibinfo {author} {\bibfnamefont
  {K.}~\bibnamefont {Nakamura}}, \bibinfo {author} {\bibfnamefont
  {H.}~\bibnamefont {Sumiya}}, \bibinfo {author} {\bibfnamefont
  {T.}~\bibnamefont {Ohshima}}, \bibinfo {author} {\bibfnamefont
  {J.}~\bibnamefont {Isoya}}, \bibinfo {author} {\bibfnamefont
  {J.}~\bibnamefont {Wrachtrup}}, \bibinfo {author} {\bibfnamefont
  {A.}~\bibnamefont {Wickenbrock}},\ and\ \bibinfo {author} {\bibfnamefont
  {D.}~\bibnamefont {Budker}},\ }\bibfield  {title} {\bibinfo {title}
  {Microwave-free vector magnetometry with nitrogen-vacancy centers along a
  single axis in diamond},\ }\href
  {https://doi.org/10.1103/PhysRevApplied.13.044023} {\bibfield  {journal}
  {\bibinfo  {journal} {Phys. Rev. Appl.}\ }\textbf {\bibinfo {volume} {13}},\
  \bibinfo {pages} {044023} (\bibinfo {year} {2020})}\BibitemShut {NoStop}%
\bibitem [{\citenamefont {Wang}\ \emph {et~al.}(2021)\citenamefont {Wang},
  \citenamefont {Liu}, \citenamefont {Zhu},\ and\ \citenamefont
  {Cappellaro}}]{Wang2021}%
  \BibitemOpen
  \bibfield  {author} {\bibinfo {author} {\bibfnamefont {G.}~\bibnamefont
  {Wang}}, \bibinfo {author} {\bibfnamefont {Y.-X.}\ \bibnamefont {Liu}},
  \bibinfo {author} {\bibfnamefont {Y.}~\bibnamefont {Zhu}},\ and\ \bibinfo
  {author} {\bibfnamefont {P.}~\bibnamefont {Cappellaro}},\ }\bibfield  {title}
  {\bibinfo {title} {Nanoscale vector ac magnetometry with a single
  nitrogen-vacancy center in diamond},\ }\href
  {https://doi.org/10.1021/acs.nanolett.1c01165} {\bibfield  {journal}
  {\bibinfo  {journal} {Nano Letters}\ }\textbf {\bibinfo {volume} {21}},\
  \bibinfo {pages} {5143} (\bibinfo {year} {2021})},\ \bibinfo {note} {pMID:
  34086471}\BibitemShut {NoStop}%
\bibitem [{\citenamefont {Fan}\ \emph {et~al.}(2015)\citenamefont {Fan},
  \citenamefont {Kumar}, \citenamefont {Sedlacek}, \citenamefont {K{\"u}bler},
  \citenamefont {Karimkashi},\ and\ \citenamefont {Shaffer}}]{Fan2015}%
  \BibitemOpen
  \bibfield  {author} {\bibinfo {author} {\bibfnamefont {H.}~\bibnamefont
  {Fan}}, \bibinfo {author} {\bibfnamefont {S.}~\bibnamefont {Kumar}}, \bibinfo
  {author} {\bibfnamefont {J.}~\bibnamefont {Sedlacek}}, \bibinfo {author}
  {\bibfnamefont {H.}~\bibnamefont {K{\"u}bler}}, \bibinfo {author}
  {\bibfnamefont {S.}~\bibnamefont {Karimkashi}},\ and\ \bibinfo {author}
  {\bibfnamefont {J.~P.}\ \bibnamefont {Shaffer}},\ }\bibfield  {title}
  {\bibinfo {title} {Atom based rf electric field sensing},\ }\href
  {https://doi.org/10.1088/0953-4075/48/20/202001} {\bibfield  {journal}
  {\bibinfo  {journal} {Journal of Physics B: Atomic, Molecular and Optical
  Physics}\ }\textbf {\bibinfo {volume} {48}},\ \bibinfo {pages} {202001}
  (\bibinfo {year} {2015})}\BibitemShut {NoStop}%
\bibitem [{\citenamefont {Brownnutt}\ \emph {et~al.}(2015)\citenamefont
  {Brownnutt}, \citenamefont {Kumph}, \citenamefont {Rabl},\ and\ \citenamefont
  {Blatt}}]{Brownutt2015}%
  \BibitemOpen
  \bibfield  {author} {\bibinfo {author} {\bibfnamefont {M.}~\bibnamefont
  {Brownnutt}}, \bibinfo {author} {\bibfnamefont {M.}~\bibnamefont {Kumph}},
  \bibinfo {author} {\bibfnamefont {P.}~\bibnamefont {Rabl}},\ and\ \bibinfo
  {author} {\bibfnamefont {R.}~\bibnamefont {Blatt}},\ }\bibfield  {title}
  {\bibinfo {title} {Ion-trap measurements of electric-field noise near
  surfaces},\ }\href {https://doi.org/10.1103/RevModPhys.87.1419} {\bibfield
  {journal} {\bibinfo  {journal} {Rev. Mod. Phys.}\ }\textbf {\bibinfo {volume}
  {87}},\ \bibinfo {pages} {1419} (\bibinfo {year} {2015})}\BibitemShut
  {NoStop}%
\bibitem [{\citenamefont {Sedlacek}\ \emph {et~al.}(2013)\citenamefont
  {Sedlacek}, \citenamefont {Schwettmann}, \citenamefont {K\"ubler},\ and\
  \citenamefont {Shaffer}}]{Sedlacek2013}%
  \BibitemOpen
  \bibfield  {author} {\bibinfo {author} {\bibfnamefont {J.~A.}\ \bibnamefont
  {Sedlacek}}, \bibinfo {author} {\bibfnamefont {A.}~\bibnamefont
  {Schwettmann}}, \bibinfo {author} {\bibfnamefont {H.}~\bibnamefont
  {K\"ubler}},\ and\ \bibinfo {author} {\bibfnamefont {J.~P.}\ \bibnamefont
  {Shaffer}},\ }\bibfield  {title} {\bibinfo {title} {Atom-based vector
  microwave electrometry using rubidium {Rydberg} atoms in a vapor cell},\
  }\href {https://doi.org/10.1103/PhysRevLett.111.063001} {\bibfield  {journal}
  {\bibinfo  {journal} {Phys. Rev. Lett.}\ }\textbf {\bibinfo {volume} {111}},\
  \bibinfo {pages} {063001} (\bibinfo {year} {2013})}\BibitemShut {NoStop}%
\bibitem [{\citenamefont {Albarelli}\ \emph {et~al.}(2019)\citenamefont
  {Albarelli}, \citenamefont {Friel},\ and\ \citenamefont
  {Datta}}]{Albarelli2019}%
  \BibitemOpen
  \bibfield  {author} {\bibinfo {author} {\bibfnamefont {F.}~\bibnamefont
  {Albarelli}}, \bibinfo {author} {\bibfnamefont {J.~F.}\ \bibnamefont
  {Friel}},\ and\ \bibinfo {author} {\bibfnamefont {A.}~\bibnamefont {Datta}},\
  }\bibfield  {title} {\bibinfo {title} {{Evaluating the Holevo
  Cram{\'{e}}r-Rao Bound for Multiparameter Quantum Metrology}},\ }\href
  {https://doi.org/10.1103/PhysRevLett.123.200503} {\bibfield  {journal}
  {\bibinfo  {journal} {Physical Review Letters}\ }\textbf {\bibinfo {volume}
  {123}},\ \bibinfo {pages} {200503} (\bibinfo {year} {2019})}\BibitemShut
  {NoStop}%
\bibitem [{\citenamefont {Tsang}\ \emph {et~al.}(2011)\citenamefont {Tsang},
  \citenamefont {Wiseman},\ and\ \citenamefont {Caves}}]{Tsang2011}%
  \BibitemOpen
  \bibfield  {author} {\bibinfo {author} {\bibfnamefont {M.}~\bibnamefont
  {Tsang}}, \bibinfo {author} {\bibfnamefont {H.~M.}\ \bibnamefont {Wiseman}},\
  and\ \bibinfo {author} {\bibfnamefont {C.~M.}\ \bibnamefont {Caves}},\
  }\bibfield  {title} {\bibinfo {title} {Fundamental quantum limit to waveform
  estimation},\ }\href {https://doi.org/10.1103/PhysRevLett.106.090401}
  {\bibfield  {journal} {\bibinfo  {journal} {Phys. Rev. Lett.}\ }\textbf
  {\bibinfo {volume} {106}},\ \bibinfo {pages} {090401} (\bibinfo {year}
  {2011})}\BibitemShut {NoStop}%
\bibitem [{\citenamefont {Bagan}\ \emph {et~al.}(2001)\citenamefont {Bagan},
  \citenamefont {Baig}, \citenamefont {Brey}, \citenamefont {Mu\~noz-Tapia},\
  and\ \citenamefont {Tarrach}}]{Bagan2001}%
  \BibitemOpen
  \bibfield  {author} {\bibinfo {author} {\bibfnamefont {E.}~\bibnamefont
  {Bagan}}, \bibinfo {author} {\bibfnamefont {M.}~\bibnamefont {Baig}},
  \bibinfo {author} {\bibfnamefont {A.}~\bibnamefont {Brey}}, \bibinfo {author}
  {\bibfnamefont {R.}~\bibnamefont {Mu\~noz-Tapia}},\ and\ \bibinfo {author}
  {\bibfnamefont {R.}~\bibnamefont {Tarrach}},\ }\bibfield  {title} {\bibinfo
  {title} {Optimal encoding and decoding of a spin direction},\ }\href
  {https://doi.org/10.1103/PhysRevA.63.052309} {\bibfield  {journal} {\bibinfo
  {journal} {Phys. Rev. A}\ }\textbf {\bibinfo {volume} {63}},\ \bibinfo
  {pages} {052309} (\bibinfo {year} {2001})}\BibitemShut {NoStop}%
\bibitem [{\citenamefont {Chiribella}\ \emph
  {et~al.}(2004{\natexlab{a}})\citenamefont {Chiribella}, \citenamefont
  {D'Ariano}, \citenamefont {Perinotti},\ and\ \citenamefont
  {Sacchi}}]{Chiribella2004a}%
  \BibitemOpen
  \bibfield  {author} {\bibinfo {author} {\bibfnamefont {G.}~\bibnamefont
  {Chiribella}}, \bibinfo {author} {\bibfnamefont {G.~M.}\ \bibnamefont
  {D'Ariano}}, \bibinfo {author} {\bibfnamefont {P.}~\bibnamefont
  {Perinotti}},\ and\ \bibinfo {author} {\bibfnamefont {M.~F.}\ \bibnamefont
  {Sacchi}},\ }\bibfield  {title} {\bibinfo {title} {{Efficient Use of Quantum
  Resources for the Transmission of a Reference Frame}},\ }\href
  {https://doi.org/10.1103/PhysRevLett.93.180503} {\bibfield  {journal}
  {\bibinfo  {journal} {Physical Review Letters}\ }\textbf {\bibinfo {volume}
  {93}},\ \bibinfo {pages} {180503} (\bibinfo {year}
  {2004}{\natexlab{a}})}\BibitemShut {NoStop}%
\bibitem [{\citenamefont {Bagan}\ \emph {et~al.}(2004)\citenamefont {Bagan},
  \citenamefont {Baig},\ and\ \citenamefont {Mu\~noz-Tapia}}]{Bagan2004}%
  \BibitemOpen
  \bibfield  {author} {\bibinfo {author} {\bibfnamefont {E.}~\bibnamefont
  {Bagan}}, \bibinfo {author} {\bibfnamefont {M.}~\bibnamefont {Baig}},\ and\
  \bibinfo {author} {\bibfnamefont {R.}~\bibnamefont {Mu\~noz-Tapia}},\
  }\bibfield  {title} {\bibinfo {title} {Quantum reverse engineering and
  reference-frame alignment without nonlocal correlations},\ }\href
  {https://doi.org/10.1103/PhysRevA.70.030301} {\bibfield  {journal} {\bibinfo
  {journal} {Phys. Rev. A}\ }\textbf {\bibinfo {volume} {70}},\ \bibinfo
  {pages} {030301(R)} (\bibinfo {year} {2004})}\BibitemShut {NoStop}%
\bibitem [{\citenamefont {Kahn}(2007)}]{Kahn2007}%
  \BibitemOpen
  \bibfield  {author} {\bibinfo {author} {\bibfnamefont {J.}~\bibnamefont
  {Kahn}},\ }\bibfield  {title} {\bibinfo {title} {{Fast rate estimation of a
  unitary operation in SU(d)}},\ }\href
  {https://doi.org/10.1103/PhysRevA.75.022326} {\bibfield  {journal} {\bibinfo
  {journal} {Physical Review A}\ }\textbf {\bibinfo {volume} {75}},\ \bibinfo
  {pages} {022326} (\bibinfo {year} {2007})}\BibitemShut {NoStop}%
\bibitem [{\citenamefont {Wasilewski}\ \emph {et~al.}(2010)\citenamefont
  {Wasilewski}, \citenamefont {Jensen}, \citenamefont {Krauter}, \citenamefont
  {Renema}, \citenamefont {Balabas},\ and\ \citenamefont
  {Polzik}}]{Wasilewski2010}%
  \BibitemOpen
  \bibfield  {author} {\bibinfo {author} {\bibfnamefont {W.}~\bibnamefont
  {Wasilewski}}, \bibinfo {author} {\bibfnamefont {K.}~\bibnamefont {Jensen}},
  \bibinfo {author} {\bibfnamefont {H.}~\bibnamefont {Krauter}}, \bibinfo
  {author} {\bibfnamefont {J.~J.}\ \bibnamefont {Renema}}, \bibinfo {author}
  {\bibfnamefont {M.~V.}\ \bibnamefont {Balabas}},\ and\ \bibinfo {author}
  {\bibfnamefont {E.~S.}\ \bibnamefont {Polzik}},\ }\bibfield  {title}
  {\bibinfo {title} {Quantum noise limited and entanglement-assisted
  magnetometry},\ }\href {https://doi.org/10.1103/PhysRevLett.104.133601}
  {\bibfield  {journal} {\bibinfo  {journal} {Phys. Rev. Lett.}\ }\textbf
  {\bibinfo {volume} {104}},\ \bibinfo {pages} {133601} (\bibinfo {year}
  {2010})}\BibitemShut {NoStop}%
\bibitem [{\citenamefont {Sewell}\ \emph {et~al.}(2012)\citenamefont {Sewell},
  \citenamefont {Koschorreck}, \citenamefont {Napolitano}, \citenamefont
  {Dubost}, \citenamefont {Behbood},\ and\ \citenamefont
  {Mitchell}}]{Sewell2012}%
  \BibitemOpen
  \bibfield  {author} {\bibinfo {author} {\bibfnamefont {R.~J.}\ \bibnamefont
  {Sewell}}, \bibinfo {author} {\bibfnamefont {M.}~\bibnamefont {Koschorreck}},
  \bibinfo {author} {\bibfnamefont {M.}~\bibnamefont {Napolitano}}, \bibinfo
  {author} {\bibfnamefont {B.}~\bibnamefont {Dubost}}, \bibinfo {author}
  {\bibfnamefont {N.}~\bibnamefont {Behbood}},\ and\ \bibinfo {author}
  {\bibfnamefont {M.~W.}\ \bibnamefont {Mitchell}},\ }\bibfield  {title}
  {\bibinfo {title} {Magnetic sensitivity beyond the projection noise limit by
  spin squeezing},\ }\href {https://doi.org/10.1103/PhysRevLett.109.253605}
  {\bibfield  {journal} {\bibinfo  {journal} {Phys. Rev. Lett.}\ }\textbf
  {\bibinfo {volume} {109}},\ \bibinfo {pages} {253605} (\bibinfo {year}
  {2012})}\BibitemShut {NoStop}%
\bibitem [{\citenamefont {Ockeloen}\ \emph {et~al.}(2013)\citenamefont
  {Ockeloen}, \citenamefont {Schmied}, \citenamefont {Riedel},\ and\
  \citenamefont {Treutlein}}]{Ockeloen2013}%
  \BibitemOpen
  \bibfield  {author} {\bibinfo {author} {\bibfnamefont {C.~F.}\ \bibnamefont
  {Ockeloen}}, \bibinfo {author} {\bibfnamefont {R.}~\bibnamefont {Schmied}},
  \bibinfo {author} {\bibfnamefont {M.~F.}\ \bibnamefont {Riedel}},\ and\
  \bibinfo {author} {\bibfnamefont {P.}~\bibnamefont {Treutlein}},\ }\bibfield
  {title} {\bibinfo {title} {Quantum metrology with a scanning probe atom
  interferometer},\ }\href {https://doi.org/10.1103/PhysRevLett.111.143001}
  {\bibfield  {journal} {\bibinfo  {journal} {Phys. Rev. Lett.}\ }\textbf
  {\bibinfo {volume} {111}},\ \bibinfo {pages} {143001} (\bibinfo {year}
  {2013})}\BibitemShut {NoStop}%
\bibitem [{\citenamefont {Arias}\ \emph {et~al.}(2019)\citenamefont {Arias},
  \citenamefont {Lochead}, \citenamefont {Wintermantel}, \citenamefont
  {Helmrich},\ and\ \citenamefont {Whitlock}}]{Arias2019}%
  \BibitemOpen
  \bibfield  {author} {\bibinfo {author} {\bibfnamefont {A.}~\bibnamefont
  {Arias}}, \bibinfo {author} {\bibfnamefont {G.}~\bibnamefont {Lochead}},
  \bibinfo {author} {\bibfnamefont {T.~M.}\ \bibnamefont {Wintermantel}},
  \bibinfo {author} {\bibfnamefont {S.}~\bibnamefont {Helmrich}},\ and\
  \bibinfo {author} {\bibfnamefont {S.}~\bibnamefont {Whitlock}},\ }\bibfield
  {title} {\bibinfo {title} {Realization of a {Rydberg}-dressed {Ramsey}
  interferometer and electrometer},\ }\href
  {https://doi.org/10.1103/PhysRevLett.122.053601} {\bibfield  {journal}
  {\bibinfo  {journal} {Phys. Rev. Lett.}\ }\textbf {\bibinfo {volume} {122}},\
  \bibinfo {pages} {053601} (\bibinfo {year} {2019})}\BibitemShut {NoStop}%
\bibitem [{\citenamefont {Facon}\ \emph {et~al.}(2016)\citenamefont {Facon},
  \citenamefont {Dietsche}, \citenamefont {Grosso}, \citenamefont {Haroche},
  \citenamefont {Raimond}, \citenamefont {Brune},\ and\ \citenamefont
  {Gleyzes}}]{Facon2016}%
  \BibitemOpen
  \bibfield  {author} {\bibinfo {author} {\bibfnamefont {A.}~\bibnamefont
  {Facon}}, \bibinfo {author} {\bibfnamefont {E.-K.}\ \bibnamefont {Dietsche}},
  \bibinfo {author} {\bibfnamefont {D.}~\bibnamefont {Grosso}}, \bibinfo
  {author} {\bibfnamefont {S.}~\bibnamefont {Haroche}}, \bibinfo {author}
  {\bibfnamefont {J.-M.}\ \bibnamefont {Raimond}}, \bibinfo {author}
  {\bibfnamefont {M.}~\bibnamefont {Brune}},\ and\ \bibinfo {author}
  {\bibfnamefont {S.}~\bibnamefont {Gleyzes}},\ }\bibfield  {title} {\bibinfo
  {title} {A sensitive electrometer based on a {Rydberg} atom in a
  {Schr{\"o}dinger}-cat state},\ }\href {https://doi.org/10.1038/nature18327}
  {\bibfield  {journal} {\bibinfo  {journal} {Nature}\ }\textbf {\bibinfo
  {volume} {535}},\ \bibinfo {pages} {262} (\bibinfo {year}
  {2016})}\BibitemShut {NoStop}%
\bibitem [{\citenamefont {Omran}\ \emph {et~al.}(2019)\citenamefont {Omran},
  \citenamefont {Levine}, \citenamefont {Keesling}, \citenamefont {Semeghini},
  \citenamefont {Wang}, \citenamefont {Ebadi}, \citenamefont {Bernien},
  \citenamefont {Zibrov}, \citenamefont {Pichler}, \citenamefont {Choi},
  \citenamefont {Cui}, \citenamefont {Rossignolo}, \citenamefont {Rembold},
  \citenamefont {Montangero}, \citenamefont {Calarco}, \citenamefont {Endres},
  \citenamefont {Greiner}, \citenamefont {Vuleti{\'c}},\ and\ \citenamefont
  {Lukin}}]{Omran2019}%
  \BibitemOpen
  \bibfield  {author} {\bibinfo {author} {\bibfnamefont {A.}~\bibnamefont
  {Omran}}, \bibinfo {author} {\bibfnamefont {H.}~\bibnamefont {Levine}},
  \bibinfo {author} {\bibfnamefont {A.}~\bibnamefont {Keesling}}, \bibinfo
  {author} {\bibfnamefont {G.}~\bibnamefont {Semeghini}}, \bibinfo {author}
  {\bibfnamefont {T.~T.}\ \bibnamefont {Wang}}, \bibinfo {author}
  {\bibfnamefont {S.}~\bibnamefont {Ebadi}}, \bibinfo {author} {\bibfnamefont
  {H.}~\bibnamefont {Bernien}}, \bibinfo {author} {\bibfnamefont {A.~S.}\
  \bibnamefont {Zibrov}}, \bibinfo {author} {\bibfnamefont {H.}~\bibnamefont
  {Pichler}}, \bibinfo {author} {\bibfnamefont {S.}~\bibnamefont {Choi}},
  \bibinfo {author} {\bibfnamefont {J.}~\bibnamefont {Cui}}, \bibinfo {author}
  {\bibfnamefont {M.}~\bibnamefont {Rossignolo}}, \bibinfo {author}
  {\bibfnamefont {P.}~\bibnamefont {Rembold}}, \bibinfo {author} {\bibfnamefont
  {S.}~\bibnamefont {Montangero}}, \bibinfo {author} {\bibfnamefont
  {T.}~\bibnamefont {Calarco}}, \bibinfo {author} {\bibfnamefont
  {M.}~\bibnamefont {Endres}}, \bibinfo {author} {\bibfnamefont
  {M.}~\bibnamefont {Greiner}}, \bibinfo {author} {\bibfnamefont
  {V.}~\bibnamefont {Vuleti{\'c}}},\ and\ \bibinfo {author} {\bibfnamefont
  {M.~D.}\ \bibnamefont {Lukin}},\ }\bibfield  {title} {\bibinfo {title}
  {Generation and manipulation of {Schr\"odinger} cat states in {Rydberg} atom
  arrays},\ }\href {https://doi.org/10.1126/science.aax9743} {\bibfield
  {journal} {\bibinfo  {journal} {Science}\ }\textbf {\bibinfo {volume}
  {365}},\ \bibinfo {pages} {570} (\bibinfo {year} {2019})},\ \Eprint
  {https://arxiv.org/abs/https://www.science.org/doi/pdf/10.1126/science.aax9743}
  {https://www.science.org/doi/pdf/10.1126/science.aax9743} \BibitemShut
  {NoStop}%
\bibitem [{\citenamefont {Robinson}\ \emph {et~al.}(2022)\citenamefont
  {Robinson}, \citenamefont {Miklos}, \citenamefont {Tso}, \citenamefont
  {Kennedy}, \citenamefont {Bothwell}, \citenamefont {Kedar}, \citenamefont
  {Thompson},\ and\ \citenamefont {Ye}}]{Robinson2022}%
  \BibitemOpen
  \bibfield  {author} {\bibinfo {author} {\bibfnamefont {J.~M.}\ \bibnamefont
  {Robinson}}, \bibinfo {author} {\bibfnamefont {M.}~\bibnamefont {Miklos}},
  \bibinfo {author} {\bibfnamefont {Y.~M.}\ \bibnamefont {Tso}}, \bibinfo
  {author} {\bibfnamefont {C.~J.}\ \bibnamefont {Kennedy}}, \bibinfo {author}
  {\bibfnamefont {T.}~\bibnamefont {Bothwell}}, \bibinfo {author}
  {\bibfnamefont {D.}~\bibnamefont {Kedar}}, \bibinfo {author} {\bibfnamefont
  {J.~K.}\ \bibnamefont {Thompson}},\ and\ \bibinfo {author} {\bibfnamefont
  {J.}~\bibnamefont {Ye}},\ }\bibfield  {title} {\bibinfo {title} {Direct
  comparison of two spin squeezed optical clocks below the quantum projection
  noise limit},\ }\bibfield  {journal} {\bibinfo  {journal} {arXiv}\ }\href
  {https://doi.org/10.48550/ARXIV.2211.08621} {10.48550/ARXIV.2211.08621}
  (\bibinfo {year} {2022})\BibitemShut {NoStop}%
\bibitem [{\citenamefont {Carr}\ \emph {et~al.}(2009)\citenamefont {Carr},
  \citenamefont {DeMille}, \citenamefont {Krems},\ and\ \citenamefont
  {Ye}}]{Carr2009}%
  \BibitemOpen
  \bibfield  {author} {\bibinfo {author} {\bibfnamefont {L.~D.}\ \bibnamefont
  {Carr}}, \bibinfo {author} {\bibfnamefont {D.}~\bibnamefont {DeMille}},
  \bibinfo {author} {\bibfnamefont {R.~V.}\ \bibnamefont {Krems}},\ and\
  \bibinfo {author} {\bibfnamefont {J.}~\bibnamefont {Ye}},\ }\bibfield
  {title} {\bibinfo {title} {Cold and ultracold molecules: science, technology
  and applications},\ }\href {https://doi.org/10.1088/1367-2630/11/5/055049}
  {\bibfield  {journal} {\bibinfo  {journal} {New Journal of Physics}\ }\textbf
  {\bibinfo {volume} {11}},\ \bibinfo {pages} {055049} (\bibinfo {year}
  {2009})}\BibitemShut {NoStop}%
\bibitem [{\citenamefont {Anderegg}\ \emph {et~al.}(2019)\citenamefont
  {Anderegg}, \citenamefont {Cheuk}, \citenamefont {Bao}, \citenamefont
  {Burchesky}, \citenamefont {Ketterle}, \citenamefont {Ni},\ and\
  \citenamefont {Doyle}}]{Loic2019}%
  \BibitemOpen
  \bibfield  {author} {\bibinfo {author} {\bibfnamefont {L.}~\bibnamefont
  {Anderegg}}, \bibinfo {author} {\bibfnamefont {L.~W.}\ \bibnamefont {Cheuk}},
  \bibinfo {author} {\bibfnamefont {Y.}~\bibnamefont {Bao}}, \bibinfo {author}
  {\bibfnamefont {S.}~\bibnamefont {Burchesky}}, \bibinfo {author}
  {\bibfnamefont {W.}~\bibnamefont {Ketterle}}, \bibinfo {author}
  {\bibfnamefont {K.-K.}\ \bibnamefont {Ni}},\ and\ \bibinfo {author}
  {\bibfnamefont {J.~M.}\ \bibnamefont {Doyle}},\ }\bibfield  {title} {\bibinfo
  {title} {An optical tweezer array of ultracold molecules},\ }\href
  {https://doi.org/10.1126/science.aax1265} {\bibfield  {journal} {\bibinfo
  {journal} {Science}\ }\textbf {\bibinfo {volume} {365}},\ \bibinfo {pages}
  {1156} (\bibinfo {year} {2019})},\ \Eprint
  {https://arxiv.org/abs/https://www.science.org/doi/pdf/10.1126/science.aax1265}
  {https://www.science.org/doi/pdf/10.1126/science.aax1265} \BibitemShut
  {NoStop}%
\bibitem [{\citenamefont {Bermudez}\ \emph {et~al.}(2011)\citenamefont
  {Bermudez}, \citenamefont {Jelezko}, \citenamefont {Plenio},\ and\
  \citenamefont {Retzker}}]{Bermudez2011}%
  \BibitemOpen
  \bibfield  {author} {\bibinfo {author} {\bibfnamefont {A.}~\bibnamefont
  {Bermudez}}, \bibinfo {author} {\bibfnamefont {F.}~\bibnamefont {Jelezko}},
  \bibinfo {author} {\bibfnamefont {M.~B.}\ \bibnamefont {Plenio}},\ and\
  \bibinfo {author} {\bibfnamefont {A.}~\bibnamefont {Retzker}},\ }\bibfield
  {title} {\bibinfo {title} {Electron-mediated nuclear-spin interactions
  between distant nitrogen-vacancy centers},\ }\href
  {https://doi.org/10.1103/PhysRevLett.107.150503} {\bibfield  {journal}
  {\bibinfo  {journal} {Phys. Rev. Lett.}\ }\textbf {\bibinfo {volume} {107}},\
  \bibinfo {pages} {150503} (\bibinfo {year} {2011})}\BibitemShut {NoStop}%
\bibitem [{\citenamefont {Yao}\ \emph {et~al.}(2012)\citenamefont {Yao},
  \citenamefont {Jiang}, \citenamefont {Gorshkov}, \citenamefont {Maurer},
  \citenamefont {Giedke}, \citenamefont {Cirac},\ and\ \citenamefont
  {Lukin}}]{Yao2012}%
  \BibitemOpen
  \bibfield  {author} {\bibinfo {author} {\bibfnamefont {N.~Y.}\ \bibnamefont
  {Yao}}, \bibinfo {author} {\bibfnamefont {L.}~\bibnamefont {Jiang}}, \bibinfo
  {author} {\bibfnamefont {A.~V.}\ \bibnamefont {Gorshkov}}, \bibinfo {author}
  {\bibfnamefont {P.~C.}\ \bibnamefont {Maurer}}, \bibinfo {author}
  {\bibfnamefont {G.}~\bibnamefont {Giedke}}, \bibinfo {author} {\bibfnamefont
  {J.~I.}\ \bibnamefont {Cirac}},\ and\ \bibinfo {author} {\bibfnamefont
  {M.~D.}\ \bibnamefont {Lukin}},\ }\bibfield  {title} {\bibinfo {title}
  {Scalable architecture for a room temperature solid-state quantum information
  processor},\ }\href {https://doi.org/10.1038/ncomms1788} {\bibfield
  {journal} {\bibinfo  {journal} {Nature Communications}\ }\textbf {\bibinfo
  {volume} {3}},\ \bibinfo {pages} {800} (\bibinfo {year} {2012})}\BibitemShut
  {NoStop}%
\bibitem [{\citenamefont {Bouchard}\ \emph {et~al.}(2017)\citenamefont
  {Bouchard}, \citenamefont {de~la Hoz}, \citenamefont {Bj\"{o}rk},
  \citenamefont {Boyd}, \citenamefont {Grassl}, \citenamefont {Hradil},
  \citenamefont {Karimi}, \citenamefont {Klimov}, \citenamefont {Leuchs},
  \citenamefont {\v{R}eh\'{a}\v{c}ek},\ and\ \citenamefont
  {S\'{a}nchez-Soto}}]{Bouchard2017}%
  \BibitemOpen
  \bibfield  {author} {\bibinfo {author} {\bibfnamefont {F.}~\bibnamefont
  {Bouchard}}, \bibinfo {author} {\bibfnamefont {P.}~\bibnamefont {de~la Hoz}},
  \bibinfo {author} {\bibfnamefont {G.}~\bibnamefont {Bj\"{o}rk}}, \bibinfo
  {author} {\bibfnamefont {R.~W.}\ \bibnamefont {Boyd}}, \bibinfo {author}
  {\bibfnamefont {M.}~\bibnamefont {Grassl}}, \bibinfo {author} {\bibfnamefont
  {Z.}~\bibnamefont {Hradil}}, \bibinfo {author} {\bibfnamefont
  {E.}~\bibnamefont {Karimi}}, \bibinfo {author} {\bibfnamefont {A.~B.}\
  \bibnamefont {Klimov}}, \bibinfo {author} {\bibfnamefont {G.}~\bibnamefont
  {Leuchs}}, \bibinfo {author} {\bibfnamefont {J.}~\bibnamefont
  {\v{R}eh\'{a}\v{c}ek}},\ and\ \bibinfo {author} {\bibfnamefont {L.~L.}\
  \bibnamefont {S\'{a}nchez-Soto}},\ }\bibfield  {title} {\bibinfo {title}
  {Quantum metrology at the limit with extremal majorana constellations},\
  }\href {https://doi.org/10.1364/OPTICA.4.001429} {\bibfield  {journal}
  {\bibinfo  {journal} {Optica}\ }\textbf {\bibinfo {volume} {4}},\ \bibinfo
  {pages} {1429} (\bibinfo {year} {2017})}\BibitemShut {NoStop}%
\bibitem [{\citenamefont {Brandt}\ \emph {et~al.}(2020)\citenamefont {Brandt},
  \citenamefont {Hiekkam\"{a}ki}, \citenamefont {Bouchard}, \citenamefont
  {Huber},\ and\ \citenamefont {Fickler}}]{Brandt2020}%
  \BibitemOpen
  \bibfield  {author} {\bibinfo {author} {\bibfnamefont {F.}~\bibnamefont
  {Brandt}}, \bibinfo {author} {\bibfnamefont {M.}~\bibnamefont
  {Hiekkam\"{a}ki}}, \bibinfo {author} {\bibfnamefont {F.}~\bibnamefont
  {Bouchard}}, \bibinfo {author} {\bibfnamefont {M.}~\bibnamefont {Huber}},\
  and\ \bibinfo {author} {\bibfnamefont {R.}~\bibnamefont {Fickler}},\
  }\bibfield  {title} {\bibinfo {title} {High-dimensional quantum gates using
  full-field spatial modes of photons},\ }\href
  {https://doi.org/10.1364/OPTICA.375875} {\bibfield  {journal} {\bibinfo
  {journal} {Optica}\ }\textbf {\bibinfo {volume} {7}},\ \bibinfo {pages} {98}
  (\bibinfo {year} {2020})}\BibitemShut {NoStop}%
\bibitem [{\citenamefont {Kitagawa}\ and\ \citenamefont
  {Ueda}(1993)}]{Kitagawa1993}%
  \BibitemOpen
  \bibfield  {author} {\bibinfo {author} {\bibfnamefont {M.}~\bibnamefont
  {Kitagawa}}\ and\ \bibinfo {author} {\bibfnamefont {M.}~\bibnamefont
  {Ueda}},\ }\bibfield  {title} {\bibinfo {title} {Squeezed spin states},\
  }\href {https://doi.org/10.1103/PhysRevA.47.5138} {\bibfield  {journal}
  {\bibinfo  {journal} {Phys. Rev. A}\ }\textbf {\bibinfo {volume} {47}},\
  \bibinfo {pages} {5138} (\bibinfo {year} {1993})}\BibitemShut {NoStop}%
\bibitem [{\citenamefont {Cerezo}\ \emph {et~al.}(2021)\citenamefont {Cerezo},
  \citenamefont {Arrasmith}, \citenamefont {Babbush}, \citenamefont {Benjamin},
  \citenamefont {Endo}, \citenamefont {Fujii}, \citenamefont {McClean},
  \citenamefont {Mitarai}, \citenamefont {Yuan}, \citenamefont {Cincio},\ and\
  \citenamefont {Coles}}]{Cerezo2021}%
  \BibitemOpen
  \bibfield  {author} {\bibinfo {author} {\bibfnamefont {M.}~\bibnamefont
  {Cerezo}}, \bibinfo {author} {\bibfnamefont {A.}~\bibnamefont {Arrasmith}},
  \bibinfo {author} {\bibfnamefont {R.}~\bibnamefont {Babbush}}, \bibinfo
  {author} {\bibfnamefont {S.~C.}\ \bibnamefont {Benjamin}}, \bibinfo {author}
  {\bibfnamefont {S.}~\bibnamefont {Endo}}, \bibinfo {author} {\bibfnamefont
  {K.}~\bibnamefont {Fujii}}, \bibinfo {author} {\bibfnamefont {J.~R.}\
  \bibnamefont {McClean}}, \bibinfo {author} {\bibfnamefont {K.}~\bibnamefont
  {Mitarai}}, \bibinfo {author} {\bibfnamefont {X.}~\bibnamefont {Yuan}},
  \bibinfo {author} {\bibfnamefont {L.}~\bibnamefont {Cincio}},\ and\ \bibinfo
  {author} {\bibfnamefont {P.~J.}\ \bibnamefont {Coles}},\ }\bibfield  {title}
  {\bibinfo {title} {Variational quantum algorithms},\ }\href
  {https://doi.org/10.1038/s42254-021-00348-9} {\bibfield  {journal} {\bibinfo
  {journal} {Nature Reviews Physics}\ }\textbf {\bibinfo {volume} {3}},\
  \bibinfo {pages} {625} (\bibinfo {year} {2021})}\BibitemShut {NoStop}%
\bibitem [{Note1()}]{Note1}%
  \BibitemOpen
  \bibinfo {note} {We refer to an interferometer with a unitary phase encoding
  that belongs to the group of SU(2) as an SU(2) interferometer.}\BibitemShut
  {Stop}%
\bibitem [{\citenamefont {Yang}\ \emph {et~al.}(2022)\citenamefont {Yang},
  \citenamefont {Huelga},\ and\ \citenamefont {Plenio}}]{Yang2022}%
  \BibitemOpen
  \bibfield  {author} {\bibinfo {author} {\bibfnamefont {D.}~\bibnamefont
  {Yang}}, \bibinfo {author} {\bibfnamefont {S.~F.}\ \bibnamefont {Huelga}},\
  and\ \bibinfo {author} {\bibfnamefont {M.~B.}\ \bibnamefont {Plenio}},\
  }\bibfield  {title} {\bibinfo {title} {Efficient information retrieval for
  sensing via continuous measurement},\ }\bibfield  {journal} {\bibinfo
  {journal} {arXiv}\ }\href {https://doi.org/10.48550/ARXIV.2209.08777}
  {10.48550/ARXIV.2209.08777} (\bibinfo {year} {2022})\BibitemShut {NoStop}%
\bibitem [{\citenamefont {Kolenderski}\ and\ \citenamefont
  {Demkowicz-Dobrzanski}(2008)}]{Kolenderski2008}%
  \BibitemOpen
  \bibfield  {author} {\bibinfo {author} {\bibfnamefont {P.}~\bibnamefont
  {Kolenderski}}\ and\ \bibinfo {author} {\bibfnamefont {R.}~\bibnamefont
  {Demkowicz-Dobrzanski}},\ }\bibfield  {title} {\bibinfo {title} {Optimal
  state for keeping reference frames aligned and the platonic solids},\ }\href
  {https://doi.org/10.1103/PhysRevA.78.052333} {\bibfield  {journal} {\bibinfo
  {journal} {Phys. Rev. A}\ }\textbf {\bibinfo {volume} {78}},\ \bibinfo
  {pages} {052333} (\bibinfo {year} {2008})}\BibitemShut {NoStop}%
\bibitem [{\citenamefont {Chiribella}\ \emph
  {et~al.}(2004{\natexlab{b}})\citenamefont {Chiribella}, \citenamefont
  {D'Ariano}, \citenamefont {Perinotti},\ and\ \citenamefont
  {Sacchi}}]{Chiribella2004b}%
  \BibitemOpen
  \bibfield  {author} {\bibinfo {author} {\bibfnamefont {G.}~\bibnamefont
  {Chiribella}}, \bibinfo {author} {\bibfnamefont {G.~M.}\ \bibnamefont
  {D'Ariano}}, \bibinfo {author} {\bibfnamefont {P.}~\bibnamefont
  {Perinotti}},\ and\ \bibinfo {author} {\bibfnamefont {M.~F.}\ \bibnamefont
  {Sacchi}},\ }\bibfield  {title} {\bibinfo {title} {Covariant quantum
  measurements that maximize the likelihood},\ }\href
  {https://doi.org/10.1103/PhysRevA.70.062105} {\bibfield  {journal} {\bibinfo
  {journal} {Phys. Rev. A}\ }\textbf {\bibinfo {volume} {70}},\ \bibinfo
  {pages} {062105} (\bibinfo {year} {2004}{\natexlab{b}})}\BibitemShut
  {NoStop}%
\bibitem [{Note2()}]{Note2}%
  \BibitemOpen
  \bibinfo {note} {Convexity of optimization problem guarantees that every
  local minimum is a global minimum.}\BibitemShut {Stop}%
\bibitem [{Note3()}]{Note3}%
  \BibitemOpen
  \bibinfo {note} {The upper bound on the POVM complexity for $d_N\sim N^3$
  dimensional Hilbert space is given by a POVM comprising $d_N^2\sim N^6$
  measurement operators of full rank. We find the OQS solution to be
  significantly simpler than the upper bound.}\BibitemShut {Stop}%
\bibitem [{\citenamefont {Vasilyev~et al.}(shed)}]{Vasilyev23}%
  \BibitemOpen
  \bibfield  {author} {\bibinfo {author} {\bibfnamefont {D.~V.}\ \bibnamefont
  {Vasilyev~et al.}},\ }\bibfield  {title} {\bibinfo {title} {Optimal
  multi-parameter quantum metrology: Vector field sensing}} (\bibinfo {year}
  {unpublished})\BibitemShut {NoStop}%
\bibitem [{\citenamefont {Chalopin}\ \emph {et~al.}(2018)\citenamefont
  {Chalopin}, \citenamefont {Bouazza}, \citenamefont {Evrard}, \citenamefont
  {Makhalov}, \citenamefont {Dreon}, \citenamefont {Dalibard}, \citenamefont
  {Sidorenkov},\ and\ \citenamefont {Nascimbene}}]{Chalopin2018}%
  \BibitemOpen
  \bibfield  {author} {\bibinfo {author} {\bibfnamefont {T.}~\bibnamefont
  {Chalopin}}, \bibinfo {author} {\bibfnamefont {C.}~\bibnamefont {Bouazza}},
  \bibinfo {author} {\bibfnamefont {A.}~\bibnamefont {Evrard}}, \bibinfo
  {author} {\bibfnamefont {V.}~\bibnamefont {Makhalov}}, \bibinfo {author}
  {\bibfnamefont {D.}~\bibnamefont {Dreon}}, \bibinfo {author} {\bibfnamefont
  {J.}~\bibnamefont {Dalibard}}, \bibinfo {author} {\bibfnamefont {L.~A.}\
  \bibnamefont {Sidorenkov}},\ and\ \bibinfo {author} {\bibfnamefont
  {S.}~\bibnamefont {Nascimbene}},\ }\bibfield  {title} {\bibinfo {title}
  {Quantum-enhanced sensing using non-classical spin states of a highly
  magnetic atom},\ }\href {https://doi.org/10.1038/s41467-018-07433-1}
  {\bibfield  {journal} {\bibinfo  {journal} {Nature Communications}\ }\textbf
  {\bibinfo {volume} {9}},\ \bibinfo {pages} {4955} (\bibinfo {year}
  {2018})}\BibitemShut {NoStop}%
\bibitem [{Note4()}]{Note4}%
  \BibitemOpen
  \bibinfo {note} {Results for small atom numbers are relevant for field
  sensing in scenarios where only few atoms can be used for the sensing task,
  e.g., in order to maintain a good spatial resolution.}\BibitemShut {Stop}%
\bibitem [{\citenamefont {Dowling}\ \emph {et~al.}(1994)\citenamefont
  {Dowling}, \citenamefont {Agarwal},\ and\ \citenamefont
  {Schleich}}]{Dowling1994}%
  \BibitemOpen
  \bibfield  {author} {\bibinfo {author} {\bibfnamefont {J.~P.}\ \bibnamefont
  {Dowling}}, \bibinfo {author} {\bibfnamefont {G.~S.}\ \bibnamefont
  {Agarwal}},\ and\ \bibinfo {author} {\bibfnamefont {W.~P.}\ \bibnamefont
  {Schleich}},\ }\bibfield  {title} {\bibinfo {title} {Wigner distribution of a
  general angular-momentum state: Applications to a collection of two-level
  atoms},\ }\href {https://doi.org/10.1103/PhysRevA.49.4101} {\bibfield
  {journal} {\bibinfo  {journal} {Phys. Rev. A}\ }\textbf {\bibinfo {volume}
  {49}},\ \bibinfo {pages} {4101} (\bibinfo {year} {1994})}\BibitemShut
  {NoStop}%
\bibitem [{\citenamefont {Bollinger}\ \emph {et~al.}(1996)\citenamefont
  {Bollinger}, \citenamefont {Itano}, \citenamefont {Wineland},\ and\
  \citenamefont {Heinzen}}]{Bollinger1996}%
  \BibitemOpen
  \bibfield  {author} {\bibinfo {author} {\bibfnamefont {J.~J.}\ \bibnamefont
  {Bollinger}}, \bibinfo {author} {\bibfnamefont {W.~M.}\ \bibnamefont
  {Itano}}, \bibinfo {author} {\bibfnamefont {D.~J.}\ \bibnamefont
  {Wineland}},\ and\ \bibinfo {author} {\bibfnamefont {D.~J.}\ \bibnamefont
  {Heinzen}},\ }\bibfield  {title} {\bibinfo {title} {Optimal frequency
  measurements with maximally correlated states},\ }\href
  {https://doi.org/10.1103/PhysRevA.54.R4649} {\bibfield  {journal} {\bibinfo
  {journal} {Phys. Rev. A}\ }\textbf {\bibinfo {volume} {54}},\ \bibinfo
  {pages} {R4649} (\bibinfo {year} {1996})}\BibitemShut {NoStop}%
\bibitem [{\citenamefont {Leibfried}\ \emph {et~al.}(2005)\citenamefont
  {Leibfried}, \citenamefont {Knill}, \citenamefont {Seidelin}, \citenamefont
  {Britton}, \citenamefont {Blakestad}, \citenamefont {Chiaverini},
  \citenamefont {Hume}, \citenamefont {Itano}, \citenamefont {Jost},
  \citenamefont {Langer}, \citenamefont {Ozeri}, \citenamefont {Reichle},\ and\
  \citenamefont {Wineland}}]{Leibfried2005}%
  \BibitemOpen
  \bibfield  {author} {\bibinfo {author} {\bibfnamefont {D.}~\bibnamefont
  {Leibfried}}, \bibinfo {author} {\bibfnamefont {E.}~\bibnamefont {Knill}},
  \bibinfo {author} {\bibfnamefont {S.}~\bibnamefont {Seidelin}}, \bibinfo
  {author} {\bibfnamefont {J.}~\bibnamefont {Britton}}, \bibinfo {author}
  {\bibfnamefont {R.~B.}\ \bibnamefont {Blakestad}}, \bibinfo {author}
  {\bibfnamefont {J.}~\bibnamefont {Chiaverini}}, \bibinfo {author}
  {\bibfnamefont {D.~B.}\ \bibnamefont {Hume}}, \bibinfo {author}
  {\bibfnamefont {W.~M.}\ \bibnamefont {Itano}}, \bibinfo {author}
  {\bibfnamefont {J.~D.}\ \bibnamefont {Jost}}, \bibinfo {author}
  {\bibfnamefont {C.}~\bibnamefont {Langer}}, \bibinfo {author} {\bibfnamefont
  {R.}~\bibnamefont {Ozeri}}, \bibinfo {author} {\bibfnamefont
  {R.}~\bibnamefont {Reichle}},\ and\ \bibinfo {author} {\bibfnamefont {D.~J.}\
  \bibnamefont {Wineland}},\ }\bibfield  {title} {\bibinfo {title} {{Creation
  of a six-atom `Schr{\"o}dinger cat' state}},\ }\href
  {https://doi.org/10.1038/nature04251} {\bibfield  {journal} {\bibinfo
  {journal} {Nature}\ }\textbf {\bibinfo {volume} {438}},\ \bibinfo {pages}
  {639} (\bibinfo {year} {2005})}\BibitemShut {NoStop}%
\bibitem [{\citenamefont {Monz}\ \emph {et~al.}(2011)\citenamefont {Monz},
  \citenamefont {Schindler}, \citenamefont {Barreiro}, \citenamefont {Chwalla},
  \citenamefont {Nigg}, \citenamefont {Coish}, \citenamefont {Harlander},
  \citenamefont {H\"ansel}, \citenamefont {Hennrich},\ and\ \citenamefont
  {Blatt}}]{Monz2011}%
  \BibitemOpen
  \bibfield  {author} {\bibinfo {author} {\bibfnamefont {T.}~\bibnamefont
  {Monz}}, \bibinfo {author} {\bibfnamefont {P.}~\bibnamefont {Schindler}},
  \bibinfo {author} {\bibfnamefont {J.~T.}\ \bibnamefont {Barreiro}}, \bibinfo
  {author} {\bibfnamefont {M.}~\bibnamefont {Chwalla}}, \bibinfo {author}
  {\bibfnamefont {D.}~\bibnamefont {Nigg}}, \bibinfo {author} {\bibfnamefont
  {W.~A.}\ \bibnamefont {Coish}}, \bibinfo {author} {\bibfnamefont
  {M.}~\bibnamefont {Harlander}}, \bibinfo {author} {\bibfnamefont
  {W.}~\bibnamefont {H\"ansel}}, \bibinfo {author} {\bibfnamefont
  {M.}~\bibnamefont {Hennrich}},\ and\ \bibinfo {author} {\bibfnamefont
  {R.}~\bibnamefont {Blatt}},\ }\bibfield  {title} {\bibinfo {title} {14-qubit
  entanglement: Creation and coherence},\ }\href
  {https://doi.org/10.1103/PhysRevLett.106.130506} {\bibfield  {journal}
  {\bibinfo  {journal} {Phys. Rev. Lett.}\ }\textbf {\bibinfo {volume} {106}},\
  \bibinfo {pages} {130506} (\bibinfo {year} {2011})}\BibitemShut {NoStop}%
\bibitem [{\citenamefont {Leroux}\ \emph {et~al.}(2010)\citenamefont {Leroux},
  \citenamefont {Schleier-Smith},\ and\ \citenamefont
  {Vuleti\ifmmode~\acute{c}\else \'{c}\fi{}}}]{Leroux2010}%
  \BibitemOpen
  \bibfield  {author} {\bibinfo {author} {\bibfnamefont {I.~D.}\ \bibnamefont
  {Leroux}}, \bibinfo {author} {\bibfnamefont {M.~H.}\ \bibnamefont
  {Schleier-Smith}},\ and\ \bibinfo {author} {\bibfnamefont {V.}~\bibnamefont
  {Vuleti\ifmmode~\acute{c}\else \'{c}\fi{}}},\ }\bibfield  {title} {\bibinfo
  {title} {Implementation of cavity squeezing of a collective atomic spin},\
  }\href {https://doi.org/10.1103/PhysRevLett.104.073602} {\bibfield  {journal}
  {\bibinfo  {journal} {Phys. Rev. Lett.}\ }\textbf {\bibinfo {volume} {104}},\
  \bibinfo {pages} {073602} (\bibinfo {year} {2010})}\BibitemShut {NoStop}%
\bibitem [{\citenamefont {Riedel}\ \emph {et~al.}(2010)\citenamefont {Riedel},
  \citenamefont {B{\"o}hi}, \citenamefont {Li}, \citenamefont {H{\"a}nsch},
  \citenamefont {Sinatra},\ and\ \citenamefont {Treutlein}}]{Riedel2010}%
  \BibitemOpen
  \bibfield  {author} {\bibinfo {author} {\bibfnamefont {M.~F.}\ \bibnamefont
  {Riedel}}, \bibinfo {author} {\bibfnamefont {P.}~\bibnamefont {B{\"o}hi}},
  \bibinfo {author} {\bibfnamefont {Y.}~\bibnamefont {Li}}, \bibinfo {author}
  {\bibfnamefont {T.~W.}\ \bibnamefont {H{\"a}nsch}}, \bibinfo {author}
  {\bibfnamefont {A.}~\bibnamefont {Sinatra}},\ and\ \bibinfo {author}
  {\bibfnamefont {P.}~\bibnamefont {Treutlein}},\ }\bibfield  {title} {\bibinfo
  {title} {Atom-chip-based generation of entanglement for quantum metrology},\
  }\href {https://doi.org/10.1038/nature08988} {\bibfield  {journal} {\bibinfo
  {journal} {Nature}\ }\textbf {\bibinfo {volume} {464}},\ \bibinfo {pages}
  {1170} (\bibinfo {year} {2010})}\BibitemShut {NoStop}%
\bibitem [{\citenamefont {Gross}\ \emph {et~al.}(2010)\citenamefont {Gross},
  \citenamefont {Zibold}, \citenamefont {Nicklas}, \citenamefont {Est{\`e}ve},\
  and\ \citenamefont {Oberthaler}}]{Gross2010}%
  \BibitemOpen
  \bibfield  {author} {\bibinfo {author} {\bibfnamefont {C.}~\bibnamefont
  {Gross}}, \bibinfo {author} {\bibfnamefont {T.}~\bibnamefont {Zibold}},
  \bibinfo {author} {\bibfnamefont {E.}~\bibnamefont {Nicklas}}, \bibinfo
  {author} {\bibfnamefont {J.}~\bibnamefont {Est{\`e}ve}},\ and\ \bibinfo
  {author} {\bibfnamefont {M.~K.}\ \bibnamefont {Oberthaler}},\ }\bibfield
  {title} {\bibinfo {title} {Nonlinear atom interferometer surpasses classical
  precision limit},\ }\href {https://doi.org/10.1038/nature08919} {\bibfield
  {journal} {\bibinfo  {journal} {Nature}\ }\textbf {\bibinfo {volume} {464}},\
  \bibinfo {pages} {1165} (\bibinfo {year} {2010})}\BibitemShut {NoStop}%
\bibitem [{Note5()}]{Note5}%
  \BibitemOpen
  \bibinfo {note} {Note that extensions of this gate set, i.e. allowing for
  twisting around arbitrary axes~\cite {Schulte2020, Thurtell2022} can lead to
  more efficient parameterization in terms of the number of gates required to
  achieve a certain performance.}\BibitemShut {Stop}%
\bibitem [{\citenamefont {Zimba}(2006)}]{Zimba2006}%
  \BibitemOpen
  \bibfield  {author} {\bibinfo {author} {\bibfnamefont {J.}~\bibnamefont
  {Zimba}},\ }\bibfield  {title} {\bibinfo {title} {Anticoherent spin states
  via the majorana representation},\ }\href@noop {} {\bibfield  {journal}
  {\bibinfo  {journal} {EJTP}\ }\textbf {\bibinfo {volume} {3}},\ \bibinfo
  {pages} {143} (\bibinfo {year} {2006})}\BibitemShut {NoStop}%
\bibitem [{\citenamefont {Giraud}\ \emph {et~al.}(2010)\citenamefont {Giraud},
  \citenamefont {Braun},\ and\ \citenamefont {Braun}}]{Giraud2010}%
  \BibitemOpen
  \bibfield  {author} {\bibinfo {author} {\bibfnamefont {O.}~\bibnamefont
  {Giraud}}, \bibinfo {author} {\bibfnamefont {P.}~\bibnamefont {Braun}},\ and\
  \bibinfo {author} {\bibfnamefont {D.}~\bibnamefont {Braun}},\ }\bibfield
  {title} {\bibinfo {title} {Quantifying quantumness and the quest for queens
  of quantum},\ }\href {https://doi.org/10.1088/1367-2630/12/6/063005}
  {\bibfield  {journal} {\bibinfo  {journal} {New Journal of Physics}\ }\textbf
  {\bibinfo {volume} {12}},\ \bibinfo {pages} {063005} (\bibinfo {year}
  {2010})}\BibitemShut {NoStop}%
\bibitem [{\citenamefont {Bj\"ork}\ \emph {et~al.}(2015)\citenamefont
  {Bj\"ork}, \citenamefont {Klimov}, \citenamefont {de~la Hoz}, \citenamefont
  {Grassl}, \citenamefont {Leuchs},\ and\ \citenamefont
  {S\'anchez-Soto}}]{Bjork2015}%
  \BibitemOpen
  \bibfield  {author} {\bibinfo {author} {\bibfnamefont {G.}~\bibnamefont
  {Bj\"ork}}, \bibinfo {author} {\bibfnamefont {A.~B.}\ \bibnamefont {Klimov}},
  \bibinfo {author} {\bibfnamefont {P.}~\bibnamefont {de~la Hoz}}, \bibinfo
  {author} {\bibfnamefont {M.}~\bibnamefont {Grassl}}, \bibinfo {author}
  {\bibfnamefont {G.}~\bibnamefont {Leuchs}},\ and\ \bibinfo {author}
  {\bibfnamefont {L.~L.}\ \bibnamefont {S\'anchez-Soto}},\ }\bibfield  {title}
  {\bibinfo {title} {Extremal quantum states and their majorana
  constellations},\ }\href {https://doi.org/10.1103/PhysRevA.92.031801}
  {\bibfield  {journal} {\bibinfo  {journal} {Phys. Rev. A}\ }\textbf {\bibinfo
  {volume} {92}},\ \bibinfo {pages} {031801(R)} (\bibinfo {year}
  {2015})}\BibitemShut {NoStop}%
\bibitem [{\citenamefont {Young}\ \emph {et~al.}(2020)\citenamefont {Young},
  \citenamefont {Eckner}, \citenamefont {Milner}, \citenamefont {Kedar},
  \citenamefont {Norcia}, \citenamefont {Oelker}, \citenamefont {Schine},
  \citenamefont {Ye},\ and\ \citenamefont {Kaufman}}]{Young2020}%
  \BibitemOpen
  \bibfield  {author} {\bibinfo {author} {\bibfnamefont {A.~W.}\ \bibnamefont
  {Young}}, \bibinfo {author} {\bibfnamefont {W.~J.}\ \bibnamefont {Eckner}},
  \bibinfo {author} {\bibfnamefont {W.~R.}\ \bibnamefont {Milner}}, \bibinfo
  {author} {\bibfnamefont {D.}~\bibnamefont {Kedar}}, \bibinfo {author}
  {\bibfnamefont {M.~A.}\ \bibnamefont {Norcia}}, \bibinfo {author}
  {\bibfnamefont {E.}~\bibnamefont {Oelker}}, \bibinfo {author} {\bibfnamefont
  {N.}~\bibnamefont {Schine}}, \bibinfo {author} {\bibfnamefont
  {J.}~\bibnamefont {Ye}},\ and\ \bibinfo {author} {\bibfnamefont {A.~M.}\
  \bibnamefont {Kaufman}},\ }\bibfield  {title} {\bibinfo {title}
  {Half-minute-scale atomic coherence and high relative stability in a tweezer
  clock},\ }\href {https://doi.org/10.1038/s41586-020-3009-y} {\bibfield
  {journal} {\bibinfo  {journal} {Nature}\ }\textbf {\bibinfo {volume} {588}},\
  \bibinfo {pages} {408} (\bibinfo {year} {2020})}\BibitemShut {NoStop}%
\bibitem [{\citenamefont {Graham}\ \emph {et~al.}(2019)\citenamefont {Graham},
  \citenamefont {Kwon}, \citenamefont {Grinkemeyer}, \citenamefont {Marra},
  \citenamefont {Jiang}, \citenamefont {Lichtman}, \citenamefont {Sun},
  \citenamefont {Ebert},\ and\ \citenamefont {Saffman}}]{Graham2019}%
  \BibitemOpen
  \bibfield  {author} {\bibinfo {author} {\bibfnamefont {T.~M.}\ \bibnamefont
  {Graham}}, \bibinfo {author} {\bibfnamefont {M.}~\bibnamefont {Kwon}},
  \bibinfo {author} {\bibfnamefont {B.}~\bibnamefont {Grinkemeyer}}, \bibinfo
  {author} {\bibfnamefont {Z.}~\bibnamefont {Marra}}, \bibinfo {author}
  {\bibfnamefont {X.}~\bibnamefont {Jiang}}, \bibinfo {author} {\bibfnamefont
  {M.~T.}\ \bibnamefont {Lichtman}}, \bibinfo {author} {\bibfnamefont
  {Y.}~\bibnamefont {Sun}}, \bibinfo {author} {\bibfnamefont {M.}~\bibnamefont
  {Ebert}},\ and\ \bibinfo {author} {\bibfnamefont {M.}~\bibnamefont
  {Saffman}},\ }\bibfield  {title} {\bibinfo {title} {Rydberg-mediated
  entanglement in a two-dimensional neutral atom qubit array},\ }\href
  {https://doi.org/10.1103/PhysRevLett.123.230501} {\bibfield  {journal}
  {\bibinfo  {journal} {Phys. Rev. Lett.}\ }\textbf {\bibinfo {volume} {123}},\
  \bibinfo {pages} {230501} (\bibinfo {year} {2019})}\BibitemShut {NoStop}%
\bibitem [{\citenamefont {Madjarov}\ \emph {et~al.}(2020)\citenamefont
  {Madjarov}, \citenamefont {Covey}, \citenamefont {Shaw}, \citenamefont
  {Choi}, \citenamefont {Kale}, \citenamefont {Cooper}, \citenamefont
  {Pichler}, \citenamefont {Schkolnik}, \citenamefont {Williams},\ and\
  \citenamefont {Endres}}]{Madjarov2020}%
  \BibitemOpen
  \bibfield  {author} {\bibinfo {author} {\bibfnamefont {I.~S.}\ \bibnamefont
  {Madjarov}}, \bibinfo {author} {\bibfnamefont {J.~P.}\ \bibnamefont {Covey}},
  \bibinfo {author} {\bibfnamefont {A.~L.}\ \bibnamefont {Shaw}}, \bibinfo
  {author} {\bibfnamefont {J.}~\bibnamefont {Choi}}, \bibinfo {author}
  {\bibfnamefont {A.}~\bibnamefont {Kale}}, \bibinfo {author} {\bibfnamefont
  {A.}~\bibnamefont {Cooper}}, \bibinfo {author} {\bibfnamefont
  {H.}~\bibnamefont {Pichler}}, \bibinfo {author} {\bibfnamefont
  {V.}~\bibnamefont {Schkolnik}}, \bibinfo {author} {\bibfnamefont {J.~R.}\
  \bibnamefont {Williams}},\ and\ \bibinfo {author} {\bibfnamefont
  {M.}~\bibnamefont {Endres}},\ }\bibfield  {title} {\bibinfo {title}
  {High-fidelity entanglement and detection of alkaline-earth rydberg atoms},\
  }\href {https://doi.org/10.1038/s41567-020-0903-z} {\bibfield  {journal}
  {\bibinfo  {journal} {Nature Physics}\ }\textbf {\bibinfo {volume} {16}},\
  \bibinfo {pages} {857} (\bibinfo {year} {2020})}\BibitemShut {NoStop}%
\bibitem [{\citenamefont {Schine}\ \emph {et~al.}(2022)\citenamefont {Schine},
  \citenamefont {Young}, \citenamefont {Eckner}, \citenamefont {Martin},\ and\
  \citenamefont {Kaufman}}]{Schine2022}%
  \BibitemOpen
  \bibfield  {author} {\bibinfo {author} {\bibfnamefont {N.}~\bibnamefont
  {Schine}}, \bibinfo {author} {\bibfnamefont {A.~W.}\ \bibnamefont {Young}},
  \bibinfo {author} {\bibfnamefont {W.~J.}\ \bibnamefont {Eckner}}, \bibinfo
  {author} {\bibfnamefont {M.~J.}\ \bibnamefont {Martin}},\ and\ \bibinfo
  {author} {\bibfnamefont {A.~M.}\ \bibnamefont {Kaufman}},\ }\bibfield
  {title} {\bibinfo {title} {Long-lived bell states in an array of optical
  clock qubits},\ }\href {https://doi.org/10.1038/s41567-022-01678-w}
  {\bibfield  {journal} {\bibinfo  {journal} {Nature Physics}\ }\textbf
  {\bibinfo {volume} {18}},\ \bibinfo {pages} {1067} (\bibinfo {year}
  {2022})}\BibitemShut {NoStop}%
\bibitem [{\citenamefont {Kaubruegger}\ \emph {et~al.}(2019)\citenamefont
  {Kaubruegger}, \citenamefont {Silvi}, \citenamefont {Kokail}, \citenamefont
  {van Bijnen}, \citenamefont {Rey}, \citenamefont {Ye}, \citenamefont
  {Kaufman},\ and\ \citenamefont {Zoller}}]{Kaubruegger2019}%
  \BibitemOpen
  \bibfield  {author} {\bibinfo {author} {\bibfnamefont {R.}~\bibnamefont
  {Kaubruegger}}, \bibinfo {author} {\bibfnamefont {P.}~\bibnamefont {Silvi}},
  \bibinfo {author} {\bibfnamefont {C.}~\bibnamefont {Kokail}}, \bibinfo
  {author} {\bibfnamefont {R.}~\bibnamefont {van Bijnen}}, \bibinfo {author}
  {\bibfnamefont {A.~M.}\ \bibnamefont {Rey}}, \bibinfo {author} {\bibfnamefont
  {J.}~\bibnamefont {Ye}}, \bibinfo {author} {\bibfnamefont {A.~M.}\
  \bibnamefont {Kaufman}},\ and\ \bibinfo {author} {\bibfnamefont
  {P.}~\bibnamefont {Zoller}},\ }\bibfield  {title} {\bibinfo {title}
  {Variational spin-squeezing algorithms on programmable quantum sensors},\
  }\href {https://doi.org/10.1103/PhysRevLett.123.260505} {\bibfield  {journal}
  {\bibinfo  {journal} {Phys. Rev. Lett.}\ }\textbf {\bibinfo {volume} {123}},\
  \bibinfo {pages} {260505} (\bibinfo {year} {2019})}\BibitemShut {NoStop}%
\bibitem [{\citenamefont {Koczor}\ \emph {et~al.}(2020)\citenamefont {Koczor},
  \citenamefont {Endo}, \citenamefont {Jones}, \citenamefont {Matsuzaki},\ and\
  \citenamefont {Benjamin}}]{Koczor2020}%
  \BibitemOpen
  \bibfield  {author} {\bibinfo {author} {\bibfnamefont {B.}~\bibnamefont
  {Koczor}}, \bibinfo {author} {\bibfnamefont {S.}~\bibnamefont {Endo}},
  \bibinfo {author} {\bibfnamefont {T.}~\bibnamefont {Jones}}, \bibinfo
  {author} {\bibfnamefont {Y.}~\bibnamefont {Matsuzaki}},\ and\ \bibinfo
  {author} {\bibfnamefont {S.~C.}\ \bibnamefont {Benjamin}},\ }\bibfield
  {title} {\bibinfo {title} {Variational-state quantum metrology},\ }\href
  {https://doi.org/10.1088/1367-2630/ab965e} {\bibfield  {journal} {\bibinfo
  {journal} {New Journal of Physics}\ }\textbf {\bibinfo {volume} {22}},\
  \bibinfo {pages} {083038} (\bibinfo {year} {2020})}\BibitemShut {NoStop}%
\bibitem [{\citenamefont {Zheng}\ \emph {et~al.}(2022)\citenamefont {Zheng},
  \citenamefont {Li}, \citenamefont {Rosen}, \citenamefont {Zhou},
  \citenamefont {Koppenh{\"o}fer}, \citenamefont {Ma}, \citenamefont {Chong},
  \citenamefont {Clerk}, \citenamefont {Jiang},\ and\ \citenamefont
  {Maurer}}]{Zheng2022}%
  \BibitemOpen
  \bibfield  {author} {\bibinfo {author} {\bibfnamefont {T.-X.}\ \bibnamefont
  {Zheng}}, \bibinfo {author} {\bibfnamefont {A.}~\bibnamefont {Li}}, \bibinfo
  {author} {\bibfnamefont {J.}~\bibnamefont {Rosen}}, \bibinfo {author}
  {\bibfnamefont {S.}~\bibnamefont {Zhou}}, \bibinfo {author} {\bibfnamefont
  {M.}~\bibnamefont {Koppenh{\"o}fer}}, \bibinfo {author} {\bibfnamefont
  {Z.}~\bibnamefont {Ma}}, \bibinfo {author} {\bibfnamefont {F.~T.}\
  \bibnamefont {Chong}}, \bibinfo {author} {\bibfnamefont {A.~A.}\ \bibnamefont
  {Clerk}}, \bibinfo {author} {\bibfnamefont {L.}~\bibnamefont {Jiang}},\ and\
  \bibinfo {author} {\bibfnamefont {P.~C.}\ \bibnamefont {Maurer}},\ }\bibfield
   {title} {\bibinfo {title} {Preparation of metrological states in
  dipolar-interacting spin systems},\ }\href
  {https://doi.org/10.1038/s41534-022-00667-4} {\bibfield  {journal} {\bibinfo
  {journal} {npj Quantum Information}\ }\textbf {\bibinfo {volume} {8}},\
  \bibinfo {pages} {150} (\bibinfo {year} {2022})}\BibitemShut {NoStop}%
\bibitem [{\citenamefont {Kruckenhauser}\ \emph {et~al.}(2022)\citenamefont
  {Kruckenhauser}, \citenamefont {van Bijnen}, \citenamefont {Zache},
  \citenamefont {Liberto},\ and\ \citenamefont {Zoller}}]{Kruckenhauser2023}%
  \BibitemOpen
  \bibfield  {author} {\bibinfo {author} {\bibfnamefont {A.}~\bibnamefont
  {Kruckenhauser}}, \bibinfo {author} {\bibfnamefont {R.}~\bibnamefont {van
  Bijnen}}, \bibinfo {author} {\bibfnamefont {T.~V.}\ \bibnamefont {Zache}},
  \bibinfo {author} {\bibfnamefont {M.~D.}\ \bibnamefont {Liberto}},\ and\
  \bibinfo {author} {\bibfnamefont {P.}~\bibnamefont {Zoller}},\ }\bibfield
  {title} {\bibinfo {title} {High-dimensional so(4)-symmetric rydberg manifolds
  for quantum simulation},\ }\href {https://doi.org/10.1088/2058-9565/aca996}
  {\bibfield  {journal} {\bibinfo  {journal} {Quantum Science and Technology}\
  }\textbf {\bibinfo {volume} {8}},\ \bibinfo {pages} {015020} (\bibinfo {year}
  {2022})}\BibitemShut {NoStop}%
\bibitem [{\citenamefont {Gilmore}\ \emph {et~al.}(2021)\citenamefont
  {Gilmore}, \citenamefont {Affolter}, \citenamefont {Lewis-Swan},
  \citenamefont {Barberena}, \citenamefont {Jordan}, \citenamefont {Rey},\ and\
  \citenamefont {Bollinger}}]{Gilmore2021}%
  \BibitemOpen
  \bibfield  {author} {\bibinfo {author} {\bibfnamefont {K.~A.}\ \bibnamefont
  {Gilmore}}, \bibinfo {author} {\bibfnamefont {M.}~\bibnamefont {Affolter}},
  \bibinfo {author} {\bibfnamefont {R.~J.}\ \bibnamefont {Lewis-Swan}},
  \bibinfo {author} {\bibfnamefont {D.}~\bibnamefont {Barberena}}, \bibinfo
  {author} {\bibfnamefont {E.}~\bibnamefont {Jordan}}, \bibinfo {author}
  {\bibfnamefont {A.~M.}\ \bibnamefont {Rey}},\ and\ \bibinfo {author}
  {\bibfnamefont {J.~J.}\ \bibnamefont {Bollinger}},\ }\bibfield  {title}
  {\bibinfo {title} {Quantum-enhanced sensing of displacements and electric
  fields with two-dimensional trapped-ion crystals},\ }\href
  {https://doi.org/10.1126/science.abi5226} {\bibfield  {journal} {\bibinfo
  {journal} {Science}\ }\textbf {\bibinfo {volume} {373}},\ \bibinfo {pages}
  {673} (\bibinfo {year} {2021})},\ \Eprint
  {https://arxiv.org/abs/https://www.science.org/doi/pdf/10.1126/science.abi5226}
  {https://www.science.org/doi/pdf/10.1126/science.abi5226} \BibitemShut
  {NoStop}%
\bibitem [{\citenamefont {Ringbauer}\ \emph {et~al.}(2022)\citenamefont
  {Ringbauer}, \citenamefont {Meth}, \citenamefont {Postler}, \citenamefont
  {Stricker}, \citenamefont {Blatt}, \citenamefont {Schindler},\ and\
  \citenamefont {Monz}}]{Ringbauer2022}%
  \BibitemOpen
  \bibfield  {author} {\bibinfo {author} {\bibfnamefont {M.}~\bibnamefont
  {Ringbauer}}, \bibinfo {author} {\bibfnamefont {M.}~\bibnamefont {Meth}},
  \bibinfo {author} {\bibfnamefont {L.}~\bibnamefont {Postler}}, \bibinfo
  {author} {\bibfnamefont {R.}~\bibnamefont {Stricker}}, \bibinfo {author}
  {\bibfnamefont {R.}~\bibnamefont {Blatt}}, \bibinfo {author} {\bibfnamefont
  {P.}~\bibnamefont {Schindler}},\ and\ \bibinfo {author} {\bibfnamefont
  {T.}~\bibnamefont {Monz}},\ }\bibfield  {title} {\bibinfo {title} {A
  universal qudit quantum processor with trapped ions},\ }\href
  {https://doi.org/10.1038/s41567-022-01658-0} {\bibfield  {journal} {\bibinfo
  {journal} {Nature Physics}\ }\textbf {\bibinfo {volume} {18}},\ \bibinfo
  {pages} {1053} (\bibinfo {year} {2022})}\BibitemShut {NoStop}%
\bibitem [{\citenamefont {Saffman}\ and\ \citenamefont
  {M\o{}lmer}(2008)}]{Saffman2008}%
  \BibitemOpen
  \bibfield  {author} {\bibinfo {author} {\bibfnamefont {M.}~\bibnamefont
  {Saffman}}\ and\ \bibinfo {author} {\bibfnamefont {K.}~\bibnamefont
  {M\o{}lmer}},\ }\bibfield  {title} {\bibinfo {title} {Scaling the
  neutral-atom rydberg gate quantum computer by collective encoding in holmium
  atoms},\ }\href {https://doi.org/10.1103/PhysRevA.78.012336} {\bibfield
  {journal} {\bibinfo  {journal} {Phys. Rev. A}\ }\textbf {\bibinfo {volume}
  {78}},\ \bibinfo {pages} {012336} (\bibinfo {year} {2008})}\BibitemShut
  {NoStop}%
\bibitem [{\citenamefont {Robicheaux}\ \emph {et~al.}(2018)\citenamefont
  {Robicheaux}, \citenamefont {Booth},\ and\ \citenamefont
  {Saffman}}]{Robicheaux2018}%
  \BibitemOpen
  \bibfield  {author} {\bibinfo {author} {\bibfnamefont {F.}~\bibnamefont
  {Robicheaux}}, \bibinfo {author} {\bibfnamefont {D.~W.}\ \bibnamefont
  {Booth}},\ and\ \bibinfo {author} {\bibfnamefont {M.}~\bibnamefont
  {Saffman}},\ }\bibfield  {title} {\bibinfo {title} {Theory of long-range
  interactions for rydberg states attached to hyperfine-split cores},\ }\href
  {https://doi.org/10.1103/PhysRevA.97.022508} {\bibfield  {journal} {\bibinfo
  {journal} {Phys. Rev. A}\ }\textbf {\bibinfo {volume} {97}},\ \bibinfo
  {pages} {022508} (\bibinfo {year} {2018})}\BibitemShut {NoStop}%
\bibitem [{\citenamefont {Patscheider}\ \emph {et~al.}(2020)\citenamefont
  {Patscheider}, \citenamefont {Zhu}, \citenamefont {Chomaz}, \citenamefont
  {Petter}, \citenamefont {Baier}, \citenamefont {Rey}, \citenamefont
  {Ferlaino},\ and\ \citenamefont {Mark}}]{Patscheider2020}%
  \BibitemOpen
  \bibfield  {author} {\bibinfo {author} {\bibfnamefont {A.}~\bibnamefont
  {Patscheider}}, \bibinfo {author} {\bibfnamefont {B.}~\bibnamefont {Zhu}},
  \bibinfo {author} {\bibfnamefont {L.}~\bibnamefont {Chomaz}}, \bibinfo
  {author} {\bibfnamefont {D.}~\bibnamefont {Petter}}, \bibinfo {author}
  {\bibfnamefont {S.}~\bibnamefont {Baier}}, \bibinfo {author} {\bibfnamefont
  {A.-M.}\ \bibnamefont {Rey}}, \bibinfo {author} {\bibfnamefont
  {F.}~\bibnamefont {Ferlaino}},\ and\ \bibinfo {author} {\bibfnamefont
  {M.~J.}\ \bibnamefont {Mark}},\ }\bibfield  {title} {\bibinfo {title}
  {Controlling dipolar exchange interactions in a dense three-dimensional array
  of large-spin fermions},\ }\href
  {https://doi.org/10.1103/PhysRevResearch.2.023050} {\bibfield  {journal}
  {\bibinfo  {journal} {Phys. Rev. Res.}\ }\textbf {\bibinfo {volume} {2}},\
  \bibinfo {pages} {023050} (\bibinfo {year} {2020})}\BibitemShut {NoStop}%
\bibitem [{\citenamefont {Barndorff-Nielsen}\ and\ \citenamefont
  {Gill}(2000)}]{Barndorff2000}%
  \BibitemOpen
  \bibfield  {author} {\bibinfo {author} {\bibfnamefont {O.}~\bibnamefont
  {Barndorff-Nielsen}}\ and\ \bibinfo {author} {\bibfnamefont {R.~D.}\
  \bibnamefont {Gill}},\ }\bibfield  {title} {\bibinfo {title} {Fisher
  information in quantum statistics},\ }\href
  {https://doi.org/10.1088/0305-4470/33/24/306} {\bibfield  {journal} {\bibinfo
   {journal} {Journal of Physics A: Mathematical and General}\ }\textbf
  {\bibinfo {volume} {33}},\ \bibinfo {pages} {4481} (\bibinfo {year}
  {2000})}\BibitemShut {NoStop}%
\bibitem [{\citenamefont {Trees}(1968)}]{Trees1968}%
  \BibitemOpen
  \bibfield  {author} {\bibinfo {author} {\bibfnamefont {H.~L.~V.}\
  \bibnamefont {Trees}},\ }\href@noop {} {\emph {\bibinfo {title} {Detection,
  Estimation and Modulation}}}\ (\bibinfo  {publisher} {Wiley},\ \bibinfo
  {address} {New York},\ \bibinfo {year} {1968})\BibitemShut {NoStop}%
\bibitem [{\citenamefont {Fulton}\ and\ \citenamefont
  {Harris}(2004)}]{Fulton2004}%
  \BibitemOpen
  \bibfield  {author} {\bibinfo {author} {\bibfnamefont {W.}~\bibnamefont
  {Fulton}}\ and\ \bibinfo {author} {\bibfnamefont {J.}~\bibnamefont
  {Harris}},\ }\href {https://doi.org/10.1007/978-1-4612-0979-9} {\emph
  {\bibinfo {title} {{Representation Theory}}}},\ \bibinfo {series} {Graduate
  Texts in Mathematics}, Vol.\ \bibinfo {volume} {129}\ (\bibinfo  {publisher}
  {Springer New York},\ \bibinfo {address} {New York, NY},\ \bibinfo {year}
  {2004})\BibitemShut {NoStop}%
\bibitem [{\citenamefont {Cirac}\ \emph {et~al.}(1999)\citenamefont {Cirac},
  \citenamefont {Ekert},\ and\ \citenamefont {Macchiavello}}]{Cirac1999}%
  \BibitemOpen
  \bibfield  {author} {\bibinfo {author} {\bibfnamefont {J.~I.}\ \bibnamefont
  {Cirac}}, \bibinfo {author} {\bibfnamefont {A.~K.}\ \bibnamefont {Ekert}},\
  and\ \bibinfo {author} {\bibfnamefont {C.}~\bibnamefont {Macchiavello}},\
  }\bibfield  {title} {\bibinfo {title} {{Optimal Purification of Single
  Qubits}},\ }\href {https://doi.org/10.1103/PhysRevLett.82.4344} {\bibfield
  {journal} {\bibinfo  {journal} {Physical Review Letters}\ }\textbf {\bibinfo
  {volume} {82}},\ \bibinfo {pages} {4344} (\bibinfo {year}
  {1999})}\BibitemShut {NoStop}%
\bibitem [{\citenamefont {Shen}\ \emph {et~al.}(2017)\citenamefont {Shen},
  \citenamefont {Diamond}, \citenamefont {Udell}, \citenamefont {Gu},\ and\
  \citenamefont {Boyd}}]{Shen2017}%
  \BibitemOpen
  \bibfield  {author} {\bibinfo {author} {\bibfnamefont {X.}~\bibnamefont
  {Shen}}, \bibinfo {author} {\bibfnamefont {S.}~\bibnamefont {Diamond}},
  \bibinfo {author} {\bibfnamefont {M.}~\bibnamefont {Udell}}, \bibinfo
  {author} {\bibfnamefont {Y.}~\bibnamefont {Gu}},\ and\ \bibinfo {author}
  {\bibfnamefont {S.}~\bibnamefont {Boyd}},\ }\bibfield  {title} {\bibinfo
  {title} {{Disciplined multi-convex programming}},\ }in\ \href
  {https://doi.org/10.1109/CCDC.2017.7978647} {\emph {\bibinfo {booktitle}
  {{2017 29th Chinese Control And Decision Conference (CCDC)}}}}\ (\bibinfo
  {publisher} {IEEE},\ \bibinfo {year} {2017})\ pp.\ \bibinfo {pages}
  {895--900}\BibitemShut {NoStop}%
\bibitem [{Note6()}]{Note6}%
  \BibitemOpen
  \bibinfo {note} {We note that due to the resulting low rank of the optimal
  measurement operators $\{M^{\star }_q\}$, it is beneficial to use the
  nonlinear programming with augmented Lagrangian algorithm~\protect \citep
  {Burer2003} for solving the SDP in subproblem~(ii).}\BibitemShut {Stop}%
\bibitem [{\citenamefont {Schulte}\ \emph {et~al.}(2020)\citenamefont
  {Schulte}, \citenamefont {Mart{\'{i}}nez-Lahuerta}, \citenamefont
  {Scharnagl},\ and\ \citenamefont {Hammerer}}]{Schulte2020}%
  \BibitemOpen
  \bibfield  {author} {\bibinfo {author} {\bibfnamefont {M.}~\bibnamefont
  {Schulte}}, \bibinfo {author} {\bibfnamefont {V.~J.}\ \bibnamefont
  {Mart{\'{i}}nez-Lahuerta}}, \bibinfo {author} {\bibfnamefont {M.~S.}\
  \bibnamefont {Scharnagl}},\ and\ \bibinfo {author} {\bibfnamefont
  {K.}~\bibnamefont {Hammerer}},\ }\bibfield  {title} {\bibinfo {title} {Ramsey
  interferometry with generalized one-axis twisting echoes},\ }\href
  {https://doi.org/10.22331/q-2020-05-15-268} {\bibfield  {journal} {\bibinfo
  {journal} {{Quantum}}\ }\textbf {\bibinfo {volume} {4}},\ \bibinfo {pages}
  {268} (\bibinfo {year} {2020})}\BibitemShut {NoStop}%
\bibitem [{\citenamefont {Thurtell}\ and\ \citenamefont
  {Miyake}(2022)}]{Thurtell2022}%
  \BibitemOpen
  \bibfield  {author} {\bibinfo {author} {\bibfnamefont {T.~G.}\ \bibnamefont
  {Thurtell}}\ and\ \bibinfo {author} {\bibfnamefont {A.}~\bibnamefont
  {Miyake}},\ }\bibfield  {title} {\bibinfo {title} {Optimizing one-axis twists
  for realistic variational bayesian quantum metrology},\ }\bibfield  {journal}
  {\bibinfo  {journal} {arXiv}\ }\href
  {https://doi.org/10.48550/ARXIV.2212.12461} {10.48550/ARXIV.2212.12461}
  (\bibinfo {year} {2022})\BibitemShut {NoStop}%
\bibitem [{\citenamefont {Burer}\ and\ \citenamefont
  {Monteiro}(2003)}]{Burer2003}%
  \BibitemOpen
  \bibfield  {author} {\bibinfo {author} {\bibfnamefont {S.}~\bibnamefont
  {Burer}}\ and\ \bibinfo {author} {\bibfnamefont {R.~D.}\ \bibnamefont
  {Monteiro}},\ }\bibfield  {title} {\bibinfo {title} {{A nonlinear programming
  algorithm for solving semidefinite programs via low-rank factorization}},\
  }\href {https://doi.org/10.1007/s10107-002-0352-8} {\bibfield  {journal}
  {\bibinfo  {journal} {Mathematical Programming}\ }\textbf {\bibinfo {volume}
  {95}},\ \bibinfo {pages} {329} (\bibinfo {year} {2003})}\BibitemShut
  {NoStop}%
\end{thebibliography}
\end{document}